\documentclass[12pt]{article}
\pdfoutput=1

\usepackage[usenames]{color}

\usepackage{colortbl}
\definecolor{lightgreen}{cmyk}{0.2, 0, 0.2, 0.2}
\definecolor{lightgray}{cmyk}{0.1,0.2,0,0.1}
\definecolor{lightgray2}{cmyk}{0.1,0.1,0,0.1}

\usepackage{booktabs}

\newcommand{\otoprule}{\midrule[\heavyrulewidth]}

\newcommand{\Feyn}[1]{#1\kern-0.45em/}

\usepackage[english]{babel}
\usepackage{amsmath,amssymb,amsbsy,amstext}
\usepackage{graphicx}
\usepackage{amsfonts}
\usepackage{amssymb}
\usepackage{subfigure}
\usepackage{color}
\usepackage{float}
\usepackage[small]{caption}

\numberwithin{equation}{section}

\def\beq{\begin{equation}}
\def\eeq{\end{equation}}

\def\bea{\begin{eqnarray}}
\def\eea{\end{eqnarray}}
\def\d{{\rm d}}
\def\dd{{\rm d}}
\def\Tr{{\rm Tr}}

\DeclareRobustCommand{\SkipTocEntry}[4]{}

\begin{document}

\begin{titlepage}

\setcounter{page}{1} \baselineskip=15.5pt \thispagestyle{empty}


\bigskip\
\begin{center}

{\LARGE  \bf D3-brane Potentials from Fluxes in AdS/CFT}
\vskip 15pt
\end{center}

\vspace{0.5cm}
\begin{center}
{\large Daniel Baumann,$^{1}$ Anatoly Dymarsky,$^{1}$ Shamit Kachru,$^{2,3,*}$}
\vskip 6pt
{\large Igor R. Klebanov,$^{4,5}$ and Liam McAllister$^{6}$}
\end{center}

\vspace{0.3cm}

\begin{center}

\textsl{${}^1$ \small School of Natural Sciences,
 Institute for Advanced Study,
Princeton, NJ 08540}



\vskip 4pt
\textsl{${}^2$ \small Kavli Institute for Theoretical Physics, Santa Barbara, CA 93106}\\

\vskip 4pt
\textsl{${}^3$ \small Department of Physics, University of California, Santa Barbara, CA 93106}\\

\vskip 4pt
\textsl{${}^4$ \small Department of Physics, Princeton University,
Princeton, NJ 08544}\\

\vskip 4pt
\textsl{${}^5$
\small Center for Theoretical Science, Princeton University,
Princeton, NJ 08544}\\

\vskip 4pt
\textsl{${}^6$ \small Department of Physics, Cornell University,
Ithaca, NY 14853}
\end{center} 


\vspace{1.2cm}
\hrule \vspace{0.3cm}
{\small  \noindent \textbf{Abstract} \\[0.2cm]
\noindent
We give a comprehensive treatment of the scalar potential for a D3-brane in a warped conifold region of a compactification with stabilized moduli.  By studying general ultraviolet perturbations in supergravity, we systematically incorporate `compactification effects' sourced by supersymmetry breaking in the compact space.  Significant contributions to the D3-brane potential, including the leading term in the infrared, arise from imaginary anti-self-dual (IASD) fluxes.
For an arbitrary Calabi-Yau cone, we determine the most general IASD fluxes in terms of scalar harmonics,
then compute the resulting D3-brane potential. Specializing to the conifold, we identify the operator
dual to each mode of flux, and for chiral operators we confirm that the potential computed in the gauge theory matches the gravity result.
The effects of four-dimensional curvature, including the leading D3-brane mass term, arise directly from the ten-dimensional equations of motion.
Furthermore, we show that gaugino condensation on D7-branes provides a local source for IASD flux.  This flux automatically and precisely encodes the nonperturbative contributions to the D3-brane potential, yielding a promising ten-dimensional representation of four-dimensional nonperturbative effects.  Our result encompasses all significant contributions to the D3-brane potential discussed in the literature, and does so in the single coherent framework of ten-dimensional supergravity.  Moreover, we identify new terms with irrational scaling dimensions that were inaccessible in prior works. By decoupling gravity in a noncompact configuration, then systematically reincorporating compactification effects as ultraviolet perturbations, we have provided an approach in which Planck-suppressed contributions to the D3-brane effective action can be computed.
This is the companion paper to~\cite{Baumann2009k}.}
 \vspace{0.3cm}
 \hrule

\vspace{0.4cm}

\begin{flushleft}
\footnotesize $^*$ On leave of absence from Stanford University and SLAC.
\end{flushleft}

\end{titlepage}

\newpage
\tableofcontents

\newpage
\section{Introduction}

Since the dawn of time, humankind has wondered, ``what is the potential on the Coulomb branch of the conifold gauge theory, and what are the consequences for models of D-brane inflation?"
In this paper, we continue this quest.

\vskip 6pt
Recent observations \cite{Komatsu2009} give striking support to the idea that there was a period of inflation in the very early universe \cite{Guth1981,Linde1982,Albrecht1982}.
In a rather economical way,
inflation explains both the large-scale homogeneity of the universe and the small-scale inhomogeneities required for the formation of galaxies~\cite{Baumann2009a,TASI}.
As a phenomenon in quantum field theory coupled to general relativity, inflation is sensitive to ultraviolet physics: the inflationary dynamics is controlled by Planck-suppressed contributions to the effective action (see {\it e.g.}~\cite{Baumann2009a,TASI,  Baumann2009b}). This strongly motivates pursuing realizations of inflation in an ultraviolet-complete theory, such as string theory,
and then computing these contributions in detail.

In practice, determining all of the significant Planck-suppressed contributions to the effective action is highly nontrivial and requires a detailed understanding of the stabilization of compactification moduli.
Nevertheless, this undertaking is an essential prerequisite for any explicit realization of inflation in string theory.
It is therefore critical to identify scenarios for inflation in string theory that enjoy a high degree of computability, so that one can hope to compute all
relevant Planck-suppressed contributions to the inflaton action.

D3-brane inflation in a warped throat geometry \cite{KKLMMT} has been the subject of considerable research (for recent reviews, see {\it e.g.}~Refs.~\cite{TASI, Baumann2009b, Cline2006, Burgess2006, McAllister2008}).   Much of the interest is due not to any intrinsic elegance of the scenario, but rather to the prospect of explicit computations of the inflaton action: the warped deformed conifold solution \cite{KS} provides a concrete arena with a known metric and known background fluxes, and the effects of moduli stabilization, {\it e.g.}~in the
scenario of \cite{KKLT}, can be incorporated in detail \cite{BDKMMM,BDKM}.
The essential simplification is
that a finite warped region may be approximated by the {\it{noncompact}} warped deformed conifold solution, for which the supergravity solution can be written explicitly; similarly, the divisors responsible for K\"ahler moduli stabilization may be approximated by noncompact divisors.

Corrections to the noncompact approximation, {\it i.e.}~contributions to the D3-brane effective action induced by objects and fields in the compact bulk, constitute the remaining Planck-suppressed contributions to the inflaton action \cite{Baumann2009}.  On general grounds, one expects that these `compactification effects' can make order-unity corrections to the inflationary slow-roll parameters, and must therefore be incorporated, or shown to be suppressed, in any explicit realization of D3-brane inflation.  Specifically, D-branes, orientifold planes, fluxes,
and quantum effects in the bulk may not preserve the same supersymmetry as a D3-brane, and their effects on the D3-brane potential must then be included. The relevant effects can be both perturbative and nonperturbative.

\vskip 6pt
In this work, we
provide a systematic treatment of these compactification corrections to the D3-brane Lagrangian.
We take the compact bulk to be a rather general solution of type IIB supergravity, and then determine how the form of this solution in the ultraviolet region of
a finite throat affects the potential for a D3-brane well inside the throat.

Our strategy is to study a noncompact solution subject to general non-normalizable deformations, as a computable proxy for a finite throat attached to a compact space.
The critical simplification is that a completely general solution in the ultraviolet is well-approximated in the infrared by a solution parameterized by the handful of modes that diminish least rapidly under radial rescaling.  In the dual
field theory, this is the familiar statement that renormalization group flow filters out highly irrelevant perturbations, so that the dominant effect in the infrared is controlled by the coefficients of the most relevant modes.  By careful study of these modes, one can determine the leading structure of the D3-brane potential.

We identify the nonlinear interactions of imaginary anti-self-dual (IASD) three-form fluxes as an important contribution to the D3-brane potential.
In fact, under certain conditions this can be the dominant effect in the infrared.
To incorporate this effect, we provide a general solution for IASD fluxes in the conifold. Our method extends to any Calabi-Yau cone, yielding the  three-form fluxes in terms of the scalar harmonics on the angular manifold.  Taking general IASD fluxes as sources,
we compute the
corresponding flux-induced
potential.  We then include an additional contribution sourced by four-dimensional scalar curvature, showing that the leading curvature correction gives rise to the well-known
`eta problem' mass term of \cite{KKLMMT} (see also \cite{Buchel2004, Buchel2006}).
The inclusion of the nonlinear effects of fluxes and curvature is a substantial step beyond the linear treatment in our previous work~\cite{Baumann2009}.

Our analysis is
simplified by a special property of D3-branes: a D3-brane couples only to a particular scalar mode,  which we denote by $\Phi_-$, and IASD flux $G_-$
is the dominant source in the equation of motion for $\Phi_-$.  We systematically expand around solutions in which $\Phi_- = G_- = 0$.  Crucially, metric and dilaton fluctuations do not couple to D3-branes at the order to which we  work.  Therefore, although the metric is distorted away from the conformally Calabi-Yau metric of the leading order solution, we do {\it{not}} need to determine its form in order to specify the D3-brane potential. Thus, although our solutions are genuinely nonlinear in fluxes, they are far simpler than the most general solutions nonlinear in all supergravity fields.  Let us stress that our analysis rests on a double expansion: in small fluctuations around solutions in which $\Phi_- = G_- = 0$, and in the ratio of energy scales between the ultraviolet, where the throat is perturbed by effects from the compact bulk, and the infrared, where the D3-brane probes the supergravity background.

\vskip 6pt
The compactification effects studied here are most efficiently described in supergravity as perturbations to the ultraviolet region of the warped deformed conifold solution.  However, a nontrivial consistency check comes from representing these non-normalizable perturbations, through the AdS/CFT correspondence \cite{Maldacena1998,Gubser1998,Witten1998}, as perturbations to the Lagrangian of the dual
conformal field theory (CFT).  Building on the work of Ceresole {\it et al.}~\cite{Ceresole2000}, we provide the operator in the conifold CFT dual to each mode of flux.  In our earlier work \cite{Baumann2009}, we considered corrections to the D3-brane potential from linearized perturbations of the CFT K\"ahler potential; here we incorporate contributions up to quadratic order in the perturbations of the superpotential.  In the case of chiral operators perturbing the superpotential, we compute the potential on the CFT side and find
agreement with the supergravity result.
For non-chiral operators, there is no reason to expect computability in the strongly-coupled CFT, and we do not attempt to match the corresponding potentials.

Ultimately, our approach is strongly reminiscent of a four-dimensional effective field theory analysis of the inflaton action.  However, the field theory that governs the D3-brane potential
is strongly coupled.  Several of the contributing operators have irrational dimensions and cannot be studied effectively on the field theory side.  Our method, which is to consider the most general perturbations of the ultraviolet region of the supergravity solution, effectively uses AdS/CFT to provide a tractable problem that
realizes the spirit of the four-dimensional effective field theory approach.

\vskip 6pt
An additional goal of this work
is a better understanding of nonperturbative contributions to the D3-brane potential.  When the K\"ahler moduli are stabilized nonperturbatively \cite{KKLT}, a critical contribution to the D3-brane potential
arises from nonperturbative effects on branes wrapping
suitable four-cycles \cite{Witten:1996bn, KKLMMT,BHK}.
In the special case in which the dominant effect comes from a divisor that
protrudes into the warped throat region,
the nonperturbative corrections to the D3-brane potential can be computed
\cite{BDKMMM, BDKM}.  However, for general compactifications
there will be non-negligible contributions from a variety of divisors, not all of which enter the throat region. An outstanding open problem
is how to characterize these contributions.

We make progress in this direction by demonstrating that for any specified superpotential $W$ for a D3-brane in a noncompact conifold geometry, there exists a ten-dimensional supergravity solution in which the scalar potential for a probe D3-brane precisely matches the scalar potential computed in the four-dimensional supersymmetric gauge theory with superpotential $W$.  This solution contains IASD three-form flux of Hodge type $(1,2)$, $G_{(1,2)}$, and amounts to an explicit example of the general result \cite{Grana2003a} that a $G_{(1,2)}$ background induces superpotential interactions for a probe D3-brane.  We show that this relation persists even in the presence of large distortions of the metric and dilaton sourced by the classical backreaction of D7-branes.  Perhaps more surprisingly, we demonstrate that for any specified superpotential for a D3-brane in a finite conifold region, the F-term potential computed in four-dimensional super{\it{gravity}} can be geometrized by a particular ten-dimensional background of IASD fluxes.

Finally, we establish that gaugino condensation on D7-branes wrapping a four-cycle $\Sigma$ provides a source term, localized to $\Sigma$, for IASD flux.  The ten-dimensional equation of motion for the flux is corrected by a term
proportional to the expectation value $\langle\lambda\lambda\rangle$ of the gaugino bilinear,
and the corresponding solution necessarily involves $G_{(1,2)}$ flux proportional to $\langle\lambda\lambda\rangle$.  We demonstrate that a probe D3-brane in this flux background experiences precisely the scalar potential computed in four dimensions with the gaugino condensate superpotential.  In this sense, the induced flux encodes four-dimensional nonperturbative effects in the ten-dimensional supergravity solution.  This result constitutes progress towards a geometric transition for D7-branes, in that it replaces four-dimensional nonperturbative effects on D7-branes by certain bulk fluxes.   However, our methods serve only to identify terms in the supergravity solution to which a D3-brane is sensitive, and as noted above, at leading order a D3-brane does not couple to perturbations of the metric.\footnote{See \cite{Koerber2007a} for an interesting related proposal that represents the nonperturbative superpotential in terms of generalized complex geometry.} We leave for the future the very interesting problem of identifying further probes of the geometry that could guide the formulation of a complete geometric transition for D7-branes~\cite{Future}. \vskip 6pt

\begin{figure}[h!]
    \centering
        \includegraphics[width=0.5\textwidth]{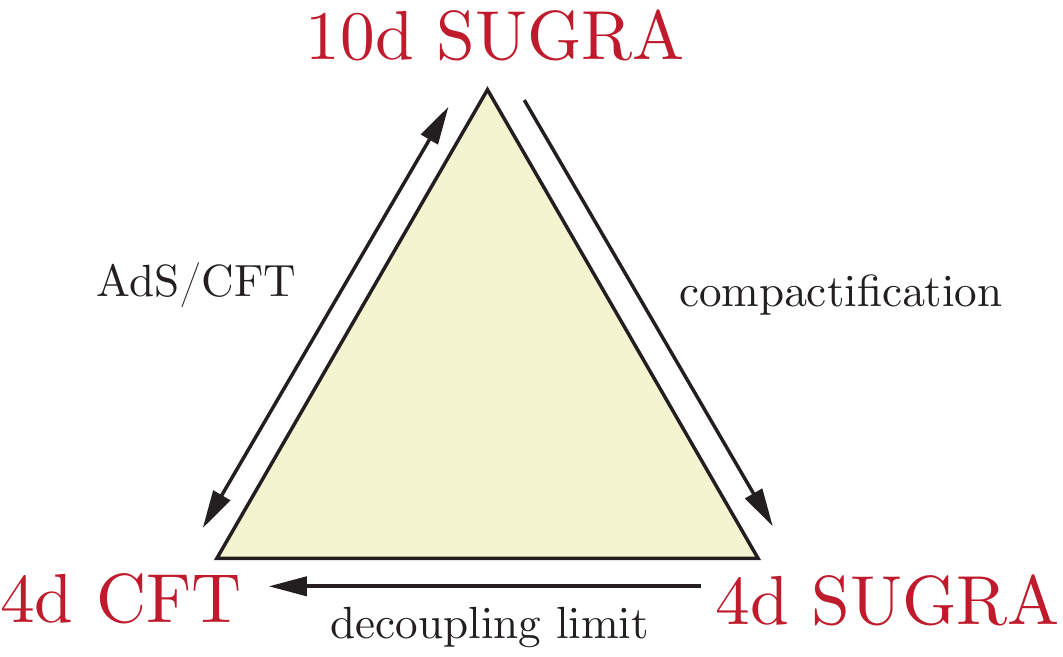}
    \caption{\small \sl Three descriptions of D3-branes in warped throats: the ten-dimensional supergravity perspective is explored in \S\ref{sec:10dSUGRA}, \S\ref{sec:flux} and \S\ref{sec:spectrum}, the dual four-dimensional conformal field theory is discussed in  \S\ref{sec:cft}, and connections to four-dimensional supergravity are made in \S\ref{sec:4dSUGRA} and \S\ref{sec:speculations}.}
    \label{fig:schematic}
\end{figure}

\vskip 6pt
In summary, in this paper we present three distinct, but complementary descriptions of the system of interest (see Fig.~\ref{fig:schematic}): {\it i}) ten-dimensional supergravity, {\it ii}) four-dimensional gauge theory, and {\it iii}) four-dimensional supergravity.  For an inflationary model, dynamical four-dimensional gravity is of course crucial, so that only the last description, which arises for a D3-brane probing a finite throat contained in a compact space,
seems of direct interest.  However, we show that one can usefully consider a decompactification limit in which a D3-brane probes a noncompact warped throat subject to suitable non-normalizable perturbations.  This theory is then connected by the AdS/CFT correspondence to an approximately-conformal four-dimensional gauge theory subject to ultraviolet perturbations of the Lagrangian.  Furthermore, the sourcing of fluxes by gaugino condensation provides a map from four dimensions to ten dimensions.
We use the noncompact ten-dimensional supergravity solution in order to determine the structure of the inflaton potential in the compact case of interest.
The internal consistency of these different approaches to computing the D3-brane potential, and the many cross-checks provided by relating 
them, give us confidence that we capture the leading contributions. \vskip 6pt

For readers familiar with our prior work \cite{BDKMMM, BDKM, Baumann2009}, we will now outline how the present analysis incorporates and extends those results.
In \cite{BDKMMM, BDKM} the D3-brane potential induced by nonperturbative effects on a stack of D7-branes was studied in four-dimensional supergravity.  The D7-branes were assumed to descend far into the throat region and
to
have limited support in the bulk, so that the effects computed explicitly in \cite{BDKMMM} would dominate over additional contributions from the compact bulk.  In
\cite{Baumann2009} we began to relax this assumption, studying more general contributions to the D3-brane action in the framework of ten-dimensional supergravity.
In the present work we study very general compactification contributions to the D3-brane action and determine the full structure of the D3-brane potential.

To compare these results, we write the D3-brane potential as
\begin{equation}
V = \sum_{i} c_{i} \phi^{\Delta_{i}} h_{i}(\Psi)\, ,
\end{equation} where $c_{i}$ are constants, $\phi$ is the canonically-normalized field describing radial motion of the D3-brane, and $h_{i}(\Psi)$ are functions of the angular directions on the
conical geometry (see \S\ref{sec:flux} for our conventions).
In this notation, the four-dimensional supergravity result of \cite{BDKM} implies that
\bea
\Delta &=& 1 \ , \ \frac{3}{2} \ , \ 2_s \ , \ 2 \ , \ \frac{5}{2} \ , 3 \ , \ \ldots
\eea
where $2_{s}$ denotes a
singlet term, and all other terms have nontrivial angular dependence.
In \cite{Baumann2009} we showed that a subset of these terms correspond to perturbations of the supergravity solution by certain scalar field harmonics
on the conifold, with
\begin{equation}
\Delta_{\cal H} \ = \  \frac{3}{2} \ , \ 2 \ , \  3 \ ,\ \ldots
\end{equation}
In this paper, we will show that the remaining terms correspond to nonlinear perturbations sourced by fluxes,
\begin{equation}
\Delta_{\rm G_-} \ = \ 1 \ , \ 2_{s} \ , \ \frac{5}{2}\ ,\  \ldots
\end{equation}
and by four-dimensional curvature ${\cal R}_{4}$,
\begin{equation}
\Delta_{{\cal{R}}} \ = \ 2_{s} \ ,  \ 3 \ , \ \ldots
\end{equation}
Moreover, we will identify the leading compactification effect  that could not be captured by the analysis of \cite{BDKM}: this is a flux perturbation dual to a non-chiral operator,
\begin{equation}
\Delta_{n\chi} \ = \ \sqrt{28}- \frac{5}{2} \ \approx \ 2.79  \, .
\end{equation}
Such a contribution is generically present, but could not be
found in~\cite{BDKM}, which incorporated only perturbations of the superpotential.\vskip 6pt

The outline of this paper is as follows: in \S\ref{sec:10dSUGRA} we recall the basic fields and equations of motion of type IIB supergravity. We define our approximation scheme for studying non-normalizable
perturbations
of backgrounds with imaginary self-dual (ISD) fluxes.
As a prerequisite for analyzing the flux-induced potential, we classify in \S\ref{sec:flux} all closed, IASD three-form perturbations
on general Calabi-Yau cones.
In \S\ref{sec:spectrum} we then discuss the spectrum of contributions to the D3-brane potential
sourced by UV deformations
of $AdS_{5}\times T^{1,1}$.
In \S\ref{sec:cft} we explain how these results can be reflected in the dual conformal field theory studied in \cite{KW}.  First, we carefully identify the operators dual to the perturbations of fluxes, building on \cite{Ceresole2000}.  We then perturb the CFT Lagrangian by these operators, allowing for explicit breaking of supersymmetry, and show that in the case of perturbations by chiral operators, the results agree with the gravity analysis.
In \S\ref{sec:4dSUGRA} we relate our results to the nonperturbatively-generated D3-brane potential in four-dimensional supergravity, showing that the effects of gaugino condensation on D7-branes can be represented in ten dimensions by suitably-chosen IASD fluxes.
Finally, in \S\ref{sec:speculations} we show that  gaugino condensation on D7-branes actually sources IASD flux, providing an intriguing link between ten-dimensional  supergravity and four-dimensional nonperturbative effects.
We present our conclusions in \S\ref{sec:conclusions}.

Two appendices contain computations supporting the results presented in the main text: in Appendix~\ref{sec:IASD} we
give the details of the classification of all closed, IASD three-form flux perturbations on
arbitrary Calabi-Yau cones,
while in Appendix~\ref{sec:running} we extend some of these results to cases with significant dilaton variations.

Throughout we use units where $c=\hbar = 1$ and $M_{\rm pl}^2 = 1/8\pi G =1$.

A condensed presentation of some of the key results of this paper has appeared in \cite{Baumann2009k}.

\section{Ten-Dimensional Supergravity}
\label{sec:10dSUGRA}

Our goal is to understand the effective action for D3-branes in a flux compactification.  Suppose that the entire compactification contains only
ISD fluxes and that all local sources saturate the inequalities described in \cite{GKP}; we will refer to such a solution as an {\it ISD compactification}.  A D3-brane in an ISD compactification is known to feel no force at leading order in $g_{s}$ and $\alpha^{\prime}$.  However, the no-scale structure that forbids a D3-brane potential at leading order simultaneously forbids a potential, again at leading order, for the breathing mode of the compactification.  Therefore, when the no-scale structure is ultimately broken to achieve K\"ahler moduli stabilization, {\it e.g.}~by nonperturbative effects, the D3-brane also experiences a potential.  We would like to describe the most general such potential in ten-dimensional supergravity by systematically expanding around an ISD solution.

We will also  make use of an expansion around the  noncompact warped deformed  conifold solution \cite{KS}.  Gluing a finite throat into a compact space requires suitable distortions of the supergravity
fields in the ultraviolet region of the throat.  Some of these distortions may involve only supergravity fields to which a D3-brane does not couple at leading order, such as the dilaton and the unwarped metric.
Examples of ISD compactifications of precisely this sort are well-understood \cite{GKP}.  However, as explained above, in general one expects that some of the distortions associated with attaching a throat into a
compact space {\it{with all moduli stabilized}} will require violations of the ISD conditions described in \cite{GKP}.  Therefore, in this section
we consider general non-normalizable perturbations of
ISD compactifications, allowing these perturbations to violate the ISD conditions.

In \S\ref{sec:EoM} we review the basic degrees of freedom of type IIB supergravity and the equations of motions that the fields satisfy.
We derive the equation of motion of the D3-brane potential and show that it is sourced by IASD fluxes and
by four-dimensional Ricci curvature.
In \S\ref{sec:UV} we explain that a filtering effect of the warped background allows us to focus on a handful of corrections, those with the smallest scaling dimensions.
Then, in \S\ref{sec:perturbation}, we describe the scheme by which we study perturbations around ISD compactifications.
(We present
these corrections in more detail in the following sections, \S\ref{sec:flux} and \S\ref{sec:spectrum}.)
Finally, in \S\ref{sec:consistency}, we describe some constraints on these backgrounds that arise when one wishes to consider a compact model; the most basic
conditions arise from avoiding destabilization of the volume modulus of the compact space, and from satisfying global tadpole constraints.

\subsection{Equations of Motion}
\label{sec:EoM}

Our starting point is the bosonic low-energy action for type IIB supergravity in Einstein frame,
\begin{eqnarray}
\label{equ:action}
S_{\rm IIB} &=& \frac{1}{2 \kappa_{10}^2} \int \dd^{10} x \sqrt{|g|} \left[ {\cal R}_{10} - \frac{\partial_M \tau \partial^M \bar \tau}{2 \, {\rm Im}(\tau)^2} - \frac{G_3 \cdot \bar G_3}{12 \, {\rm Im}(\tau)} - \frac{\tilde F_5^2}{4 \cdot 5!} \right] \nonumber \\
&& +\ \frac{1}{8 i \kappa_{10}^2} \int \frac{C_4 \wedge G_3 \wedge \bar G_3}{{\rm Im}(\tau)} + S_{\sf local}\, ,
\end{eqnarray}
where $\kappa_{10}^2 \equiv \frac{1}{2} (2\pi)^7 \alpha'^4$ is the ten-dimensional gravitational coupling (in the conventions of \cite{GKP}).
Here, $\tau \equiv C_0 + i e^{- \phi}$ is the axio-dilaton field and $G_3 \equiv F_3 - \tau H_3$ is a combination of the R-R and NS-NS three-form fluxes $F_3 \equiv \dd C_2$ and $H_3 \equiv \dd B_2 $.
The five-form $\tilde F_5 \equiv F_5 - \frac{1}{2} C_2 \wedge H_3 + \frac{1}{2} B_2 \wedge F_3$ is self-dual, $\tilde F_5 = \star_{10} \tilde F_5$,
where $\star_{10}$ is the ten-dimensional Hodge star operator.
Finally,
${\cal R}_{10}$ is the ten-dimensional Ricci scalar and $S_{\sf local}$ denotes localized contributions from D-branes and orientifold planes.

For the warped line element, we take the ansatz\footnote{To determine the full effective action governing time-dependent solutions, a more general line element is required \cite{GM,STUD}. In this work we exclusively study the scalar potential as a function of the D3-brane position, for which the much simpler line element (\ref{equ:warping}) suffices.}
\beq
\label{equ:warping}
\d s^2 = e^{2A(y)} g_{\mu \nu} \d x^\mu \d x^\nu + e^{- 2 A(y)} g_{mn} \d y^m \d y^n\, .
\eeq
The metric $g_{mn}$ of the internal space will not be Calabi-Yau in the configurations of interest, but it is useful to think of this metric as being Calabi-Yau at leading order in a certain perturbative expansion, as we shall explain in \S\ref{sec:perturbation}.
For the five-form flux, we assume
\beq
\label{equ:5flux}
\tilde F_5 =  (1+ \star_{10})\, \d \alpha(y) \wedge \sqrt{-{\rm det}g_{\mu\nu}}\ \d x^0 \wedge \d x^1 \wedge \d x^2 \wedge \d x^3\, .
\eeq
The warp factor $e^{4A(y)}$ and the scalar function $\alpha(y)$ in Eqns.~(\ref{equ:warping}) and (\ref{equ:5flux}) will play crucial roles in the potential felt by probe D-branes.
The Einstein-frame action for a Dp-brane wrapping a ($p-3$) cycle $\Sigma$ is the sum of Dirac-Born-Infeld (DBI) and Chern-Simons (CS) terms,
\beq
S_{\rm Dp} = - \int\limits_{R^4\times \Sigma}\hspace{-0.1cm} \d^{p+1} \xi\ T_p \sqrt{-g_{\rm ind}} \ +\ \mu_p \int\limits_{R^4 \times \Sigma} \hspace{-0.1cm}C_{p+1}\, ,
\eeq
where $g_{\rm ind}$ is the metric induced on the Dp-brane, and
\beq
T_p = |\mu_p| e^{(p-3) \phi(y)/4}\qquad {\rm with} \qquad |\mu_p| = (2\pi)^{-p} (\alpha')^{-(p+1)/2}\, .
\eeq
Notice the special status of D3-branes, whose action decouples from fluctuations of the dilaton $\phi(y)$ and of the internal unwarped metric $g_{mn}(y)$.
In these variables a D3-brane experiences the potential
\beq
V_{\rm D3} = T_3 \left( e^{4A} - \alpha \right) \equiv T_3 \Phi_- \, .
\eeq
In the following we are therefore interested in perturbations of the scalar quantity $\Phi_- \equiv e^{4A} - \alpha$. We also define $\Phi_+ \equiv e^{4A} + \alpha$. Furthermore,
it will be convenient to use the following parametrization of the three-form fluxes:
\beq
G_\pm \equiv (\star_6 \pm i) G_3\, ,
\eeq
where $\star_6$ is the six-dimensional Hodge star operator on the internal manifold.
Then, $G_+$ is the ISD component of the three-form flux $G_3$, while $G_-$ is
its IASD component.
Combining the external Einstein equations
with the Bianchi identity for the five-form flux ($\d \tilde F_5 = H_3 \wedge F_3 + {\sf local}$) we find
\beq
\label{equ:PhiEoM}
 \nabla^2 \Phi_- = \frac{e^{8A+ \phi}}{24}  |{G_-}|^2 + {\cal R}_4 + e^{-4 A} | \nabla \Phi_-|^2  + {\cal S}_{\sf local}\, ,
\eeq
where  ${\cal R}_4$ denotes the four-dimensional Ricci scalar, and $\nabla^2$ is constructed from $g_{mn}$.  This result is a straightforward generalization of Eqn.~(2.30) of \cite{GKP}, with the
difference\footnote{We thank David Marsh and Gang Xu for discussions of this point.} that here we have allowed $g_{\mu\nu}$ to be the metric of a maximally-symmetric four-dimensional spacetime, while with the more restrictive assumptions of \cite{GKP}, $g_{\mu\nu}=\eta_{\mu\nu}$ is required, and ${\cal R}_4=0$.

The equation of motion for the three-form flux is
\beq
\label{equ:flux0}
\d \Lambda + \frac{i}{2} \frac{ \d \tau}{{\rm Im}(\tau)} \wedge (\Lambda + \bar \Lambda) = 0 \, ,
\eeq
where we have defined
\beq
\Lambda \equiv \Phi_+ G_- + \Phi_- G_+ \, .
\eeq
This must be supplemented by the Bianchi identity, which in the absence of sources reads
\beq
\label{equ:Bianchi}
\d G_3 = \frac{1}{2i} \d (G_+ - G_-) = - \d\tau\wedge H_{3} \,.
\eeq

\subsection{UV Perturbations and RG Filtering}
\label{sec:UV}

\begin{figure}[h!]
    \centering
        \includegraphics[width=0.8\textwidth]{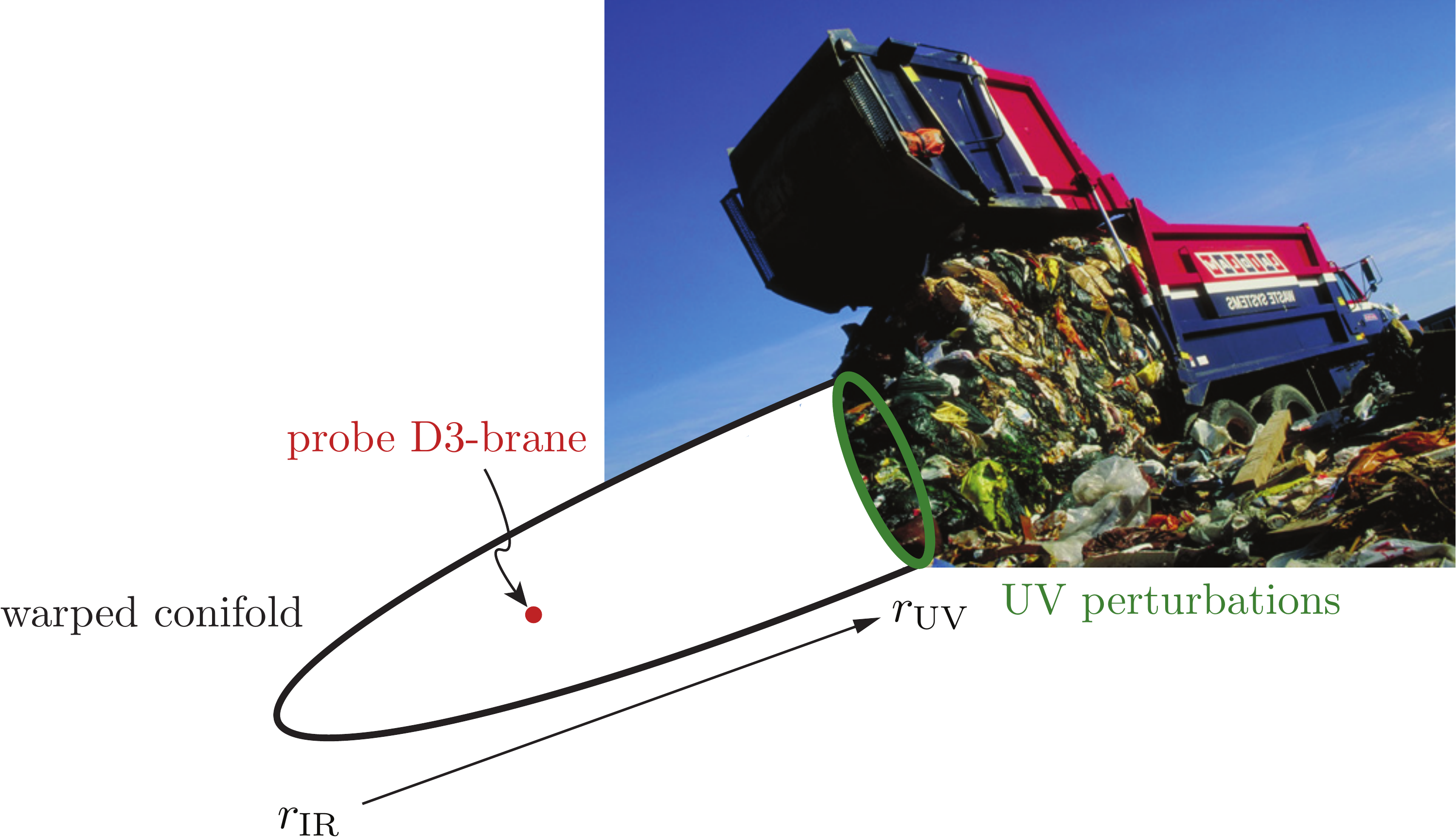}
    \caption{\small \sl Compactification can induce very general  UV perturbations of the warped conifold solution, but in the infrared only  the lowest-dimension perturbations contribute meaningfully to the D3-brane potential.}
    \label{fig:UV}
\end{figure}

In an ideal world,
one would determine the precise D3-brane potential in a fully-specified compactification, in terms of fluxes, D-brane positions,  and closed string moduli vevs.  With present methods this is difficult to achieve, except perhaps in a toroidal orientifold setting such as \cite{BHK, Haack2009}.
In this paper, our goal is to determine the general structure of the potential arising from UV deformations of the background,
\beq
\label{equ:structure}
V(\phi)= \sum_i c_i \, \frac{\phi^{\Delta_i}}{M_{\rm UV}^{\Delta_i - 4}}\ ,
\eeq
where $\phi$ is the canonically-normalized field related to the D3-brane position and $M_{\rm UV}$ is a UV mass scale (related to $r_{\rm UV}$, the ultraviolet location at which the throat is glued into the compact bulk; see Fig.~\ref{fig:UV}).
In terms of the parametrization of Eqn.~(\ref{equ:structure}), our
primary task is to compute the scaling dimensions $\Delta_i$,
while leaving the coefficients of individual terms, $c_i$, undetermined.  This undertaking is a necessary precursor to any calculation that does obtain the Wilson coefficients in a concrete model.   For comparison, in Ref.~\cite{BDKM} we argued that in certain  special circumstances, the dominant Planck-suppressed contribution to the D3-brane potential comes from interactions with nonperturbative effects on a divisor in the  conifold.  In the present paper, we are addressing the more general situation in which multiple compactification effects make important contributions to the potential.

Our interest is in the {\it{leading}} terms in the potential for a D3-brane that is well-separated from the UV cutoff $r_{\rm UV}$.  The dominant terms  come from  the Kaluza-Klein modes with the smallest AdS masses, {\it i.e.}~the modes dual to the most relevant operators in the CFT.  Highly irrelevant perturbations are filtered out by the RG flow; in gravity language, these perturbations are described by higher-order terms in a multipole expansion and are subleading at long distances from $r_{\rm UV}$.

The dominant terms at small radial position $r$ are, of course, those with the smallest $\Delta_i$ in Eqn.~(\ref{equ:structure}).  We choose to work to order $r^{4}$ and consistently neglect higher-order terms.  Although we are formally  expanding in small $r$, we also assume that the D3-brane is far above the infrared location $r_{\rm IR}$ where the duality cascade \cite{KS} terminates and the conifold is deformed.  That is, we take $r_{\rm IR} \ll r \ll r_{\rm UV}$.

Throughout this paper we restrict our attention to {\it{non-normalizable}} perturbations -- corresponding to deformations of the gauge theory Lagrangian -- sourced by effects in the compact bulk.  However, perturbations by normalizable modes of the supergravity fields, corresponding to perturbations of the state of the dual CFT, can also make important contributions to the D3-brane potential.  For example, the Coulomb potential sourced by an anti-D3-brane at the tip of the warped deformed conifold is described in supergravity as a normalizable mode of $\Phi_-$ \cite{DKM} and plays an important role in D3-brane inflation (see \cite{Baumann2009} for further details).  We focus on non-normalizable perturbations because these encode the effects of Planck-scale physics in the compact bulk, whereas normalizable perturbations are dictated by the better-understood physics in the infrared.  Extending the analysis presented here to incorporate any given set of normalizable perturbations is, however, entirely straightforward.

\subsection{Perturbative Expansion}
\label{sec:perturbation}

We now define our perturbative expansion scheme. Our method is general and can be performed in an expansion around any background in which $\Phi_-=G_{-}=0$,
although we will
sometimes specialize to perturbations around the background $AdS_5 \times T^{1,1}$.

We consider perturbations of all fields,
\beq
\label{equ:X}
X = X_{(0)} + X_{(1)} + X_{(2)}+ \dots\, , \qquad X \equiv \{\Phi_-, \Phi_+, G_-, G_+, \phi,  g_{mn} \}\, ,
\eeq
where $X_{(0)}$ is the background, $X_{(1)}$ is the first-order perturbation, $X_{(2)}$ is the second-order perturbation, etc.
We assume that all perturbations are small.\footnote{For
results in the case of large deviations of the metric and dilaton, see Appendix~\ref{sec:running}.}
Importantly, $\Phi_-$ and $G_-$ vanish in ISD backgrounds,
\beq
\Phi_-^{(0)} = G_-^{(0)} = 0\, .
\eeq
Let us now
systematically expand the equations of motion in small perturbations.

\vskip 6pt
\noindent
{\sl Equation of motion for $\Phi_-$.} \hskip 8pt
For simplicity we begin by studying the noncompact limit ($M_{\rm pl} \to \infty$) which extracts fluxes as the only source term in the equation of motion for $\Phi_-$,
\beq
\nabla^2 \Phi_- = \frac{e^{8A + \phi}}{24} |G_-|^2 + e^{-4A} |\nabla \Phi_-|^2 \,.
\eeq
We will incorporate the curvature contribution ${\cal{R}}_{4}$ for finite $M_{\rm pl}$ in \S\ref{sec:subR}.
Metric perturbations propagate into perturbations of the Laplacian,
\beq
\nabla^2 = \nabla^2_{(0)} + \nabla^2_{(1)} + \dots
\eeq
At first order in all perturbations we get
\beq
\label{equ:linear}
\nabla^2_{(0)} \Phi_-^{(1)} = 0 \, ,
\eeq
while at second order we find
\beq
\label{equ:second}
\nabla_{(1)}^2 \Phi_-^{(1)} + \nabla_{(0)}^2 \Phi_-^{(2)} = \frac{g_s}{96} \bigl(\Phi_+^{(0)}\bigr)^2 \bigl|G_-^{(1)}\bigr|^2 + 2 \bigl|\nabla_{(0)} \Phi_-^{(1)} \bigr|^2/\Phi_+^{(0)}\, .
\eeq
Here, we have defined the string coupling $g_s = e^{\phi_{(0)}}$, where $\phi_{(0)} = const$ is the asymptotic background value of the dilaton.
Clearly, the flux source term on the r.h.s.~becomes important only at second order.
Therefore, we are faced with two very different regimes, both of physical interest:
\begin{itemize}
\item {\sl Case I}
\beq
\Phi_-^{(1)} \neq 0
\eeq
In this case the linearized equation of motion (\ref{equ:linear}) suffices to determine the leading solution. We have analyzed this limit in \cite{Baumann2009}.
 \item {\sl Case II}
 \beq
 \label{equ:CaseII}
 \Phi_-^{(1)} =0
 \eeq
In this case flux-induced second-order terms can be important. This limit is the focus of the present paper.
 In the limit (\ref{equ:CaseII}), the first-order equation (\ref{equ:linear}) is identically satisfied, and the second-order equation (\ref{equ:second}) simplifies to
 \beq
\label{equ:second2}
 \nabla_{(0)}^2 \Phi_-^{(2)} = \frac{g_s}{96} (\Phi_+^{(0)})^2 \bigl|G_-^{(1)}\bigr|^2 \, .
\eeq
In \S\ref{sec:cft} we will provide a further physical justification for the perturbative expansion scheme we have proposed.  We will see that in a spurion analysis of supersymmetry breaking in the compact bulk, perturbations of $G_{-}$ can arise at linear order in the (small) spurion vacuum expectation value, while perturbations of the homogeneous mode of $\Phi_-$ require two spurion insertions.  Therefore, it is natural to consider cases in which $\Phi_-^{(1)}=0$ but $G_-^{(1)}\neq 0$.

In solving (\ref{equ:second2}), one must in general include harmonic $\Phi_-^{(2)}$ perturbations, in addition to the $\Phi_-^{(2)}$ solution sourced by the $G_-^{(1)}$ flux.
These terms are of comparable size in concrete scenarios, such as that of \cite{BDKM},  and indeed our techniques suffice to reproduce the potential of \cite{BDKM} as a
special
case (see \S\ref{sec:BDKM}).
 \end{itemize}
Note that metric and dilaton perturbations do not appear in either of Eqn.~(\ref{equ:linear}) and Eqn.~(\ref{equ:second2}): their effects on $\Phi_-$ are subleading in both cases.  Thus, although the metric and  dilaton must obey their own equations of motion, the corresponding solutions for these fields are not required in order to determine the leading contributions to $\Phi_-$.  Therefore, we will not pursue explicit solutions of the metric and dilaton equations of motion in this paper.
We remark in passing that although we will take $g_{mn}^{(0)}$ to be a Calabi-Yau metric, the perturbed metric $g_{mn}$ will in general not be Calabi-Yau, {\it{cf.}}~\cite{Butti}.

\vskip 6pt
\noindent
{\sl Flux equation of motion.} \hskip 8pt
Next we consider the flux equation of motion,
\beq \label{equ:fluxeom}
{\rm d} \Lambda + \frac{i}{2} \frac{{\rm d} \tau}{{\rm Im}(\tau)} \wedge (\Lambda + \bar \Lambda) = 0\, ,
\eeq
where
\beq
\Lambda \equiv \Phi_+ G_- + \Phi_- G_+\, .
\eeq
Since the $\Phi_-$ equation (\ref{equ:second2}) is second order in the fluxes, it suffices to solve (\ref{equ:fluxeom}) at first order.
We get
\beq
\label{equ:2}
{\rm d} \Lambda_{(1)} = 0\, ,
\eeq
where
\beq
\label{equ:Lambda1x}
\Lambda_{(1)} = \Phi_+^{(0)} G_-^{(1)} + \Phi_-^{(1)} G_+^{(0)}\, .
\eeq
The flux-induced contributions to $\Phi_-$ are only important when $\Phi_-^{(1)}=0$, so that we may take
\beq
\Lambda_{(1)} \approx \Phi_+^{(0)} G_-^{(1)} \, .
\eeq
This is precisely the source term in Eqn.~(\ref{equ:second2}). We can therefore write
 \beq
\label{equ:second3}
 \nabla_{(0)}^2 \Phi_-^{(2)} = \frac{g_s}{96} \bigl|\Lambda_{(1)}\bigr|^2 \, .
\eeq

\vskip 6pt
\noindent
{\sl IASD condition.} \hskip 8pt
In general, metric perturbations induce changes in the definition of IASD fluxes, by perturbing the Hodge star operator.  However, since $\Lambda_{(0)} = 0$,
the relevant IASD condition at first order is
\beq
\label{equ:3}
\star_6^{(0)} \Lambda_{(1)} = - i \Lambda_{(1)}\, ,
\eeq
{\it i.e.}~$\Lambda_{(1)}$ is IASD with respect to the background metric.  Therefore, one does not need the explicit form of the perturbed metric in order to determine the leading IASD flux solution, a substantial simplification analogous to that occurring in Eqn.~(\ref{equ:second2}).
\vskip 8pt

Equations (\ref{equ:2}),  (\ref{equ:second3}) and (\ref{equ:3}) form the basis for our exploration of flux-induced corrections to the D3-brane potential in warped throats with UV deformations (\S\ref{sec:flux} and \S\ref{sec:spectrum}).

\subsection{Consistency Requirements}
\label{sec:consistency}

\noindent
{\sl Stability of the background.} \hskip 8pt
Let us comment on the stability of the throat solutions in the presence of UV perturbations.  We will ultimately allow non-normalizable perturbations dual to {\it{relevant}} operators.  The corresponding supergravity profiles grow in the infrared, and given enough RG evolution, these modes could become uncontrollably large perturbations of the proposed background solution.  We will now argue that this instability is under control whenever the bulk supersymmetry breaking is small enough that decompactification does not ensue (see also Appendix~A of Ref.~\cite{Baumann2009}).

The configuration of interest is a finite throat in a stabilized compactification, with supersymmetry broken controllably in the bulk, and with a moduli potential that provides a finite barrier preventing decompactification. When this system is perturbed so that a positive four-dimensional potential energy is induced, this energy shifts the metastable minimum of the compactification toward larger volume. Sufficiently large perturbations create a decompactification instability. We will insist on studying configurations that remain metastable
and hence
must impose an upper bound on the four-dimensional potential energy.

As argued in \cite{Baumann2009}, the requirement of metastability implies that the bulk supersymmetry breaking is in fact not large in units of the infrared scale of the throat. A priori these scales were completely unrelated, but demanding an adequate barrier in the moduli potential, and assuming that effects in the infrared region of the throat ({\it e.g.}, an anti-D3-brane) suffice to uplift to a de Sitter solution, 
one finds the condition \cite{Baumann2009},
\begin{equation}
\Phi_-(r) < \Phi_+^{(0)}(r_{\rm IR}) \le \Phi_+^{(0)}(r)\, .
\end{equation}
The consequence is that the relevant deformations, evaluated in the ultraviolet, have exponentially small coefficients that are no larger than $(M_{\rm IR}/M_{\rm UV})^2$, for modes of flux
$\Lambda$, and no larger than $(M_{\rm IR}/M_{\rm UV})^4$ for harmonic modes of $\Phi_-$.  The different scaling for these two classes of modes may be understood to arise from the condition (\ref{equ:CaseII}).

Quite generally, relevant perturbations of the form
\begin{equation}
\delta {\cal{L}} = c\, M_{\rm UV}^{4-\Delta}{\cal{O}}_{\Delta}\,,
\end{equation} with $c \ll 1$, lead to important instabilities after RG evolution to a scale $M_{\rm crit}$ obeying
$\frac{M_{\rm crit}}{M_{\rm UV}} < c^{\frac{1}{4-\Delta}}$.  Above we have argued that in the cases of interest, requiring metastability implies that the coefficients (for operators dual to modes of flux,
whose proper treatment is the novelty in this paper) should obey $c \lesssim (M_{\rm IR}/M_{\rm UV})^2$. Then, for operators with $\Delta>2$, the RG evolution does not persist long enough for the relevant perturbations to have unit size.\footnote{There is {\it e.g.}~an operator with $\Delta = \frac{3}{2}$, but it is dual to a harmonic  mode of $\Phi_-$ and hence its coefficient should satisfy the stronger constraint  $c \leq (M_{\rm IR}/M_{\rm UV})^4$~\cite{Baumann2009}.
The leading flux perturbation has dimension $\Delta = \frac{5}{2}$ (see \S\ref{sec:flux}).}
\vskip 6pt
\noindent
{\sl Constraints from compactness.} \hskip 8pt
Finally, let us briefly comment on the consistency of our approach when global constraints are taken into account. It is well-understood that beginning with a compact ISD solution \cite{GKP}, the addition of IASD fluxes  alone, with no other new ingredients, is inconsistent with the integrated Einstein equation and Bianchi identity.   A consistent compact solution containing both ISD sources (such as ISD  fluxes, D7-branes, and O3/O7 orientifold planes) and IASD fluxes requires some additional sources in order to obey, {\it e.g.}, Eqn.~(2.30) of Ref.~\cite{GKP}.   These sources could be be additional classical brane sources (for instance, anti-O3-planes), or
may arise from quantum effects.  In this work we do not explicitly specify any additional sources in the ten-dimensional solution, but anticipate that the contributions of nonperturbative effects will render these solutions consistent, as is strongly suggested by the four-dimensional analyses of \cite{KKLT} {\it{et seq.}}  The situation is therefore precisely the same as in studies of supersymmetry breaking from an anti-D3-brane, and of soft terms from IASD fluxes.

\section{Fluxes in the Conifold}
 \label{sec:flux}

One of the main results
of this paper is a comprehensive treatment of the contributions of IASD fluxes
to the D3-brane potential.
To this end, in this section we will present, in closed form, the most general solution to the flux equations of motion,
\beq
\label{equ:fluxEoM}
\fbox{$\displaystyle
\d \Lambda =0 $} \, , \quad {\rm where} \quad \fbox{$\displaystyle \star_6 \Lambda = - i \Lambda $}\, ,
\eeq
on general Calabi-Yau cones, including the conifold.
The classification of all {\it closed, IASD three-forms $\Lambda$} on Calabi-Yau cones---Eqn.~(\ref{equ:fluxEoM})---is a well-defined mathematical problem whose solution is presented
in detail in Appendix~\ref{sec:IASD}.
Here, we summarize the results, first outlining the problem in terms of a harmonic expansion on the base manifold $X_5$ (\S\ref{sec:nonlinear}), and then giving the flux solutions in closed form (\S\ref{sec:explicitsolution}).

In \S\ref{sec:spectrum} we will integrate the $\Phi_-$ equation of motion,
\beq
\label{equ:MasterX}
\fbox{$\displaystyle
\nabla^2 \Phi_- = \frac{g_s}{96} | \Lambda|^2 + {\cal R}_4 $}\, ,
\eeq
incorporating four-dimensional curvature and the flux solutions of \S\ref{sec:flux} as sources.

\subsection{The Conifold}
\label{sec:conifold}

Although all of our results in principle apply to arbitrary Calabi-Yau cones, when computing spectral data we will specialize to the conifold.
We therefore begin by briefly setting our notation for the conifold (more details may be found in \cite{Herzog2001b, BDKM}).

\begin{figure}[h!]
    \centering
        \includegraphics[width=0.45\textwidth]{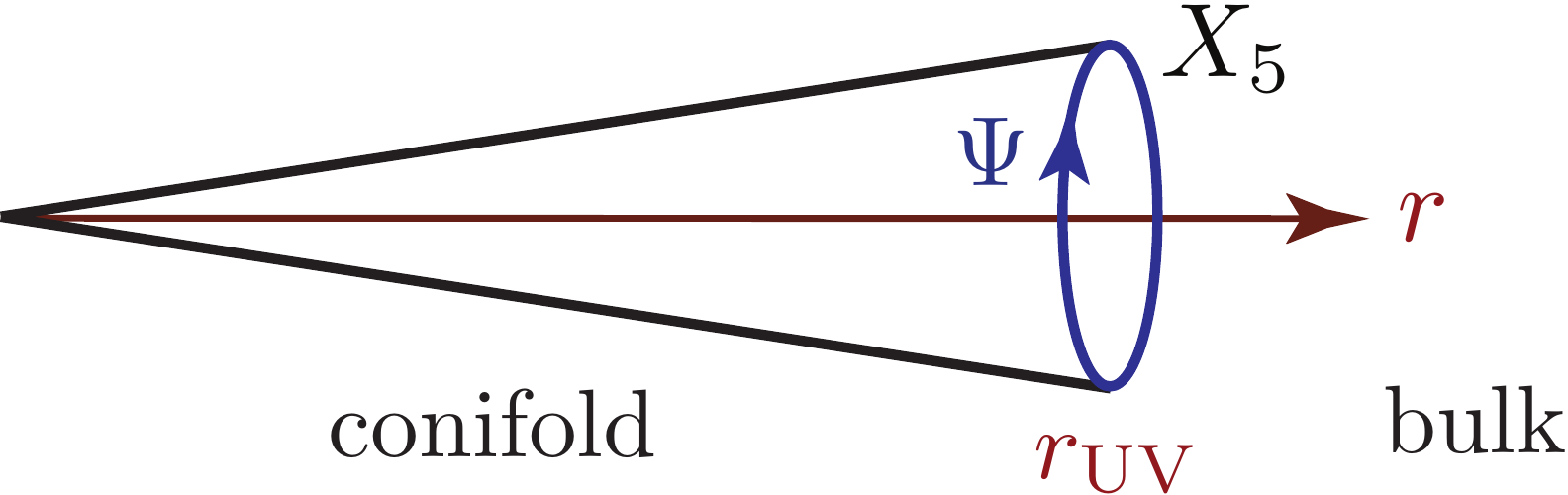}
    \caption{\small \sl Schematic of the conifold geometry. The five angular coordinates on the base $X_5=T^{1,1}$ are denoted by $\Psi =\{\theta_i, \phi_i, \psi \}$. The geometry is noncompact, but we imagine that eventually this space smoothly attaches to a compact bulk space at $r_{\rm UV}$; cf.~Fig.~\ref{fig:BDKMMM}.}
    \label{fig:conifold}
\end{figure}

The conifold is a singular noncompact Calabi-Yau threefold defined in $\mathbb{C}^4$ by
the constraint equation
\beq
\label{equ:conifold0}
\sum_{a=1}^4 z_a^2 = 0\, , \qquad z_a \in \mathbb{C}\, .
\eeq
Eqn.~(\ref{equ:conifold0}) describes a cone over the five-dimensional manifold $T^{1,1}$,
 \beq
 \label{equ:tildeg}
 g_{mn}^{(0)}\, \d y^m \d y^n =
\d r^2 + r^2 \d \Omega_{T^{1,1}}^2 \, ,
\eeq
where we have introduced the real coordinates $r^3 \equiv \bigl(\frac{3}{2}\bigr)^{3/2}\sum_a |z_a|^2$ and $\Psi =\{\theta_i, \phi_i, \psi \}$, with $i=1,2$.

 A stack of $N$ D3-branes placed at the singularity $z_a=0$ backreacts on the geometry, producing the ten-dimensional warped line element
\beq
\label{equ:metric0}
\d s^2 = e^{2A_{(0)}(r)} \eta_{\mu \nu} \d x^\mu \d x^\nu + e^{-2A_{(0)}(r)} (\d r^2 + r^2 \d \Omega_{T^{1,1}}^2)\, ,
\eeq
where
\beq
\label{equ:eA0}
e^{-4A_{(0)}(r)} = \frac{L^4}{r^4}\qquad {\rm and} \qquad L^4 \equiv \frac{27\pi}{4} g_s N (\alpha')^2\, .
\eeq
This ISD solution---$AdS_5 \times T^{1,1}$---is summarized in Table~\ref{tab:KW}.
In the following we will study small perturbations around this background.

\begin{table}[h!] \footnotesize
\caption{\sl Supergravity fields and the $AdS_5 \times T^{1,1}$ solution.}
\label{tab:KW}
\vspace{-0.5cm}
\begin{center}
\begin{tabular}{ c  c }
\toprule %
 { \bf Fields} & {\bf $\mathbf{AdS_5 \times T^{1,1}}$} \\ \otoprule %
$e^{4A}$ & Eqn.~(\ref{equ:eA0}) \\ \midrule
$\alpha$ & $e^{4A}$ \\
\midrule $g_{m n}$ & Eqn.~(\ref{equ:tildeg})\\
\midrule $\Phi_-$ & 0 \\
\midrule $\partial_r \phi$ & 0 \\
\midrule$C_0$ & 0 \\
\midrule $G_3$ & 0 \\
\midrule $G_-$ & 0 \\ \midrule
$G_+$ & 0 \\ \bottomrule
\hline
\end{tabular}
\end{center}
\end{table}%

An essential element of our solutions will be harmonic functions on the conifold, {\it i.e.}~solutions to the Laplace equation, $\nabla^2 f = 0$.
Expanding $f$ in angular harmonics on $T^{1,1}$ \cite{Ceresole2000}, we have
\beq \label{equ:harmX}
f(r, \Psi) = \sum_{L, M} f_{LM} \left(\frac{r}{r_{\rm UV}} \right)^{\Delta_f(L)} Y_{LM}(\Psi) \ +\ c.c.\, ,
\eeq
where $f_{LM}$ are constant coefficients, $L \equiv (j_1, j_2, R_f)$ and $M \equiv (m_1, m_2)$ label the $SU(2)\times SU(2)\times U(1)_R$ quantum numbers under the isometries of $T^{1,1}$, and the radial scaling dimensions $\Delta_f(L)$ are related to the eigenvalues of the angular Laplacian,
\beq
\label{equ:DeltafX}
\Delta_f(L) \equiv - 2 + \sqrt{H(j_1, j_2, R_f)+4}\, ,
\eeq
where
\beq
\label{equ:HX}
H(j_1,j_2,R_f) \equiv 6 \left[j_1(j_1+1) + j_2 (j_2+1) - R_f^2/8  \right] \, .
\eeq
Group-theoretic selection rules restrict the allowed quantum numbers \cite{Ceresole2000}. The lowest eigenvalues, {\it i.e.}~smallest scaling dimensions, are shown in Table~\ref{tab:f}. For chiral modes, $j_1 =j_2 = \frac{1}{2} R_f$, we have $\Delta_f = \frac{3}{2} R_f$.
\begin{table}[h!]\footnotesize
\caption{\sl Spectrum of lowest-dimension harmonic functions on the conifold. }
\label{tab:f}
\vspace{-0.5cm}
\begin{center}
\begin{tabular}{>{\columncolor{lightgray}\raggedright}c c  c c }
\toprule
$\Delta_f$ & $j_1$ & $j_2$  & $R_f$ \\ \otoprule
 $\frac{3}{2}$ & $\frac{1}{2}$ & $\frac{1}{2}$  & $1$  \\
\midrule
$2$ & $0$ & $1$  & $0$ \\
\midrule $2$ & $1$ & $0$  & $0$ \\
\midrule $3$ & $1$ &  $1$  & $2$ \\
\midrule $\sqrt{28}-2$ & $1$ & $1$ & $0$ \\
 \bottomrule
\hline
\end{tabular}
\end{center}
\end{table}%

\subsection{Harmonic Expansion of Flux Perturbations}
\label{sec:nonlinear}

We begin with a harmonic expansion of flux perturbations on Calabi-Yau cones.

We recall that the (linearized) flux equation of motion
in an
 $AdS_5 \times X_5$ background is
\beq
\label{equ:FEoM}
\d \Lambda = \d (e^{4A} G_-) = 0\, , \quad {\rm where} \quad G_- = (\star_6 - i)G_3 \quad {\rm and} \quad e^{4A} = \frac{r^4}{L^4}\, .
\eeq
This must be supplemented by the Bianchi identity,
\beq
\label{equ:BGx}
\d G_3 = - \d \tau \wedge H_3 \, .
\eeq
We solve Eqn.~(\ref{equ:FEoM}) by expanding the three-form $G_3$ in harmonics on $X_5$.
We introduce the Laplace-Beltrami operator $\star_5 \d$ which acts on two-forms on $X_5$,
\beq
\star_5 \d \, \Omega_2 = i \delta \, \Omega_2\, .
\eeq
The eigenvalues $\delta$ are real and the spectrum is invariant under $\delta \to -\,  \delta$.
When the dilaton is constant,
$G_3$ is closed\footnote{We consider non-constant dilaton in Appendix~\ref{sec:running}.}---{\it cf.}~Eqn.~(\ref{equ:BGx})---and can therefore be expressed locally as a sum of exact forms labeled by an index $\alpha$,
\beq
G_3 = \sum_\alpha \d(r^\alpha \Omega^{(\alpha)}_2)\, .
\eeq
Since Eqn.~(\ref{equ:FEoM}) is linear, we can focus on a single exact form,
\beq
\label{equ:G3exact}
G_3 = \d (r^\alpha \Omega_2^{(\alpha)})\, .
\eeq
There is no need to consider terms like $\d(r^{\#} \d r \wedge \Omega_1)$ as these are gauge equivalent to Eqn.~(\ref{equ:G3exact}).

For $\delta \ne 0$, the IASD forms $G_-$ and the ISD  forms $G_{+}$ are given by
\beq
\label{equ:IASD}
G_\mp =   -i \, \frac{\alpha \pm \delta}{\delta}\, r^\alpha \left( \d \Omega_2^{(\alpha)} \pm \delta\, \frac{\d r}{r} \wedge \Omega_2^{(\alpha)}\right)\, .
\eeq
The three-form $G_-$ in Eqn.~(\ref{equ:IASD}) satisfies Eqn.~(\ref{equ:FEoM}) when
\beq
(\alpha + \delta) (\alpha + 4 - \delta) \, r^{\alpha +4} \, \frac{\d r}{r} \wedge \d \Omega_2^{(\alpha)} = 0\, ,
\eeq
which holds for $\alpha = \delta - 4$ and $\alpha = -\, \delta$.
The form of the resulting linearized perturbation,
\beq
G_3 = \d (r^{\delta-4} \Omega_2 + r^{-\delta} \Omega_2)\, ,
\eeq
indicates that the two-form $\Omega_2$
corresponds, via AdS/CFT
(see \S\ref{sec:cft}), to an operator of dimension $\Delta=\delta$ for $\delta>2$, and to an operator of dimension $\Delta=4-\delta$ for $\delta<2$.
We are interested only in the {\it{non-normalizable}} mode $r^{\Delta-4}\Omega_2$, which corresponds to a perturbation of the field theory Lagrangian.  Moreover, we notice that for $\delta < 2$ the IASD three-form in Eqn.~(\ref{equ:IASD}) vanishes and the corresponding fluctuation
($G_{-}=0,~G_+ \ne 0$) does not affect a probe D3-brane.
Therefore, we restrict attention to the modes with $\delta > 2$, for which one has a non-normalizable perturbation containing IASD flux,
\bea
G_3 &=& \d(r^{\delta-4} \Omega_2) \, ,\\
G_- &=&  - 2 i\, \frac{\delta-2}{\delta}\, r^{\delta-4} \left( \d \Omega_2 + \delta\, \frac{\d r}{r} \wedge \Omega_2\right)\, .
\eea

The results above are general, {\it i.e.}~valid for arbitrary Calabi-Yau cones, but not entirely explicit. In any concrete example, such as
the warped conifold, $AdS_5 \times T^{1,1}$, we still need to obtain the eigenfunctions $\Omega_{2}^{(\alpha)}$ of the Laplace-Beltrami operator, as well as the corresponding spectrum of eigenvalues $\delta$.
This is achieved most easily along a slightly different route, as we explain in  \S\ref{sec:explicitsolution}.
There we derive explicit solutions for three series of IASD flux modes ($k = {\rm I}$, {\rm II}$, {\rm III}$) with positive $\delta_k$---Eqns.~(\ref{equ:ll1}), (\ref{equ:ll2}) and (\ref{equ:ll3})---for which the dimensions of the dual operators are $\Delta_k = \delta_k$.
We note that for each series~${\rm I, II, III}$ with positive $ \delta_{\rm I, II, III }$ there is a complementary series ${\rm \overline{I}, \overline{II}, \overline{III}}$ with negative $\delta_{\rm \overline{I}, \overline{II}, \overline{III}} = - \delta_{\rm I, II, III } $.
The dimensions of the dual operators in Series ${\rm \overline{I}, \overline{II}, \overline{III}}$ are
\bea
\label{series negative}
\Delta_{\overline{ k}}=4+ \delta_k\ .
\eea
These operators are dual to $G_+$ perturbations, and are therefore of limited interest for our considerations, whereas the operators in Series ${\rm I, II, III}$ are dual to mixtures of $G_-$ and $G_+$ perturbations and give rise to a D3-brane potential.

For completeness, we remark that if $X_5$ has a cohomologically nontrivial two-form with $\delta = 0$, the above harmonic analysis yields a special case with vanishing $G_3$ flux.  This mode does not affect the potential of a D3-brane, but is nontrivial in the sense that it corresponds to a perturbation by an operator in the field theory. For example, in the case of $X_5=T^{1,1}$ there is a single Betti two-form $\Omega_2= \omega_2$ with
\beq
\label{special3form}
G_3 = \d \omega_2 = 0 \qquad {\rm and} \qquad
G_- = 0\, .
\eeq
This mode changes the difference of the inverse coupling constants $g_1^{-2} - g_2^{-2}$ in the gauge theory.

\subsection{Explicit Flux Solutions}
\label{sec:explicitsolution}

In this section we give the explicit solutions for flux perturbations on arbitrary Calabi-Yau cones (see Appendix~\ref{sec:IASD} for details).
Furthermore, for the special case of the conifold background we derive the spectral dimensions of the perturbations.

\subsubsection{Building Blocks}

Our approach is simple.  For any Calabi-Yau cone, we directly construct the most general solution to
Eqn.~(\ref{equ:fluxEoM}) using the K\"ahler form $J$, the holomorphic $(3,0)$ form $\Omega$, and harmonic functions $f$ on the Sasaki-Einstein base $X_5$ as building blocks.  We will later specialize to $X_5=T^{1,1}$, in which case the harmonic functions $f$ are known in detail---{\it cf.}~Eqn.~(\ref{equ:harmX}).

The components of the K\"ahler form are
\beq
J_{\alpha \bar \beta} = i g_{\alpha \bar \beta}\, ,
\eeq
where $g_{\alpha \bar \beta} \equiv  \partial_\alpha \partial_{\bar \beta} k$ is the K\"ahler metric.
The holomorphic $(3,0)$ form has components
\beq
\Omega_{\alpha \beta \gamma} = q\, \epsilon_{\alpha \beta \gamma}\, ,
\eeq
where $q$ is a holomorphic function satisfying $q \bar q = \det g$.

Given these basic elements, we are ready to assemble the most general solution for IASD flux on a Calabi-Yau cone.

\subsubsection{Classification of Fluxes}

Three distinct types of closed, IASD three-forms can be constructed
using the ingredients of the previous section.
We now describe these solutions, leaving detailed derivations to Appendix~\ref{sec:IASD}.

\subsubsection*{Series I: $(1,2)$ Flux}

The  first and simplest flux series is of Hodge type $(1,2)$ with components,
\beq
\label{equ:Series1}
(\Lambda_{\rm I})_{\alpha \bar \beta \bar \gamma} = \nabla_\alpha \nabla_{\sigma} f_1 \, g^{\sigma \bar \zeta}\, \bar \Omega_{\bar \zeta \bar \beta \bar \gamma}\, ,
\eeq
where $f_1$ is a harmonic function and $\nabla_\alpha$ denotes the covariant derivative
with respect to the K\"ahler metric.
More compactly, this
flux can be written as
\beq
\label{equ:LambdaI}
\Lambda_{\rm I} = \nabla \nabla f_1 \cdot \bar \Omega\, .
\eeq
In Appendix~\ref{sec:IASD} we prove that $\Lambda_{\rm I}$ is indeed closed and
therefore satisfies the supergravity equations of motion.

We stress that Eqn.~(\ref{equ:LambdaI}) is valid for general Calabi-Yau manifolds.
However,
to quantify the radial scaling of the flux solution for a concrete example we now specialize to the conifold background.
We note that the radial scaling of the flux solution descends from the scaling dimension
$\Delta_{f}$ of the harmonic function $f_1$, defined in Eqn.~(\ref{equ:DeltafX}) in terms of the quantum numbers $j_1$, $j_2$, $R_f$.
The form $\Lambda_{\rm I}$ has the same $SU(2)\times SU(2)$ quantum numbers $j_1$, $j_2$
as $f_{1}$, but the $R$-charge is $R=R_f-2$. The $R$-charge is shifted by two because the anti-holomorphic three-form $\Bar \Omega$ has $R_{\Omega}= -2$. Given that $g^{\alpha \bar \beta}$ scales as $r^{-2}$ and $\Omega_{\alpha \beta \gamma}$ scales as $r^3$, the three-form $\Lambda_{\rm I}$ scales as
\beq
\Lambda_{\rm I} \sim r^4 G_{-}  \sim r^{\delta_{\rm I}}\, ,
\eeq
with
\beq
\label{equ:ll1}
\delta_{\rm I} = 1 + \Delta_f = - 1 + \sqrt{H(j_1,j_2,R+2)+4}\, .
\eeq
According to the AdS/CFT correspondence, $\delta_{\rm I}$ is the dimension of the dual field theory operator (see \S\ref{sec:cft}).
The
dual operator is chiral if the function $f_1$ is chiral, {\it i.e.}~obeys $j_1=j_2= \frac{1}{2}R_f$.
The lowest-dimension modes
of Series~I flux
are given in Table~\ref{tab:seriesI}.

 \begin{table}[h!]
\caption{\sl Series I: lowest modes of $(1,2)$ flux. }
\label{tab:seriesI}
\vspace{-0.5cm}
\begin{center}
\begin{tabular}{>{\columncolor{lightgray}\raggedright} c c c  c c c}
\toprule
 $\delta_{\rm I}$ & $j_1$ & $j_2$ & $R$ & {\footnotesize \bf Type} \\
\otoprule
 {\footnotesize $\frac{5}{2}$} & {\footnotesize $\frac{1}{2}$} & {\footnotesize $\frac{1}{2}$} & {\footnotesize $-1$} & {\footnotesize chiral} \\
\midrule
{\footnotesize $4$} & {\footnotesize $1$} & {\footnotesize $1$} & {\footnotesize $0$} & {\footnotesize chiral} \\
\midrule {\footnotesize $ \sqrt{28}-1$} & {\footnotesize $1$} & {\footnotesize $1$} & {\footnotesize $-2$} & {\footnotesize non-chiral} \\
\midrule{\footnotesize $\frac{9}{2}$} & {\footnotesize $\frac{1}{2}$} & {\footnotesize $\frac{3}{2}$} & {\footnotesize $-1$} & {\footnotesize non-chiral} \\
 {\footnotesize $\frac{9}{2}$} & {\footnotesize $\frac{3}{2}$} & {\footnotesize $\frac{1}{2}$} & {\footnotesize $-1$} & {\footnotesize non-chiral} \\
\midrule {\footnotesize $ \sqrt{40}-1$} & {\footnotesize $0$} & {\footnotesize $2$} & {\footnotesize $-2$} & {\footnotesize non-chiral} \\
{\footnotesize $ \sqrt{40}-1$} & {\footnotesize $2$} & {\footnotesize $0$} & {\footnotesize $-2$} & {\footnotesize non-chiral} \\
\midrule {\footnotesize $\frac{11}{2}$} & {\footnotesize $\frac{3}{2}$} & {\footnotesize $\frac{3}{2}$} & {\footnotesize $1$} & {\footnotesize chiral} \\
\bottomrule
\end{tabular}
\end{center}
\end{table}

\subsubsection*{Series II: $(2,1)_{\rm NP}+(1,2)$ Flux}

The second flux series is a mixture of fluxes with different Hodge types, non-primitive\footnote{Let us remark that if our analysis were extended to a compact Calabi-Yau space, there would be no non-primitive $G_{(2,1)}$ that is nontrivial in cohomology.} $(2,1)$, denoted $(2,1)_{\rm NP}$, and $(1,2)$,
\beq
\label{equ:LambdaII}
\Lambda_{\rm II} =( \partial +\bar \partial ) \Bigl(f_2 + \frac{1}{2} k^\alpha \partial_\alpha f_2 \Bigr) \wedge J + \partial(\bar \partial f_2 \wedge \bar \partial k) \, ,
\eeq
where $k^{\alpha} = g^{\alpha\bar\beta}\nabla_{\bar\beta}k$ is holomorphic (see Appendix~\ref{sec:IASD}).

The $R$-charge of the three-form $\Lambda_{\rm II}$ is the same as the $R$-charge of the harmonic function $f_2$, {\it i.e.}~$R=R_f$, while the radial scaling (and hence the dimension of the dual field theory operator) is shifted by two because $J_{\alpha \bar \beta} = i g_{\alpha \bar \beta} \sim r^2$, so
 \beq
 \label{equ:ll2}
\delta_{\rm II} = 2 + \Delta_f =  \sqrt{H(j_1,j_2,R)+4}\, .
\eeq
The lowest-dimension modes
of Series~II flux
are given in Table~\ref{tab:seriesII}.

 \begin{table}[h!]
 \caption{\sl Series II: lowest modes of $(2,1)_{\rm NP}+(1,2)$ flux. }
\label{tab:seriesII}
\vspace{-0.5cm}
\begin{center}
\begin{tabular}{ >{\columncolor{lightgray}\raggedright}  c c c  c c c}
\toprule
$\delta_{\rm II}$ & $j_1$ & $j_2$ & $R$ & {\footnotesize \bf Type} \\
\otoprule
 {\footnotesize $\frac{7}{2}$} & {\footnotesize $\frac{1}{2}$} & {\footnotesize $\frac{1}{2}$} & {\footnotesize $1$} & {\footnotesize chiral} \\
\midrule {\footnotesize $4$} & {\footnotesize $0$} & {\footnotesize $1$} & {\footnotesize $0$} & {\footnotesize non-chiral} \\
{\footnotesize $4$} & {\footnotesize $1$} & {\footnotesize $0$} & {\footnotesize $0$} & {\footnotesize non-chiral} \\
\midrule {\footnotesize $5$} & {\footnotesize $1$} & {\footnotesize $1$} & {\footnotesize $2$} & {\footnotesize chiral} \\
\midrule {\footnotesize $\sqrt{28}$} & {\footnotesize $1$} & {\footnotesize $1$} & {\footnotesize $0$} & {\footnotesize non-chiral} \\
\midrule {\footnotesize $\frac{11}{2}$} & {\footnotesize $\frac{1}{2}$} & {\footnotesize $\frac{3}{2}$} & {\footnotesize $1$} & {\footnotesize non-chiral} \\
{\footnotesize $\frac{11}{2}$} & {\footnotesize $\frac{3}{2}$} & {\footnotesize $\frac{1}{2}$} & {\footnotesize $1$} & {\footnotesize non-chiral} \\
{\footnotesize $\frac{11}{2}$} & {\footnotesize $\frac{1}{2}$} & {\footnotesize $\frac{3}{2}$} & {\footnotesize $-1$} & {\footnotesize non-chiral} \\
{\footnotesize $\frac{11}{2}$} & {\footnotesize $\frac{3}{2}$} & {\footnotesize $\frac{1}{2}$} & {\footnotesize $-1$} & {\footnotesize non-chiral} \\
\bottomrule
\end{tabular}
\end{center}
\end{table}

\subsubsection*{Series III: $(3,0)+(2,1)_{\rm NP}+(1,2)$ Flux}
The third flux series is a mixture of fluxes with three different Hodge types, $(3,0)$, $(2,1)_{\rm NP}$, and $(1,2)$,
\beq
\label{equ:LambdaIII}
\Lambda_{\rm III} =(2h + k^\alpha \partial_\alpha h) \Omega +(\bar \partial h \cdot \omega) \wedge J + \bar \partial (\bar \partial f_3 \cdot \omega) \wedge \bar \partial k\, ,
\eeq
where we have defined the following auxiliary forms,
\beq
\omega^{\bar \alpha}_{~\beta} \equiv \Omega^{\bar \alpha}_{\ \beta\gamma} k^{\gamma} \, , \qquad k^{\gamma} \equiv g^{\gamma \bar \zeta} \partial_{\bar \zeta} k\, ,
\qquad {\rm and} \qquad
h \equiv 3 f_3 + k^{\alpha} \partial_\alpha f_3 \, .
\eeq
The $R$-charge of the three-form $\Lambda_{\rm III}$ is $R=R_f+2$ and its radial scaling dimension is
 \beq
 \label{equ:ll3}
\delta_{\rm III} = 3 + \Delta_f = 1 + \sqrt{H(j_1,j_2,R-2)+4}\, .
\eeq
The lowest-dimension modes
of Series~III flux are given in Table~\ref{tab:seriesIII}.
This completes our classification of all closed, IASD fluxes on the singular conifold.

 \begin{table}[h!]
  \caption{\sl Series III: lowest modes of $(3,0)+(2,1)_{\rm NP} + (1,2)$ flux.}
\label{tab:seriesIII}
\vspace{-0.5cm}
\begin{center}
\begin{tabular}{ >{\columncolor{lightgray}\raggedright}  c c c  c c c}
\toprule
$\delta_{\rm III}$ & $j_1$ & $j_2$ & $R$ & {\footnotesize \bf Type} \\
\otoprule
 {\footnotesize $3$} & {\footnotesize $0$} & {\footnotesize $0$} & {\footnotesize $2$} & {\footnotesize chiral} \\
 \midrule
 {\footnotesize $\frac{9}{2}$} & {\footnotesize $\frac{1}{2}$} & {\footnotesize $\frac{1}{2}$} & {\footnotesize $3$} & {\footnotesize chiral} \\
 {\footnotesize $\frac{9}{2}$} & {\footnotesize $\frac{1}{2}$} & {\footnotesize $\frac{1}{2}$} & {\footnotesize $1$} & {\footnotesize non-chiral} \\
 \midrule
{\footnotesize $5$} & {\footnotesize $0$} & {\footnotesize $1$} & {\footnotesize $2$} & {\footnotesize non-chiral} \\
 {\footnotesize $5$} & {\footnotesize $1$} & {\footnotesize $0$} & {\footnotesize $2$} & {\footnotesize non-chiral} \\
\bottomrule
\end{tabular}
\end{center}
\end{table}

\subsubsection*{Chiral Modes}

For chiral perturbations the harmonic functions $f_1$, $f_2$, and $f_3$ are in fact holomorphic functions.
For holomorphic $f_2$ and $f_3$ the
fluxes of Series~II and III become a pure $(2,1)$ mode and a pure $(3,0)$ mode, respectively; {\it i.e.}
\bea
\Lambda_{\rm I}^{(1,2)} &=& \nabla^2 {\sf f}_1 \cdot \Omega\, , \label{equ:L1}\\
\Lambda_{\rm II}^{(2,1)} &=& \partial {\sf f}_2 \wedge J\, , \\
\Lambda_{\rm III}^{(3,0)} &=&  {\sf f}_3 \, \Omega\, , \label{equ:L3}
\eea
where ${\sf f}_1 \equiv f_1$, ${\sf f}_2 \equiv f_2 + \frac{1}{2} k^\alpha \partial_\alpha f_2$ and ${\sf f}_3 \equiv 6 f_3 + 5 k^\alpha \partial_\alpha f_3 + k^\alpha \partial_\alpha (k^\beta \partial_\beta f_3)$ are holomorphic functions related to $f_1$, $f_2$ and $f_3$.

\vskip 8pt
Next, we discuss in detail the effects that these fluxes have on D3-branes.

\newpage
\section{Spectrum of the D3-brane Potential}
\label{sec:spectrum}

In this section we present the complete spectrum of corrections to the D3-brane potential arising from compactification effects.
In \S\ref{sec:solutions} we compute the flux-induced corrections sourced by the closed, IASD three-form flux perturbations $\Lambda$, constructed in Appendix~\ref{sec:IASD} and reviewed in the previous section.
In
\S\ref{sec:subR}  we include ${\cal R}_4$ curvature corrections in the analysis.
We summarize the leading correction terms and discuss their physical significance in \S\ref{sec:summaryS}.

\subsection{Flux-Induced Corrections}
\label{sec:solutions}

As we explained above, in the noncompact limit ($M_{\rm pl} \to \infty$) the equation of motion for $\Phi_-$ contains only flux source terms,
\beq
\label{equ:PhiMinus0}
\nabla^2 \Phi_- = \frac{g_s}{96} |\Lambda|^2\, .
\eeq
We will discuss the contribution from four-dimensional curvature ${\cal{R}}_{4}$ for finite $M_{\rm pl}$ in \S\ref{sec:subR}.

\subsubsection{Green's Function Solution}
The Green's function solution to Eqn.~(\ref{equ:PhiMinus0}) is
\beq
\label{equ:PhiMinus00}
\Phi_-(y) = \frac{g_s}{96} \int \d^6 y' \, G(y;y') \, |\Lambda|^2(y')\, + \Phi_{\cal H}(y) ,
\eeq
where
\beq
\nabla_y^2 \,G(y;y') = \delta(y-y')\, ,
\eeq
and the homogeneous solution $\Phi_{\cal H}(y)$ is an arbitrary harmonic function, {\it i.e.}~a solution of
\beq
\nabla_y^2 \,\Phi_{\cal H}(y) = 0\, ,
\eeq
whose general form is given in Eqn.~(\ref{equ:harmX}).
The spectrum of the homogeneous solution was studied in Ref.~\cite{Baumann2009} (see Table~\ref{tab:f}):
\bea
\label{equ:DeltaH}
\Delta_{\cal H} &=& \frac{3}{2}\ , \ 2 \ , 3 \ ,\ ...
\eea
The Green's function on the singular conifold was presented in Refs.~\cite{Klebanov2007, BDKMMM},
\beq
\label{equ:Green}
G(y;y') = \sum_{L,M} Y_{LM}(\Psi) Y_{LM}^*(\Psi') g_L(r;r') \, ,
\eeq
where
\beq
g_L(r;r') \equiv - \frac{1}{2\Delta(L)+4} \left\{ \begin{array}{c l} \frac{1}{r'^4} \Bigl( \frac{r}{r'}\Bigr)^{\Delta(L)} & r \le r' \\
\frac{1}{r^4} \Bigl( \frac{r'}{r}\Bigr)^{\Delta(L)} & r \ge r'
 \end{array}\right. \, . \label{equ:gL}
\eeq
In Eqn.~(\ref{equ:gL}) the scaling $\Delta(L)$ is defined as in Eqn.~(\ref{equ:DeltafX}).
The D3-brane potential can always be written in the form of Eqn.~(\ref{equ:PhiMinus00}), but in the particularly interesting special case of chiral perturbations, significant simplifications of the final answer can be achieved.

\subsubsection{Chiral Modes}

In Eqns.~(\ref{equ:L1}) -- (\ref{equ:L3}) we have seen that for chiral perturbations each flux series is of distinct Hodge type.
Hence, no mixed terms of the different flux series appear in the flux-squared source term, \beq
\label{equ:LambdaSqr}
|\Lambda|^2 = |\Lambda_{\rm I}|^2 + |\Lambda_{\rm II}|^2 + |\Lambda_{\rm III}|^2 \, ,
\eeq
where
\bea
 |\Lambda_{\rm I}|^2 &=& 6\, g^{\alpha \bar \alpha} g^{\beta \bar \beta} \nabla^2_{\alpha \beta} {\sf f}_1 \,\overline{\nabla^2_{\alpha \beta} {\sf f}_1}\, , \\
 |\Lambda_{\rm II}|^2 &=& 12\, g^{\alpha \bar \alpha} \nabla_{\alpha} {\sf f}_2 \, \overline{\nabla_{\alpha} {\sf f}_2} \, ,\\
|\Lambda_{\rm III}|^2 &=& 6\, |{\sf f}_3|^2\, .
\eea
The flux-induced potential then becomes
\beq \label{phiminus}
\Phi_- \ = \ \frac{g_s}{96} \Bigl[ 3 g^{\alpha \bar \alpha} \nabla_\alpha {\sf f}_1 \, \overline{\nabla_\alpha {\sf f}_1} + 12 |{\rm Re}({\sf f}_2)|^2 + 6 \nabla^{-2}|{\sf f}_3|^2 \Bigr] \ + \ {\rm harmonic}\, .
\eeq

\subsubsection{General Solution}

In general, we should allow the functions $f_i$ to be harmonic rather than just holomorphic. In this case the
fluxes of  Series~II and III are not of pure Hodge type and the overlap of flux modes from
Series~I, II, and III is
nonzero and can lead to new terms in the D3-brane potential.

We now discuss the
resulting spectrum of $\Phi_-$ up to $r^4$,
\beq
\label{equ:Deltaij}
\Phi_- = \sum_{\delta_i, \delta_j} r^{\Delta(\delta_i, \delta_j)} h_{(\delta_i, \delta_j)}(\Psi)\, ,
\eeq
where $h_{(\delta_{i}, \delta_j)}(\Psi)$ are angular wavefunctions describing the overlap of the different flux modes.
The explicit forms of the functions $h_{(\delta_{i}, \delta_j)}(\Psi)$ can be inferred from Eqn.~(\ref{equ:Green}), but we will not write them out here.
Instead, we focus on the radial scaling dimensions in Eqn.~(\ref{equ:Deltaij}), which are
\beq
\Delta \equiv \delta_i + \delta_j -4\, ,
\eeq
where $\delta_i$ and $\delta_j$ are the scaling dimensions of the fluxes $\Lambda_i$ and $\Lambda_j$.
Using the scaling dimensions in Tables~\ref{tab:seriesI}, \ref{tab:seriesII}, and \ref{tab:seriesIII} and recalling that for chiral modes the overlaps between different flux series vanish, we may infer the smallest scaling dimensions of the flux-induced potential:
\bea
\label{equ:DeltaLL}
\Delta_\Lambda &=& 1\ , \ 2\ ,\ \frac{5}{2}\ , \ \sqrt{28} - \frac{5}{2} \  , \ \dots
\eea

We notice that the  square of a $\delta = \frac{5}{2}$ mode results in a linear term in the potential, $\Phi_- \propto r$. This is the same $r$-scaling as the leading term in the nonperturbatively-generated D3-D7 potential of Ref.~\cite{BDKM}. We will expand on this correspondence in the following sections.  In fact, in \S\ref{sec:4dSUGRA} we will argue that the full nonperturbative potential of Ref.~\cite{BDKM} can be ``geometrized" by turning on appropriate fluxes.

The wavefunctions of the $\delta=\frac{5}{2}$ and $\delta=3$ modes have zero overlap since the corresponding modes are chiral perturbations in different flux series, {\it i.e.}
\beq
h_{(\frac{5}{2},3)}(\Psi) = 0\, .
\eeq
Hence, there is no $\Delta = \frac{3}{2}$ contribution to the flux-induced potential.

The overlap of the $\delta= \frac{5}{2}$ chiral mode and the
 non-chiral $\delta = \sqrt{28} - 1$ mode (corresponding to the leading contribution of an operator in a long multiplet) results in a potential term with irrational scaling dimension $\sqrt{28}-5/2\approx 2.79$ .
This term is not protected by supersymmetry or by a global symmetry\footnote{In \cite{Baumann2009} we considered non-chiral operators with $\Delta=2$, but these operators were related by supersymmetry to global symmetry currents and hence were protected.}, and hence would be inaccessible in a field theory analysis at large 't~Hooft coupling or in a four-dimensional supergravity analysis
such as
 \cite{BDKMMM,BDKM} in which only superpotential interactions are computed.  In a general compactification one expects this mode of flux to contribute to the D3-brane potential, and a substantial advantage of the supergravity approach of this paper is the ability to capture such terms.

\subsection{Coupling to the Ricci Scalar}
\label{sec:subR}

The potential for a D3-brane in a noncompact warped conifold perturbed by
non-normalizable modes of IASD flux is given by the
solution (\ref{equ:PhiMinus00}) of (\ref{equ:PhiMinus0}).  However, a D3-brane in a warped throat region of a {\it compact} space receives an additional contribution to its potential when the four-dimensional Ricci scalar is nonvanishing: one must then solve
\beq
\label{equ:Master1}
\nabla^2 \Phi_- = \frac{g_s}{96} |\Lambda|^2 + {\cal R}_4\, .
\eeq
In particular, if we assume that the four-dimensional theory is approximately de Sitter, then the Ricci scalar is given by the Friedmann equation as
\beq
\label{equ:Ricci}
{\cal R}_4 = 12 H^2 \approx \frac{4 }{M_{\rm pl}^2} V = \frac{4}{M_{\rm pl}^2} \Bigl(V_0 + T_3\Phi_-\Bigr)\, ,
\eeq
where
$V_0$, a constant
independent of the D3-brane position $r$, has been extracted from the potential.  We have incorporated the fact that
the D3-brane potential $T_3\Phi_-$ contributes to the four-dimensional energy density, but there may be additional contributions from other sectors included in $V_{0}$.

\subsubsection{The Eta Problem}

For simplicity we first study Eqn.~(\ref{equ:Master1}) in the absence of IASD flux perturbations,
\beq
\label{equ:Master2}
\nabla^2 \Phi_- = {\cal R}_4\, .
\eeq
 This case is sufficient to understand the appearance of a generic eta problem arising from the curvature coupling of the inflaton in D3-brane models \cite{KKLMMT}
 (see also \cite{Buchel2004, Buchel2006}). Combining Eqns.~(\ref{equ:Ricci}) and (\ref{equ:Master2}) we find
 \beq
 \label{equ:q}
 \nabla^2 Q = \lambda\, Q \, , \qquad \lambda \equiv \frac{4 T_3}{M_{\rm pl}^2}\ ,
 \eeq
where
\beq
Q(r, \Psi) \equiv \Phi_- + \frac{V_0}{T_3}\, .
\eeq
We solve Eqn.~(\ref{equ:q}) by separation of variables,
\beq
Q(r,\Psi)= \sum_L Q_L(r) h_L(\Psi)\, ,
\eeq
where the radial functions satisfy
\beq
\label{equ:qL}
\frac{d^2 Q_{L}}{dr^2} + \frac{5}{r} \frac{d Q_L}{d r} - \frac{H(L)}{r^2} Q_L = \lambda\, Q_L\, .
\eeq
Here, $H(L)$ are the eigenvalues of the angular Laplacian defined in Eqn.~(\ref{equ:HX}).
Eqn.~(\ref{equ:qL}) may be solved exactly in terms of (modified) Bessel functions,
\beq
 Q_L =  x^{-2} \left[ c_1 I_{n}(x) + c_2 K_n(x)\right]\, ,
\eeq
where
\beq
x \equiv \sqrt{\lambda} r = 2 \frac{\phi}{M_{\rm pl}} \qquad {\rm and} \qquad n^2(L) \equiv  H(L) + 4\, .
\eeq
The functions $I_n(x) = i^{-n} J_n(i x)$ and $K_n(x)$ are modified Bessel functions of the first and second kind, respectively. $K_n$ diverges for small $x$, so we choose $c_2 = 0$. The second integration constant is fixed by the requirement $\Phi_-(x=0) \equiv 0$ to be
\beq
c_1 = \frac{8 V_0}{T_3}\, .
\eeq
Hence, we find
\beq
\label{equ:InFull}
V(\phi)\ = \ V_0 + T_3 \Phi_- \ =\  2 V_0 \,  \frac{M_{\rm pl}^2}{\phi^2}\,   \sum_L I_{n(L)} \Bigl(2 \frac{\phi}{M_{\rm pl}} \Bigr) h_L(\Psi) \, .
\eeq
The leading terms in the potential follow from the small $x$ expansion\footnote{\label{expansion}The bound of Ref.~\cite{Baumann2007}, $x = 2 \frac{\phi}{M_{\rm pl}} < \frac{4}{\sqrt{N}}< 1$, in an $AdS_5 \times X_5$ geometry with D3-brane charge $N \gg 1$, implies that $x$ is a good expansion parameter.},
\beq
\label{equ:InSmall}
I_n(x) = \Bigl(\frac{x}{2}\Bigr)^n \sum_{k=0}^\infty \frac{\bigl( \frac{x}{2} \bigr)^{2k}}{ k! \, \Gamma(n+k+1)}\, \quad {\rm for} \quad n \in \mathbb{R}\, .
\eeq
One finds that the leading contribution arises from the term with $L=\{0\}$, and takes the form
\beq
I_2(x) = \frac{1}{8} x^2 \Bigl( 1 + \frac{1}{12} x^2 + \cdots \Bigr)\, .
\eeq
This implies
\beq
V_s \ \equiv \ T_3\, Q_{L=\{0\}} \ =\ V_0\, \Bigl( 1 + \frac{1}{3} \frac{\phi^2}{M_{\rm pl}^2} + \cdots \Bigr)\, ,
\eeq
and
\beq
\eta \ =\ M_{\rm pl}^2 \frac{V''}{V} \ = \ \frac{2}{3} \ + \ \cdots
\eeq
The leading curvature correction therefore precisely explains the `eta problem' mass term found in~\cite{KKLMMT}. Higher-order contributions from modes with non-trivial angular dependence may be obtained from Eqns.~(\ref{equ:InFull}) and (\ref{equ:InSmall}).

\subsubsection{Higher-Order Corrections}

Next, we include IASD flux perturbations in the analysis.
The equation of motion for $\Phi_-$ now has the form
\begin{equation}\label{equ:iter}
\nabla^2 \Phi_- = \rho(y) + \lambda\Phi_{-}  \, ,
\end{equation} where
\begin{equation}
\rho(y) \equiv \frac{g_s}{96} |\Lambda|^2(y) + \frac{4}{M_{\rm pl}^2} V_0 \, .
\end{equation}
One can easily verify that Eqn.~(\ref{equ:iter}) is solved by
\beq
\label{iterative}
\Phi_- = \sum_{n=0}^\infty \Phi_-^{[n]}\, ,
\eeq
where
\beq
\label{equ:zero}
\Phi_-^{[0]}(y) = \int \d^6 y' \, G(y;y')\, \rho(y') + \Phi_{\cal H}(y) \, ,
\eeq
and
\beq
\label{equ:one}
\Phi_-^{[1]}(y) = \lambda \int \d^6 y' \, G(y;y')\, \Phi_-^{[0]}(y') \quad , \quad \cdots \quad , \quad
\Phi^{[n]}_-(y) = \lambda \int \d^6 y'\, G(y;y')\, \Phi_-^{[n-1]}(y') \, .
\eeq
Eqn.~(\ref{equ:zero}) describes the solution in the absence of Ricci curvature---{\it cf.}~Eqn.~(\ref{equ:PhiMinus00})---plus the mass term of the previous section associated with the constant $V_0$,
\beq
\label{equ:phi[0]}
\Phi_-^{[0]} = \sum_\alpha r^{\alpha} h_\alpha^{[0]}(\Psi)\ , \quad {\rm with} \quad \alpha = \{ \Delta_\Lambda , \Delta_{\cal H} , 2_{s} \}\ ,
\eeq
where $h^{[0]}_\alpha(\Psi)$ are angular wavefunctions determined by (\ref{equ:zero}).
 Eqn.~(\ref{equ:one})
provides the iterative inclusion of higher-order corrections induced by the Ricci curvature.
Using Eqns.~(\ref{equ:Green}) and (\ref{equ:phi[0]}) to perform
the Green's function integrals in Eqn.~(\ref{equ:one}), we find
\beq
\Phi_-^{[n]} = (\lambda r^2)^n \sum_\alpha r^\alpha h_\alpha^{[n]}(\Psi)\, ,
\eeq
with angular wavefunctions $h_\alpha^{[n]}$ determined by the above.
Evidently,
Eqn.~(\ref{iterative}) is an expansion in the dimensionless parameter $x \equiv \lambda r^2 = 4 \frac{\phi^2}{M_{\rm pl}^2} < 1$ (see Footnote~\ref{expansion}).
The iterative inclusion of curvature contributions effectively dresses each term in the potential with factors $r^{2n}$, $n=1,2, \ \cdots$ .
We therefore infer from Eqns.~(\ref{equ:DeltaH}) and (\ref{equ:DeltaLL}) that the corrections induced by the coupling to the Ricci scalar are of the form $\Phi_- \sim r^\Delta$ with
\bea
\Delta_{\cal R} &=& 2_s\ ,\ 3\ ,\ \frac{7}{2}\ ,\ 4\ , \ \cdots \
\eea

\subsection{Summary of Supergravity Perturbations}
\label{sec:summaryS}

Let us briefly summarize the contributions to the D3-brane potential. The potential is given by $T_3\Phi_-$, where $\Phi_-$ is a general solution (\ref{iterative}) of the equation of motion (\ref{equ:iter}).  This solution includes a homogeneous solution $\Phi_{\cal H}$, which is an arbitrary harmonic function on the conifold, and also includes inhomogeneous contributions
sourced by $|\Lambda|^2$ and by ${\cal R}_4$.\vskip 8pt

We are interested in the dominant terms in the infrared, {\it i.e.}~the most relevant contributions for a D3-brane far from the ultraviolet region where the throat is attached to the compact bulk.\footnote{Recall
that we also assume that the D3-brane is well above any infrared deformations of the geometry as occur {\it e.g.}~in the warped deformed conifold solution \cite{KS}.}
We write the solution in terms of the canonical coordinate $\phi^2 = T_3 r^2$.
The leading terms from the homogeneous solution take the form \cite{Baumann2009}
\begin{equation}
V_{\cal H}(\phi)\ = \ V_0\ + \ a_{3/2}\, \phi^{3/2} h_{3/2}(\Psi)\ +\ a_2\, \phi^2 h_{2}(\Psi) \ + \ \dots \ \,,
\end{equation}
where $h_\alpha(\Psi)$ are angular harmonics on $T^{1,1}$
and $a_\alpha$ are constants.
We have seen in \S\ref{sec:solutions} that the leading terms from the flux-sourced solution take the form
\begin{equation}
V_\Lambda(\phi)\ = \ b_1\, \phi^1 j_{1}(\Psi) \ + \ b_2\, \phi^2 j_{2}(\Psi) \ + \ b_{5/2} \, \phi^{5/2} j_{5/2}(\Psi) \ + \ b_{2.79} \, \phi^{2.79} j_{2.79}(\Psi)  \ + \  \dots \ \,,
\end{equation}
where the $j_{\alpha}(\Psi)$ are angular functions determined by the analysis of \S\ref{sec:solutions}.
Finally, the leading term from the curvature-sourced solution, as obtained in \S\ref{sec:subR}, takes the form
\begin{equation}
V_{\cal R}(\phi)\ = \ c_2\, \phi^2 \ + \ \dots \,,
\end{equation}
where $c_2 \equiv  H_0^2 = \frac{1}{3} \frac{V_0}{M_{\rm pl}^2}$ is a fixed coefficient. The Wilson coefficients $a_\Delta$, $b_\Delta$ and $c_\Delta$ are normalized at the UV cutoff, so it is convenient to extract the energy scaling as follows
\beq
\{\, a_\Delta \, , \, b_\Delta \, , \, c_\Delta \, \} =  \{\, \widetilde{a}_\Delta \, , \, \widetilde{b}_\Delta \, , \, \widetilde{c}_\Delta \, \} \cdot M_{\rm UV}^{4-\Delta}\ ,
\eeq
so that $\widetilde{a}_\Delta$, $\widetilde{b}_\Delta$ and $\widetilde{c}_\Delta$ are dimensionless coefficients.

Finally,
 the general form of the potential is
\begin{eqnarray}
\label{equ:OurAnswerAtLast}
V(\phi) &=&  V_0 \ + \ b_1\,j_{1}(\Psi) \, \phi^1  \ + \ a_{3/2}\,  h_{3/2}(\Psi)\, \phi^{3/2} \ +\ \Bigl(c_2 + a_2\, h_{2}(\Psi)+ b_2\, j_{2}(\Psi)\Bigr)\, \phi^2 \nonumber\\
&& \hspace{0.65cm} + \  b_{5/2}\, j_{5/2}(\Psi)\, \phi^{5/2}  \ +\  b_{2.79}\, j_{2.79}(\Psi)\, \phi^{2.79}  \ + \ \dots
\end{eqnarray}
This is one of the main results of this paper.

Notice the disparate origin of the various competing terms in Eqn.~(\ref{equ:OurAnswerAtLast}): IASD fluxes, harmonic perturbations of $\Phi_-$, and coupling to the four-dimensional
spacetime
curvature. The curvature coupling was identified in Ref.~\cite{KKLMMT} and further investigated in Refs.~\cite{Buchel2004, Buchel2006}.  The harmonic perturbations of $\Phi_-$ were studied in Ref.~\cite{Baumann2009}.  The flux contribution $\phi^{1}$, which actually dominates at small $\phi$,
has not been previously identified in ten-dimensional supergravity.
Also novel is the contribution $\Delta = \sqrt{28} - \frac{5}{2} \approx 2.79$, which was inaccessible in analyses such as \cite{BDKM} that incorporated only corrections to the superpotential. Clearly, such a mode would be very difficult to guess in field theory, but as we see here it can make an important contribution to the D3-brane potential.

We remark that the leading curvature contribution has a fixed coefficient, leading to the generic eta problem. However, the other contributions have tunable Wilson coefficients that in principle allow a small effective mass term if the different contributions locally cancel against each other, as described in four-dimensional supergravity in \cite{Baumann2007a,Krause2008,BDKM}.   The results obtained above provide a dictionary between physical effects in the compactification and specific terms in the inflaton potential, {\it e.g.}~the linear term arises only from $G_{(1,2)}$ flux.   This characterization of the physical origin of each term is a necessary precursor to any attempt at explicit fine-tuning of the potential.

 \section{Conformal Field Theory}
\label{sec:cft}

Having understood the structure of the D3-brane potential  induced by non-normalizable perturbations of the supergravity solution, it is natural to ask whether additional insights, or cross-checks, can be obtained by mapping these results into the dual conformal field theory.  As usual when applying the AdS/CFT correspondence at large 't Hooft coupling, the gauge theory is strongly coupled and most computations are far simpler in the gravity picture.
However, understanding the D3-brane potential on the gauge theory side will help us to make contact with four-dimensional reasoning, by clarifying the relation between our analysis and a Wilsonian treatment in the four-dimensional effective theory.

\vskip 4pt
We begin by reviewing
the field content of the Klebanov-Witten (KW) theory and
listing all protected operators~(\S\ref{sec:KW}).
We then discuss the operators dual to IASD flux perturbations in more detail (\S\ref{sec:CFTflux}).
We first present all
chiral operators (\S\ref{sec:chiral}), explicitly identifying three towers of operators as duals to the three
series of holomorphic flux.
Next, we discuss the complete AdS/CFT spectroscopy (\S\ref{sec:nonchiral}) including non-chiral operators.
Finally, we present arguments that allow a comparison between the scalar potentials computed in the gauge theory and in supergravity (\S\ref{sec:potential}).

 \subsection{Review of the KW CFT}
\label{sec:KW}

A canonical class of examples of the AdS/CFT correspondence
consists of ${\cal N}=1$ superconformal gauge theories dual to string theory on $AdS_5 \times X_5$, where $X_5$ is a Sasaki-Einstein space~\cite{KW, KT, KS}.  These theories arise from the near-horizon limit of a stack of
$N$ D3-branes placed at the tip of a six-dimensional Calabi-Yau cone $M_6$ with base manifold $X_5$. The supergravity results of the previous sections were general enough to capture all $AdS_5 \times X_5$ backgrounds. In principle, the gauge theory analysis could be formulated equally generally,
but in practice we will specialize our CFT results to the dual of $AdS_5 \times T^{1,1}$.

\vskip 6pt
{\sl Basic symmetries and degrees of freedom.}
\hskip 6pt
The CFT dual of $AdS_5 \times T^{1,1}$ is an ${\cal N}=1$ supersymmetric Yang-Mills theory
with gauge group ${\cal G} = SU(N) \times SU(N)$ and continuous global symmetries $G = SU(2) \times SU(2) \times U(1)_{R}$
(inherited from the isometries of $T^{1,1}$) \cite{KW}.
The matter content consists of two doublets of chiral superfields $A_i$ and $B_j$ ($i,j=1,2$) which are in the ($N$, $\bar N$) and ($\bar N$, $N$) of ${\cal G}$ and in the $(2, 0, \frac{1}{2})$ and $(0,2,\frac{1}{2})$ of $G$.
Introducing the $SU(N)$ vector superfields $V_i$, the chiral gauge field strength superfields for the two gauge symmetries are
\bea
W_\alpha^{(1)} &=& \bar D \bar D(e^{V_1} D_\alpha e^{-V_1}) \, , \\
W_\alpha^{(2)} &=& \bar D \bar D(e^{V_2} D_\alpha e^{-V_2}) \, .
\eea
 The $W_\alpha^{(i)}$ fields have dimension $\Delta = \frac{3}{2}$ and $R$-charge $R=1$, while the $A$ and $B$ fields have  $\Delta = \frac{3}{4}$ and $R=\frac{1}{2}$.
 The interactions are encoded in the superpotential
 \beq
 W \propto \epsilon^{ij} \epsilon^{kl} \, {\rm Tr}(A_i B_k A_j B_l)\, ,
 \eeq
with $\Delta=3$ and $R=2$.  We now use these results to enumerate the low-dimension operators in the CFT, following Ceresole {\it{et al.}}~\cite{Ceresole2000,Ceresole1999}.

{\sl Operators with protected dimensions.}
\hskip 6pt We are interested in determining the most relevant operators in the CFT that contribute to the potential on the Coulomb branch.
We first consider operators whose dimensions are protected in the gauge theory;
unprotected operators generically acquire large anomalous dimensions in the limit of large 't Hooft coupling and have their dual description only in the full string theory, not in supergravity.
Later, in \S\ref{sec:nonchiral}, we will include the very interesting case of operators that appear unprotected in the gauge theory analysis, but in fact have fixed -- and in many cases, irrational -- dimensions at large 't Hooft coupling \cite{Gubser}.  For simplicity we refer to such operators as unprotected.

For the CFT in question there are
three series of protected operators \cite{Ceresole2000}:
\begin{enumerate}
\item {\it Chiral}
\begin{alignat}{3}
S^k &= {\rm Tr}(AB)^k\ : &\qquad \Delta_k &= \frac{3}{2} k &\qquad  R_k &=k\, , \label{equ:FirstOperator}\\
T^k_\alpha &= {\rm Tr}[W_\alpha (AB)^k]\ : &\qquad \Delta_k &= \frac{3}{2} + \frac{3}{2} k &\qquad R_k &=k+1\, ,\\
\Phi^k &= {\rm Tr}[W^\alpha W_\alpha (AB)^k]\ : &\qquad \Delta_k &= 3 + \frac{3}{2} k &\qquad R_k &=k+2\, . \label{equ:C3}
\end{alignat}

\item {\it Conserved}
\begin{alignat}{3}
J^k &= {\rm Tr}[J (AB)^k)] \ : &\qquad \Delta_k &= 2 + \frac{3}{2} k &\qquad R_k &= k\, , \\
J^k_{\alpha \dot \alpha} &= {\rm Tr}[J_{\alpha \dot \alpha} (AB)^k]\ : &\qquad \Delta_k &= 3 + \frac{3}{2} k &\qquad R_k &= k\, , \\
I^k &= {\rm Tr}[J W^2 (AB)^k] \ : &\qquad \Delta_k &= 5 + \frac{3}{2} k &\qquad R_k &= k+2\, ,
\end{alignat}
where $J = \{ J_a, J_b\}$, $J_a \equiv A e^V \bar A e^{-V}$, $J_b \equiv B e^V \bar B e^{-V}$ (with $V \equiv V_1+V_2$), $J_{\alpha \dot \alpha} \equiv W_\alpha L_{\dot \alpha}$,
and $L_{\dot \alpha} \equiv e^V \bar W_{\dot \alpha} e^{-V}$.

\item {\it Semi-conserved}
\begin{alignat}{3}
L^{1, k}_{\dot \alpha} &= {\rm Tr}[L_{\dot \alpha} (AB)^k]\ : &\qquad \Delta_k &= \frac{3}{2} + \frac{3}{2}k &\qquad R_k &= k-1\, , \\
L^{2, k}_\alpha &= {\rm Tr}[W_\alpha J (AB)^k]\ : &\qquad \Delta_k &= \frac{7}{2} + \frac{3}{2} k  &\qquad R_k &= k+1\, , \\
L^{3, k}_{\dot \alpha} &= {\rm Tr}[L_{\dot \alpha} W^2 (AB)^k] \ : &\qquad \Delta_k &= \frac{9}{2} + \frac{3}{2} k &\qquad R_k &= k+1  \, . \label{equ:LastOperator}
\end{alignat}
\end{enumerate}

The operators here are written a bit schematically: the trace is over the color degrees of freedom, and we have suppressed  the $SU(2) \times SU(2)$ flavor indices.
Furthermore, the KW theory has two gauge groups, and naively there should be two distinct series for the operators $T^k_{\alpha}$ and $\Phi^k$, one with $W_\alpha=W^{(1)}_\alpha$ and the other with $W_\alpha=W^{(2)}_\alpha$. However, the commutators of the matter fields and $W_\alpha$ vanish in the chiral ring \cite{Cachazo2002},
\begin{equation}
AW_\alpha^{(1)}-W_\alpha^{(2)} A\ \sim \ W_\alpha^{(1)} B-BW_\alpha^{(2)} \ \sim \ 0  \, .
\end{equation}
Hence, there is only one chiral mode with protected dimension, ${\rm Tr}[(W^{(1)}_\alpha+W^{(2)}_\alpha) (AB)^k]$, while the twisted mode ${\rm Tr}[(W^{(1)}_\alpha - W^{(2)}_\alpha) (AB)^k]$ vanishes in the chiral ring and has infinite dimension in the large $N$ limit. Similarly, the chiral primary operator $\Phi^k$ should be written as ${\rm Tr}[ (W_{(1)}^2+W_{(2)}^2) (AB)^k]$. Finally, there is an exceptional chiral operator with protected dimension,
\begin{equation}
\label{specchiral}
{\rm Tr}[W_{(1)}^2-W_{(2)}^2] \, ,
\end{equation}
that does not belong to any tower.

We define the conformal dimension and $R$-charge of a supermultiplet as the conformal dimension and $R$-charge of the bottom ($\theta = 0$) component of the corresponding superfield.
The conformal dimensions and $R$-charges of descendants are then shifted in half-integer and integer steps, respectively.
In the following we will refer to the individual components of a superfield by the notation
\beq
F = [F]_{\rm b} + \dots + [F]_{\theta^n \bar \theta^m} \, \theta^n \bar \theta^m + \cdots \ .
\eeq
This completes the bookkeeping required for this section.

\subsection{CFT Duals of IASD Fluxes}
\label{sec:CFTflux}

Having enumerated the lowest-dimension protected supermultiplets of operators, we now focus on the components of these supermultiplets that generate a potential on the Coulomb branch.

\subsubsection{Chiral Operators}
\label{sec:chiral}

We consider first the three sets of chiral operators in Eqns.~(\ref{equ:FirstOperator}) -- (\ref{equ:C3}):
\bea
{\cal O}_{\rm I} &=& \Tr(AB)^k \, , \label{equ:cc1}\\
{\cal O}_{\rm II}^\alpha &=& \Tr[W^\alpha_+ (AB)^k]\, , \\
{\cal O}_{\rm III} &=& \Tr [W^2_+ (AB)^k]\, , \label{equ:cc3}
\eea
where we have introduced the
notation $W^\alpha_\pm \equiv W^\alpha_{(1)} \pm W^\alpha_{(2)}$ and $W^2_\pm \equiv W^2_{(1)} \pm W^2_{(2)}$.
The integer $k$ obeys $k \ge 1$ for Series ${\rm I}$ and ${\rm II}$ and $k \ge 0$ for Series ${\rm III}$.

Our interest is in the components of the above supermultiplets that induce a D3-brane potential, {\it i.e.}~the components dual to $\Phi_-$ and $G_-$ perturbations.
The bottom component of ${\cal O}_{\rm I}$ corresponds to a perturbation of  $\Phi_-$ \cite{Baumann2009}:
\begin{table}[h!]\footnotesize
\begin{center}
\begin{tabular}{c c c c c c}
\toprule
\hskip 5pt $\mathbf{\Phi_-}$ \hskip 10pt  &{\bf Operator} & \hskip 4pt $\mathbf{\Delta}$ \hskip 4pt & $R$ & $j_1$ & $j_2$\hskip 5pt \\ \otoprule
\hskip 5pt$f$ \hskip 10pt & \hskip 2pt $[{\rm Tr}(AB)^k]_{\rm b}$ & \hskip 2pt $\frac{3}{2} k$ \hskip 2pt & \hskip 2pt $k$ \hskip 2pt & $\frac{1}{2}k$ & $\frac{1}{2} k$ \hskip 5pt \\
\bottomrule
\end{tabular}
\end{center}
\end{table}

\vspace{-0.6cm}
\noindent
while the top, middle and bottom components, respectively, of ${\cal O}_{\rm I}$, ${\cal O}_{\rm II}^\alpha$, and ${\cal O}_{\rm III}$ are dual to $G_-$ perturbations:
\begin{table}[h!]\footnotesize
\begin{center}
\begin{tabular}{l l c c c c}
\toprule
\hspace{0.15cm} {\bf Flux}   &\hspace{.7cm}  {\bf Operator} & \hskip 4pt $\mathbf{\Delta}$ \hskip 4pt & $R$ & $j_1$ & $j_2$\hskip 2pt \\ \otoprule

\hskip 2pt $\nabla \nabla {\sf f}_1 \cdot \bar \Omega$ & \hskip 2pt $[{\rm Tr}(AB)^k]_{\theta^2}$ & \hskip 2pt $\frac{3}{2} k + 1$ \hskip 2pt & \hskip 2pt $k-2$ \hskip 2pt & $\frac{1}{2}k$ & $\frac{1}{2} k$ \hskip 2pt \\
\midrule \hskip 2pt  $\partial {\sf f}_2 \wedge J$  &  \hskip 2pt $[{\rm Tr}[W^\alpha_+ (AB)^k]]_\theta$ \hskip 4pt&
$\frac{3}{2} k + 2$& $k$ & $\frac{1}{2} k$ & $\frac{1}{2} k$ \hskip 2pt\\
\midrule \hskip 2pt ${\sf f}_3  \Omega$ &  \hskip 2pt  $[{\rm Tr}[(W^2_+ (AB)^k]]_{\rm b}$  \hskip 4pt  & $\frac{3}{2} k + 3$ & $k+2$ & $\frac{1}{2} k$ & $\frac{1}{2} k$ \hskip 2pt\\
\bottomrule
\end{tabular}
\end{center}
\end{table}

\vspace{-0.3cm}
Evidently, these three towers of chiral operators are dual to the three series of `holomorphic' flux perturbations found in \S\ref{sec:flux}.
A few comments are in order:
\begin{itemize}
\item {\sl Series I operators.}
The perturbations $\int d^2 \theta \, {\cal O}_{\rm I} = [{\cal O}_{\rm I}]_{\theta^2}$ are  {\it{superpotential perturbations}}, consistent with the supersymmetry of the unperturbed CFT.
These operators are dual to perturbations by $G_{(1,2)}$ fluxes in Series~I, Eqn.~(\ref{equ:LambdaI}), with holomorphic functions ${\sf f}_1\sim (AB)^k$.  This is consistent with the results of \cite{Grana2003a}, where it was found that $G_{(3,0)}$ and non-primitive $G_{(2,1)}$ fluxes are incompatible with the original supersymmetry of the ISD background, while $G_{(1,2)}$ flux generates superpotential interactions for a D3-brane.

\item {\sl Series II operators.} The operators $\int d \theta_\alpha\,  {\cal O}_{\rm II}^\alpha = [{\cal O}_{\rm II}]_{\theta}$ are chiral but not supersymmetric. They correspond to the non-primitive $G_{(2,1)}$ flux of Series~II, Eqn.~(\ref{equ:LambdaII}), with holomorphic ${\sf f}_2\sim (AB)^k$. This is in agreement with the result of \cite{Grana2003a, Camara2004} where it was shown that non-primitive $(2,1)$ flux couples the gaugino to the fermions from the chiral superfields.

\item {\sl Series III operators.} Finally, the operators $[{\cal O}_{\rm III}]_{\rm b}$ are mapped to fluxes in Series~III, Eqn.~(\ref{equ:LambdaIII}), with holomorphic ${\sf f}_3\sim (AB)^k$. Again this is supported by the fact that $(3,0)$ flux creates a mass term for the gaugino \cite{Grana2003a, Camara2004}.

\end{itemize}

\subsubsection{Non-Chiral Operators}
\label{sec:nonchiral}

Next, we study generalizations of the chiral operators of the previous section:
\bea
{\cal O}_{\rm I} &=& \Tr f_1 \, , \\
{\cal O}_{\rm II}^\alpha &=& \Tr[W^\alpha_+ f_2]\, , \\
{\cal O}_{\rm III} &=& \Tr [W^2_+ f_3]\, ,
\eea
where $f_i \equiv f_i(A, B, \bar A, \bar B)$ are harmonic,
but not holomorphic, functions of the matter fields $A$ and $B$.
The low-lying operators corresponding to $\Phi_-$ and $G_-$ perturbations are collected in Tables~\ref{tab:Phi} and \ref{tab:G}.\footnote{The subscripts `$a$' and `$b$' on $J_{a, b}$ and $f_{a,b}$ indicate that these are functions
of $A$ only or of $B$ only. The dimensions of operators dual to fluxes are denoted by $\delta$ (rather than $\Delta$) to agree with the notation used in~\S\ref{sec:flux}.}

\begin{table}[h!]
\caption{\sl Matching between supergravity $\Phi_-$ modes and CFT operators \cite{Baumann2009}.}
\label{tab:Phi}
\vspace{-0.5cm}
\begin{center}
\begin{tabular}{>{\columncolor{lightgray}\raggedright}c c  c c >{\columncolor{lightgray2}\raggedright}c >{\columncolor{lightgray2}\raggedright}c c c}
\toprule
$\Delta$ & $j_1$ & $j_2$  & $R$ & \multicolumn{2}{c}{\cellcolor{lightgray2} \footnotesize \bf Operator} & {\footnotesize \bf Multiplet} & {\footnotesize \bf Type}  \\
\otoprule
$\frac{3}{2}$ & $\frac{1}{2}$ & $\frac{1}{2}$  & $1$ & {\footnotesize $[S^1]_{\rm b}$}& {\footnotesize $[\Tr(AB)]_{\rm b}$} & {\footnotesize V.I} & {\footnotesize chiral} \\ \midrule
$2$ & $0$ & $1$  & $0$ & {\footnotesize $[{}_a J^0]_{\rm b} $} & {\footnotesize $[\Tr J_a]_{\rm b}$} &  {\footnotesize V.I} &   {\footnotesize semi-long}\\ \midrule
$2$ & $1$ & $0$  & $0$ & {\footnotesize $[{}_b J^0]_{\rm b} $}  & {\footnotesize $[\Tr J_b]_{\rm b}$} &  {\footnotesize V.I} &  {\footnotesize semi-long} \\ \midrule
$3$ & $1$ &  $1$  & $2$ & {\footnotesize $[S^2]_{\rm b}$} & {\footnotesize $[\Tr(AB)^2]_{\rm b}$}  & {\footnotesize V.I} &  {\footnotesize chiral} \\
\bottomrule
\end{tabular}
\end{center}
\end{table}%

\begin{table}[h!]
\caption{\sl Matching between supergravity $G_-$ flux modes and CFT operators.}
\label{tab:G}
\vspace{-0.5cm}
\begin{center}
\begin{tabular}{>{\columncolor{lightgray}\raggedright}c c  c c >{\columncolor{lightgray2}\raggedright}c >{\columncolor{lightgray2}\raggedright}c c c c}
\toprule
$\delta$ & $j_1$ & $j_2$  & $R$ & \multicolumn{2}{c}{\cellcolor{lightgray2} \footnotesize \bf Operator} & {\footnotesize \bf Multiplet} & {\footnotesize \bf Type} & {\footnotesize \bf Flux Series} \\ \otoprule
$\frac{5}{2}$ & $\frac{1}{2}$ & $\frac{1}{2}$  & $-1$ & {\footnotesize $[S^1]_{\theta^2}$} & {\footnotesize $[{\rm Tr}(AB)]_{\theta^2}$}  & {\footnotesize V.I} & {\footnotesize chiral} &I \\ \midrule
$3$ & $0$ & $0$  & $2$ & {\footnotesize $[\Phi^0_+]_{\rm b}$} & {\footnotesize $[{\rm Tr}(W_{(1)}^2 + W_{(2)}^2)]_{\rm b}$}  & {\footnotesize V.IV} & {\footnotesize chiral} & III \\ \midrule
$\frac{7}{2}$ & $\frac{1}{2}$ & $\frac{1}{2}$  & $1$ & {\footnotesize $[T^1_\alpha]_{\theta}$} & {\footnotesize $[{\rm Tr}(W_\alpha (AB))]_{\theta}$} & {\footnotesize G.I} & {\footnotesize chiral} & II\\ \midrule
$4$ & $0$ & $0$  & $0$ & {\footnotesize $[\Phi^0_-]_{\theta^2}$} & {\footnotesize $[{\rm Tr}(W_{(1)}^2 - W_{(2)}^2)]_{\theta^2}$} & {\footnotesize V.III} & {\footnotesize chiral} & $\star$\\
$4$ & $0$ & $1$  & $0$ & {\footnotesize $[{}_a L^{2,0}_\alpha]_{\theta}$} & {\footnotesize $[{\rm Tr}(W_\alpha  J_a)]_{\theta}$} & {\footnotesize G.I+G.III} & {\footnotesize semi-long} & II \\
$4$ & $1$ & $0$  & $0$ &  {\footnotesize $[{}_b L^{2,0}_\alpha]_{\theta}$ } & {\footnotesize $[{\rm Tr}(W_\alpha  J_b)]_{\theta}$} & {\footnotesize G.I+G.III} & {\footnotesize semi-long} & II \\
$4$ & $1$ & $1$  & $0$ & {\footnotesize $[S^2]_{\theta^2}$} & {\footnotesize $[{\rm Tr}(AB)^2]_{\theta^2}$} &{\footnotesize V.I} & {\footnotesize chiral} & I \\ \midrule
{\footnotesize $ \sqrt{28}-1 $} & $1$ & $1$  & $-2$ & -- & {\footnotesize $[\Tr(f)]_{\theta^2}$} & {\footnotesize V.I} & {\footnotesize long} & I \\ \midrule
$\frac{9}{2}$ & $\frac{1}{2}$ & $\frac{1}{2}$  & $3$ & {\footnotesize $[\Phi^1_+]_{\rm b}$} & {\footnotesize $[{\rm Tr}(W_{(1)}^2 + W_{(2)}^2) (AB)]_{\rm b}$}  & {\footnotesize V.IV} &{\footnotesize chiral} &  III \\
$\frac{9}{2}$ & $\frac{1}{2}$ & $\frac{1}{2}$  & $1$ & {\footnotesize $[\bar \Phi^1_+]_{\rm b}$} & {\footnotesize $[{\rm Tr}(W_{(1)}^2 + W_{(2)}^2) (\overline{AB})]_{\rm b}$}  & {\footnotesize V.IV} & -- & III \\
$\frac{9}{2}$ & $\frac{1}{2}$ & $\frac{3}{2}$  & $-1$ & {\footnotesize $[{}_a J^1]_{\theta^2}$} & {\footnotesize $[\Tr (J_a (AB))]_{\theta^2}$} & {\footnotesize V.I} & {\footnotesize semi-long} &  I \\
$\frac{9}{2}$ & $\frac{3}{2}$ & $\frac{1}{2}$  & $-1$ & {\footnotesize $[{}_b J^1]_{\theta^2}$} & {\footnotesize $[\Tr (J_b (AB))]_{\theta^2}$} & {\footnotesize V.I} & {\footnotesize semi-long} &  I \\ \midrule
$5$ & $1$ & $1$  & $2$ &  {\footnotesize $[T^2_\alpha]_\theta$} & {\footnotesize $[{\rm Tr}(W_\alpha (AB)^2)]_\theta$} & {\footnotesize G.I} & {\footnotesize chiral} & II \\
$5$ & $0$ & $1$  & $2$ & {\footnotesize $[{}_a I^0]_{\rm b}$} &  {\footnotesize $[\Tr((W_{(1)}^2+W_{(2)}^2) J_a)]_{\rm b}$} & {\footnotesize V.IV} & {\footnotesize semi-long} & III \\
$5$ & $1$ & $0$  & $2$ &{\footnotesize $[{}_b I^0]_{\rm b}$} &  {\footnotesize $[\Tr((W_{(1)}^2+W_{(2)}^2) J_b)]_{\rm b}$} & {\footnotesize V.IV} &  {\footnotesize semi-long} & III \\ \midrule
{\footnotesize $ \sqrt{28} $} & $1$ & $1$  & $0$ & -- & {\footnotesize $[\Tr(W_\alpha f)]_{\theta}$} &  {\footnotesize G.I+G.III}&  {\footnotesize long} & II \\ \midrule
{\footnotesize $ \sqrt{40} -1$} & $0$ & $2$  & $-2$ & -- & {\footnotesize $[\Tr(f_a)]_{\theta^2}$} & {\footnotesize V.I} &  {\footnotesize long} & I \\
{\footnotesize $ \sqrt{40} -1$} & $2$ & $0$  & $-2$ & -- &  {\footnotesize $[\Tr(f_b)]_{\theta^2}$}  & {\footnotesize V.I} &  {\footnotesize long} & I \\ \midrule
$\frac{11}{2}$ & $\frac{1}{2}$ & $\frac{3}{2}$  & $1$ & {\footnotesize $[{}_a L^{2,1}_\alpha]_{\theta}$} & {\footnotesize $[{\rm Tr}(W_\alpha  J_a (AB))]_{\theta}$} & {\footnotesize G.I+G.III} &  {\footnotesize semi-long} & II \\
$\frac{11}{2}$ & $\frac{3}{2}$ & $\frac{1}{2}$  & $1$ & {\footnotesize $[{}_b L^{2,1}_\alpha]_{\theta}$} & {\footnotesize $[{\rm Tr}(W_\alpha  J_b (AB))]_{\theta}$} & {\footnotesize G.I+G.III} & {\footnotesize semi-long} & II \\
$\frac{11}{2}$ & $\frac{1}{2}$ & $\frac{3}{2}$  & $-1$ & {\footnotesize $[{}_a \bar L^{2,1}_\alpha]_{\theta}$} & {\footnotesize $[{\rm Tr}(W_\alpha  J_a (\overline{AB}))]_{\theta}$} & {\footnotesize G.I+G.III} & {\footnotesize -- } & II \\
$\frac{11}{2}$ & $\frac{3}{2}$ & $\frac{1}{2}$  & $-1$ & {\footnotesize $[{}_b \bar L^{2,1}_\alpha]_{\theta}$} & {\footnotesize $[{\rm Tr}(W_\alpha  J_b (\overline{AB}))]_{\theta}$} & {\footnotesize G.I+G.III} & {\footnotesize -- } & II \\
$\frac{11}{2}$ & $\frac{3}{2}$ & $\frac{3}{2}$  & $1$ &  {\footnotesize $[S^3]_{\theta^2}$} & {\footnotesize $[{\rm Tr}(AB)^3]_{\theta^2}$}  & {\footnotesize V.I} & {\footnotesize chiral} & I\\
\bottomrule
\end{tabular}
\end{center}
\end{table}%

Let us make a few comments about each of these operators and their supermultiplet structure:
\begin{itemize}
\item ${\cal O}_{\rm I } =  \Tr {f_1}$:

These superfields correspond to Vector Multiplet I of \cite{Ceresole2000} (see Table~7 of \cite{Ceresole2000}).
The bottom components are operators made out of scalar fields only; they are dual to perturbations
that are certain combinations of the four-form potential $C_4$ and the trace of the metric (denoted by $b$ in \cite{Ceresole2000}).
Such fluctuations induce $\Phi_-$ perturbations
that contribute to the D3-brane potential at linear order,
as discussed in our previous paper \cite{Baumann2009} (see Table~\ref{tab:Phi} above). The $\theta^2$ components of ${\cal O}_{\rm I}$ are operators bilinear in fermions with $R$-charge $R-2$, and
correspond to the $G_-$ perturbations from Series~I. Here we disagree with Table~7 of \cite{Ceresole2000},
which states that these operators are dual to metric perturbations.

We may further consider the superfields
\beq
{\widetilde{\cal O}}_{\rm I} =  \Tr [\overline{W}_+^2 {f_1}]\, ,
\eeq
where $\overline{W}^2_+ \equiv \overline{W}^2_{(1)} + \overline{W}^2_{(2)}$.
The $\bar\theta^2 \theta^2$ components of these superfields are operators with dimensions $\Delta+4$. As shown in \S\ref{sec:nonlinear},
these additional operators arise from Series~$\overline{\rm I}$ $G_+$ perturbations.

\item ${\cal O}_{\rm II}^\alpha =  \Tr [W^\alpha_+ {f_2}]$:

These superfields correspond to Gravitino Multiplet I (Table 3 of \cite{Ceresole2000}). In particular, the $\theta$ components correspond to $G_-$ perturbations from Series~II, as can be confirmed by comparing their dimensions $\Delta$.
This agrees with the field assignment in \cite{Ceresole2000}.

We may further consider the superfields
\beq
{\widetilde{\cal O}}_{\rm II}^\alpha =  \Tr [W_\alpha \overline{W}_+^2 {f_2}]\, .
\eeq
The $\bar\theta^2 \theta$ components of these superfields are operators with dimensions $\Delta+4$,
corresponding to Series~$\overline{\rm II}$ $G_+$ perturbations. More generally, we note that these superfields correspond to Gravitino Multiplet~II (Table 4 of \cite{Ceresole2000}).

\item ${\cal O}_{\rm III} =  \Tr [ W_+^2 {f_3}]$:

These superfields correspond to Vector Multiplet IV (Table 10 of \cite{Ceresole2000}). The bottom components,
bilinear in the gauginos, correspond to $G_-$ perturbations from Series III. The $\theta^2$ components instead correspond to the dilaton and RR scalar of type~IIB theory.

We may further consider the superfields
\beq
\label{vectortwo}
{\widetilde{\cal O}}_{\rm III} =  \Tr [W_+^2 \overline{W}_+^2 {f_3}]
\ .\eeq
The $\bar\theta^2 $ components of these superfields are operators whose dimensions are higher by 4 than those of the operators in the previous paragraph.
As shown in \S\ref{sec:nonlinear},
these additional operators arise from Series~$\overline{\rm III}$ $G_+$ perturbations.
We find that the superfields (\ref{vectortwo}) correspond to Vector Multiplet~II (Table 8 of \cite{Ceresole2000}).
The bottom components correspond to metric deformations squashing the $S^1$ fiber of $T^{1,1}$ relative to the base, while the $\theta^2 \bar \theta^2$ components correspond to well-known operators of the form $F^4 f_3$ that appear as a combination of fluctuations of $C_4$ and of the trace of the metric. These fluctuations induce $\Phi_+$ perturbations. These assignments agree with Table~8 of \cite{Ceresole2000}.
However, in disagreement with that table we find that
the $\theta^2$ and $\bar \theta^2$ components of these superfields are dual to $G_+$ perturbations.

\end{itemize}

In summary, we have matched the three series of flux modes with holomorphic $f$ (see \S\ref{sec:flux}) to the three series of chiral operators.
Similarly,
the three corresponding series of non-chiral operators obtained by taking $f_i \equiv f_i(A, B, \bar A, \bar B)$ to be harmonic but not holomorphic match the three series of flux modes with harmonic, but non-holomorphic $f$.
The apparent complexity of Table~\ref{tab:G} is therefore reduced to three distinct
series of flux or three different types of operator perturbations.

There is also one special case, denoted $\star$ in the table, which corresponds to the vanishing three-form flux (\ref{special3form}).
This mode changes the difference between the complex coupling constants
$g_1^{-2}- g_2^{-2}$ in the KW theory and hence corresponds to the top component of the chiral operator (\ref{specchiral}). The corresponding bottom component ${\rm Tr}(\lambda_1^2-\lambda_2^2)$, with $R$-charge 2 and dimension 3, corresponds to a traceless perturbation of the metric on $T^{1,1}$.
As this mode is not dual to a $G_{-}$ or $\Phi_-$ perturbation, the operator ${\rm Tr}(\lambda_1^2-\lambda_2^2)$ does not create a D3-brane potential at quadratic order.

Notice that some of the non-chiral operators described above reside in long multiplets and have irrational dimensions.  In an analysis conducted exclusively in the gauge theory, it would be rather difficult to determine the dimensions of these operators, but on the gravity side of the correspondence it is straightforward to identify these contributions.  Moreover, these long operators can have non-negligible effects on the Coulomb branch potential: {\it e.g.}~an interference term between fluxes dual to the long operator with $\delta = \sqrt{28}-1$ and the chiral operator with $\delta = \frac{5}{2}$
gives rise to a term in the scalar potential of the form $r^{\sqrt{28}-\frac{5}{2}}$, which can substantially affect the structure of the scalar potential.

\subsection{Potential on the Coulomb Branch}
\label{sec:potential}

Having identified the leading operators dual to perturbations of $\Phi_-$ and $G_-$, we can now write the leading perturbations to the CFT Lagrangian and compute the resulting potential on the Coulomb branch.

\subsubsection{Perturbed Lagrangian and SUSY Breaking}

We begin by introducing an efficient representation for the perturbed Lagrangian in terms of spurion fields.  Some of the perturbations of interest are consistent with the supersymmetry of the gauge theory, while others break this supersymmetry explicitly.\footnote{We must distinguish spontaneous breaking of supersymmetry in the conifold gauge theory from breaking that is spontaneous in a larger supersymmetric theory, but appears as explicit breaking in the conifold gauge theory.  If dynamics in a distant region of the compactification ({\it e.g.}~gauge dynamics on D-branes at some local singularity distant from the conifold) leads to mass splittings in bulk supermultiplets, this spontaneous breaking of supersymmetry will manifest itself as an explicit breaking of the supersymmetry of the conifold gauge theory after we integrate out the distant physics.}
To characterize the bulk supersymmetry breaking, and more generally the bulk sourcing of perturbations of the gauge theory, we introduce spurion fields $X$
and $Y_{\alpha}$.
We will take $X$ to be a
chiral superfield external to the gauge theory, each of whose components will in general have a nonvanishing expectation value,
\begin{equation}
X = [X]_{\rm b} + [X]_{\theta}\,\theta + [X]_{\theta^2}\,\theta^2 \equiv x + x^{\alpha}\theta_{\alpha}+ F_{X}\theta^2\,.
\end{equation}
Similarly, for $Y_{\alpha}$ we have
\begin{equation}
Y_{\alpha} = [Y_{\alpha}]_{\rm b} + [Y_{\alpha}]_{\theta}\,\theta + [Y_{\alpha}]_{\theta^2}\,\theta^2 \equiv y_{\alpha} + y \, \theta_{\alpha}+ F_{Y_{\alpha}}\theta^2\,.
\end{equation}

The perturbations to the CFT Lagrangian involving {\it{chiral}} supermultiplets ${\cal O}_{\rm I}, {\cal O}_{\rm II}^{\alpha}, {\cal O}_{\rm III}$ may then be written as
\beq
\Delta {\cal L} = \int d^2 \theta \Bigl[ {\cal O}_{\rm I} X  +{\cal O}_{\rm II}^{\alpha} Y_{\alpha}  +{\cal O}_{\rm III} X \Bigr] \ + \ c.c. \ .
\eeq
Among the resulting terms, some correspond to (the chiral subset of)
$G_-$ perturbations,
\beq
\Delta {\cal L}_{\rm flux} = [{\cal O}_{\rm I}]_{\theta^2}[X]_{\rm b} + [{\cal O}_{\rm II}^{\alpha}]_{\theta} [Y_{\alpha}]_{\theta} + [{\cal O}_{\rm III}]_{\rm b}[X]_{\theta^2} \ + \ c.c. \ ,
\eeq while the remaining terms are not dual to $G_{-}$ perturbations and can be neglected for
the present purposes.

Similarly, the perturbations involving non-chiral supermultiplets ${\cal O}_{\rm I}, {\cal O}_{\rm II}^\alpha, {\cal O}_{\rm III}$ may be written as
\beq
\Delta {\cal L} = \int d^4 \theta \Bigl[ {\cal O}_{\rm I} X X^{\dagger} +  {\cal O}_{\rm II}^{\alpha} Y_{\alpha} X^{\dagger}  + {\cal O}_{\rm III} X X^{\dagger} \Bigr] \ + \ c.c.\ ,
\eeq with the operators dual to $G_{-}$ taking the form
\beq
\Delta {\cal L}_{\rm flux} = [{\cal O}_{\rm I}]_{\theta^2}[X]_{\rm b}[X^{\dagger}]_{\bar\theta^2} + [{\cal O}_{\rm II}^{\alpha}]_{\theta} [Y_{\alpha}]_{\theta}[X^{\dagger}]_{\bar\theta^2} + [{\cal O}_{\rm III}]_{\rm b}[X]_{\theta^2}[X^{\dagger}]_{\bar\theta^2} \ +\ c.c. \ ,
\eeq
and the operators dual to harmonic $\Phi_-$ modes taking the form
\beq
\Delta {\cal L}_{\Phi_-} = [{\cal O}_{\rm I}]_{\rm b}[X]_{\theta^2}[X^{\dagger}]_{\bar\theta^2} \ + \ c.c.\ .
\eeq

We remark that the spurion analysis presented here gives further justification for the expansion scheme proposed in
{\sl Case~II} in \S\ref{sec:perturbation}.  Chiral modes, including the leading modes of $G_{-}$, are proportional to a single power of the small spurion expectation values $[X]_{\rm b}, [X]_{\theta}, [X]_{\theta^2}$, and $[Y_{\alpha}]_{\rm b}, [Y_{\alpha}]_{\theta}, [Y_{\alpha}]_{\theta^2} $, whereas the harmonic modes of $\Phi_-$ \cite{Baumann2009} require two spurion insertions.
This motivates considering the case $\Phi_-^{(1)}=0$ in which harmonic perturbations of $\Phi_-$ are neglected at linear order, while perturbations of $G_{-}$ are retained at linear order.
The leading potential, $\Phi_-^{(2)}$, then receives important contributions both from harmonic $\Phi_-^{(2)}$ perturbations and from the particular $\Phi_-^{(2)}$ solution sourced by $G_{-}^{(1)}$, as in the analysis of \S\ref{sec:spectrum}.

\subsubsection{Scalar Potential in the Gauge Theory}

To
give further evidence for the correspondence between the operators of this section and the flux modes of the previous sections, let us discuss\footnote{We thank Zohar Komargodski, Juan Maldacena and Nathan Seiberg for extensive
discussions of these issues.} how one can arrive at the scalar potential (\ref{phiminus}) in the gauge theory:
\begin{itemize}
\item
The first term in (\ref{phiminus}), $g^{\alpha \bar \beta} \partial_\alpha {\sf f}_1 \partial_{\bar \beta} \bar {\sf f}_1$, is simply the F-term potential due to the superpotential perturbation $\int d^2\theta \, {{\sf f}_1}$, for holomorphic ${\sf f}_1 \sim (AB)^k$.\footnote{As explained in \S\ref{sec:perturbation}, we consistently omit corrections to the inverse metric $g^{\alpha\bar \beta}$, taking it to be the unperturbed conifold metric $g^{\alpha\bar \beta}_{(0)}$.}

\item
To derive the second term in (\ref{phiminus}), $({\rm Re}\, {\sf f}_2)^2$, we consider the chiral operator dual to the flux in Series~II
as a superpotential correction,
\bea
\label{supII}
\int d^2\theta \, {\mathcal O}_{\rm II}^\alpha Y_\alpha \sim  y\, [{\cal O}_{\rm II}^\alpha]_{\theta_{\alpha}} \ .
\eea
Utilizing the component expansion of $W_{\alpha}$, we obtain a D-term of the form
\beq
y\, \Tr [(D_{(1)}+D_{(2)}) {\sf f}_2]\ +\ c.c. \ ,
\eeq
where ${\sf f}_2 \sim (AB)^k$.
After the $D$ fields are integrated out, the resulting potential is proportional to $({\rm Re}\, {\sf f}_2)^2$.
The exact proportionality coefficient depends on the coupling constants at the IR fixed point, but at leading order in perturbations this can be taken to be the KW value.
Although the coupling (\ref{supII}) changes the renormalization of the K\"ahler potential, there is no corresponding K\"ahler potential contribution to the scalar potential at first or second order in $y$:
because $y$ is accompanied by only one power of $\theta$ in $Y_\alpha$, terms from the K\"ahler potential $\int d^4\theta \, K(Y_\alpha,\bar Y_\alpha,A,B,\bar A,\bar B)$ that are linear or quadratic in $y$ could contribute {\it e.g.}~to fermion masses, but not to the scalar potential.

\item
The third term in (\ref{phiminus}), $\nabla^{-2}|{\sf f}_3|^2$, is more involved.  This term corresponds to the contribution of the operator
${\mathcal O}_{\rm III}$
dual to flux in Series~III, and
is very difficult to calculate
in field theory,
even for chiral ${\mathcal O}_{\rm III}$. The problem is that if we represent this operator as a perturbation of the superpotential, $\int d^2 \theta \, {\mathcal O}_{\rm III} X$, the scalar potential will receive a new contribution proportional to K\"ahler potential corrections quadratic in~$X$.

Although we will not
calculate the resulting potential in field theory, we can provide an interesting consistency check.
The perturbation
\begin{equation}\label{newinteraction}
\int d^2 \theta\, {\cal{O}}_{\rm III} X
\end{equation} gives rise both to
\begin{equation}\label{dilatonmode}
[{\cal{O}}_{\rm III}]_{\theta^2} [X]_{\rm b}
\end{equation} and to
\begin{equation}\label{gauginomode}
[{\cal{O}}_{\rm III}]_{\rm b} [X]_{\theta^2} = [{\cal{O}}_{\rm III}]_{\rm b} \, F_{X} \,.
\end{equation}
The former is a perturbation of the gauge coupling function, and is well-known to be dual to a perturbation of the axio-dilaton in supergravity, $\delta\tau = [X]_{\rm b} {\sf f}_3$.
However, (\ref{gauginomode}) is a gaugino mass term and is dual to Series III flux. Supersymmetry manifestly relates (\ref{dilatonmode}) and (\ref{gauginomode}).

Next, we note that adding the interaction (\ref{newinteraction}) changes the renormalization of the K\"ahler potential, so that
\beq
\label{Kpotperturb}
K_0(A, \bar A, B, \bar B)\ \ \to \ \  K_0(A, \bar A, B, \bar B) + \delta K(X,\bar X, A, \bar A, B, \bar B) \, .
\eeq
We treat the chiral spurion $X$ as a dynamical field with a large mass and a nonzero vev acquired through the superpotential $\int d^2 \theta\, w(X)$. Before the coupling $\int d^2\theta \, {\mathcal O}_{\rm III}X$ is introduced, $X$
does not couple to the KW theory, and has diagonal metric $g^{(0)}_{X\bar X}$. After the coupling $\int d^2\theta\, {\mathcal O}_{\rm III}X$ is introduced, the $F$-term potential for $X$,
\begin{equation}
V_{X} = g^{X\bar X} \left|\frac{\partial w}{\partial X}\right|^2 \, ,
\end{equation}
acquires a correction because Eqn.~(\ref{Kpotperturb}) induces a change in the metric for $X$,
\begin{equation}
\delta V = -\left|\frac{\partial w}{\partial X}\right|^2 (g^{(0)}_{X\bar X})^2\frac{\partial^2}{\partial X \partial \bar X}\, \delta K \,.
\end{equation}
At the same time, there is a correction to the metric of the KW fields,
\begin{equation}
\delta g_{\alpha\bar\beta} = \frac{\partial^2}{\partial z^{\alpha}\partial z^{\bar \beta}}\, \delta K \,,
\end{equation}
where we have changed from $A,B$ to the geometric notation $z^{\alpha}$ for the coordinates on the conifold.  Therefore, supersymmetry provides a relationship between the perturbations to the metric and the potential,
\begin{equation}\label{consistency}
\partial_{\alpha}\partial_{\bar\beta} \, \delta V \propto  - \,\delta g_{\alpha\bar\beta}\, ,
\end{equation} where the unspecified constant of proportionality is positive.  Moreover, from the definition of the Ricci tensor, $R_{\alpha \bar \beta} = - \partial_\alpha \partial_{\bar \beta} \ln \det g$, we have
\beq \label{logdet}
\delta R_{\alpha\bar \beta}= - \partial_\alpha\partial_{\bar \beta}\,\nabla^2 \delta K \, .
\eeq
One then easily shows that
\beq \label{requiredrelationship}
\partial_{\alpha}\partial_{\bar\beta} \, \nabla^2 \delta V \propto \delta R_{\alpha\bar \beta}\, .
\eeq
We will now verify that the supergravity solution implied by our analysis
satisfies the relationship (\ref{requiredrelationship}), providing a strong consistency check.
To do this, we compute the perturbation of the Ricci tensor induced by the dilaton perturbation (\ref{dilatonmode}) and show that this is related by (\ref{requiredrelationship}) to the perturbed potential (\ref{phiminus}) computed in supergravity in the presence of the flux perturbation (\ref{gauginomode}).
We recall that the dilaton perturbation sources a perturbation of the metric via the Einstein equation ({\it{cf.}} Appendix \ref{sec:running}),
\beq \label{dilatonRicci}
\delta R_{\alpha\bar \beta}= \frac{1}{4({\rm Im} \,\tau)^2} \partial_\alpha \delta \tau \, \partial_{\bar \beta} \delta \bar \tau\, .
\eeq
Substituting $\delta \tau \propto {\sf f}_3$ into Eqn.~(\ref{dilatonRicci}) and inserting $\delta V =\nabla^{-2}|{\sf f}_3|^2$ into Eqn.~(\ref{requiredrelationship}),
we confirm that our result (\ref{phiminus}) obeys the relationship (\ref{requiredrelationship}) required by supersymmetry.
\end{itemize}

\newpage
\subsection{Summary of CFT Perturbations}

Let us recapitulate our results from the CFT viewpoint.
The simplest and most important perturbations,
\begin{equation}
\int d^2 \theta \, \Tr (AB)^k\, ,
\end{equation} are perturbations of the CFT superpotential. This tower of perturbations is dual to the {\it{chiral}} subseries, in Series ~I, of $G_{(1,2)}$ flux perturbations.
The $k=1$ mode is relevant in the RG sense, and is dual to the lowest-dimension $(\delta = \frac{5}{2})$ mode of flux; it is responsible for the {\it{leading}} $r^{1}$ contribution to the D3-brane potential.
The bottom component of the same supermultiplet is a linear $\Phi_-$ perturbation, and corresponds to the
$r^{3/2}$ term of \cite{Baumann2009}.

Non-chiral generalizations of Series~I operators are of the form
\beq
\int d^2 \theta\, f_1\Bigl|_{\bar\theta^2=0}\, ,
\eeq
for $f_1$ a harmonic function of the matter fields $A$ and $B$. The lowest-dimension non-chiral operator of this form has irrational dimension $\sqrt{28}-1$. Adding it to the Lagrangian, together with $\int d^2 \theta\, \Tr (AB)$, leads to a term $r^{\sqrt{28} - \frac{5}{2}} \approx r^{2.79}$
in the scalar potential.

Similarly, the Series~II and III chiral operators are
\begin{equation}
\int d \theta_\alpha \, [W^\alpha_+ \Tr (AB)^k]
\qquad {\rm and} \qquad
[W^2_+ \Tr (AB)^k ]_{\rm b}\ ,
\end{equation}
with the non-chiral extensions
\begin{equation}
\int d \theta_\alpha \, [W^\alpha_+ f_2]\Bigl|_{\bar \theta^2=0}
\qquad {\rm and} \qquad
[W^2_+ f_3]_{\rm b}\ .
\end{equation}
The leading contributions to the D3-brane potential from chiral Series~II and III modes scale as $r^{3}$ and $r^{2}$, respectively.
Furthermore, the leading non-chiral Series~II mode ($\delta=4$) can interfere with the leading chiral Series~I mode ($\delta=\frac{5}{2}$) to give a term scaling as $r^{5/2}$.

\newpage
\section{D3-brane Superpotentials from Fluxes}
\label{sec:4dSUGRA}

We have seen that perturbations of the supergravity solution by chiral modes of flux in Series~I,
\beq
(\Lambda_{\rm I})_{\alpha \bar \beta \bar \gamma} = \nabla_\alpha \nabla_{\sigma} f \, g^{\sigma \bar \zeta}\, \bar \Omega_{\bar \zeta \bar \beta \bar \gamma}\, ,
\eeq with $f$ a holomorphic function on the conifold, correspond to superpotential perturbations in the CFT,
\begin{equation}
\Delta W \sim f(A,B) \,.
\end{equation}
In this section we will apply this correspondence to derive a useful representation of nonperturbative superpotentials in terms of fluxes.

\subsection{Superpotentials in Global Supersymmetry}
\label{sec:global}

First, we observe that the D3-brane potential\footnote{For simplicity, in this subsection we take all derivatives to be with respect to the canonically-normalized field $\phi_\alpha \equiv \sqrt{T_3} z_\alpha$.} 
in global supersymmetry,
\begin{equation}\label{equ:Vr1}
V=g^{\alpha\bar\beta}\nabla_\alpha W\overline{\nabla_\beta W} \,,
\end{equation}
is reproduced by the supergravity calculation in the corresponding flux background.  Specifically, given any holomorphic function $W$ on the conifold, we turn on the flux
\begin{equation}\label{g121}
 \Lambda_{\alpha\bar\beta\bar\gamma}= 
 \nabla_{\alpha} \nabla_{\sigma}W \,g^{\sigma\bar\rho}\,\bar\Omega_{\bar\rho\bar\beta\bar\gamma} \equiv \nabla \nabla W\cdot\bar\Omega\, .
\end{equation}  Solving (\ref{equ:PhiMinus0}) in the background (\ref{g121}), we obtain the desired
potential,
\begin{equation}\label{equ:withgradient}
T_3 \Phi_- = g^{\alpha\bar\beta}\nabla_\alpha W\overline{\nabla_\beta W} \,.
\end{equation}
Therefore, for any superpotential interaction added to the conifold gauge theory, there is a $G_{(1,2)}$ flux that geometrizes this superpotential.  This
is implicit in the analysis of \cite{GranaPolchinski}, in which it was found that $G_{(1,2)}$ fluxes induce superpotential interactions for D3-branes.  In Appendix~\ref{sec:running}, we show that this
correspondence can be extended to backgrounds with large dilaton gradients and Ricci curvature sourced by D7-branes.

We now turn to an interesting application of this representation
of superpotentials by fluxes.

\subsection{Superpotentials from D7-branes}

In the presence of a nonperturbative superpotential on wrapped D7-branes, a D3-brane feels a nontrivial potential \cite{BDKMMM}. In this section, we briefly review this result, then observe that the corresponding potential can be represented in ten dimensions by a background of three-form fluxes.

\begin{figure}[h!]
    \centering
        \includegraphics[width=0.6\textwidth]{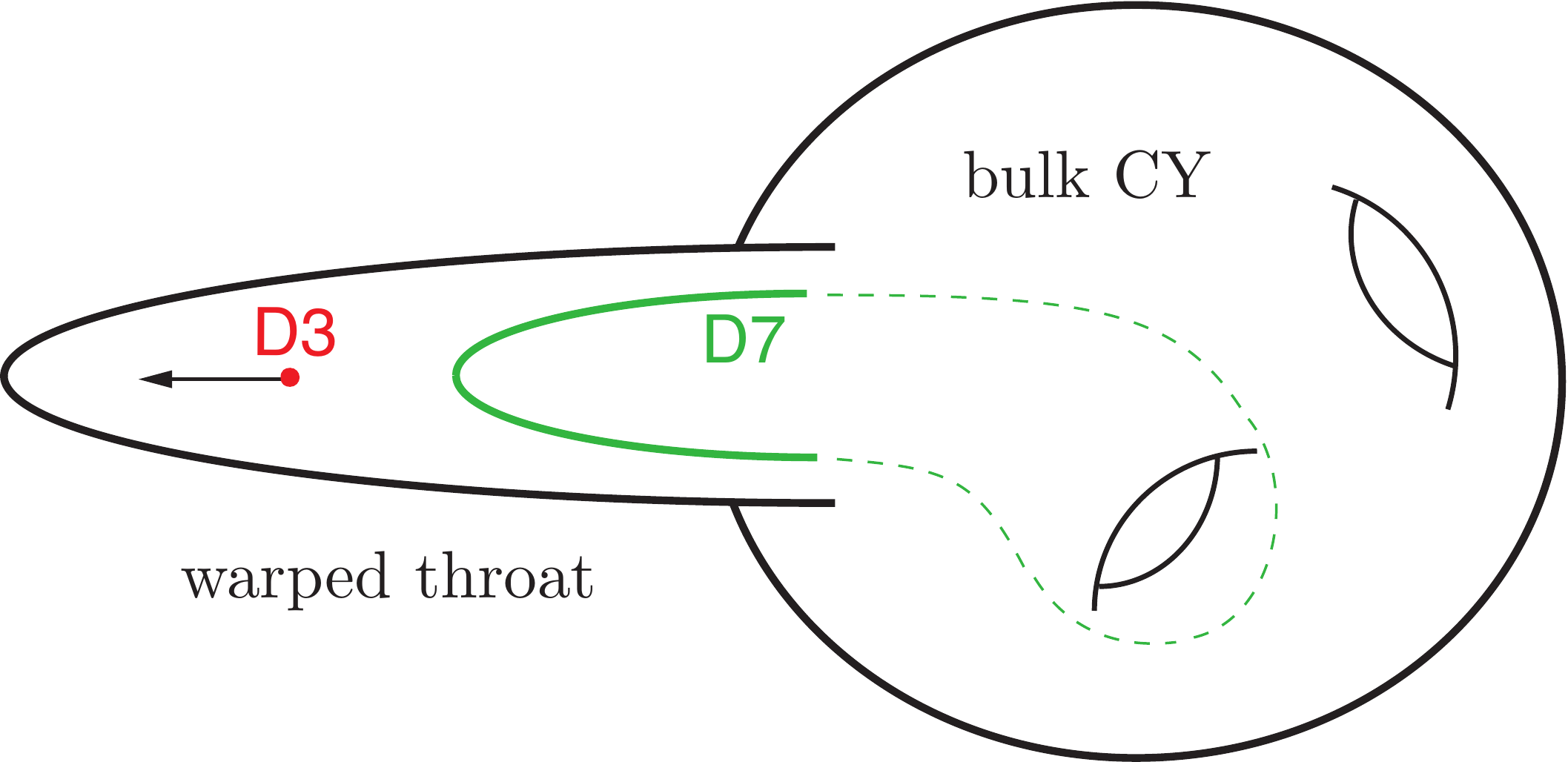}
          \caption{\small \sl Schematic of a finite throat with an embedded stack of D7-branes wrapping a four-cycle. The four-cycle is compact and resides partially in the bulk and partially in the throat. Gaugino condensation on the D7-branes induces a potential for the D3-brane.}
        \label{fig:BDKMMM}
\end{figure}

The scalar potential for a D3-brane in a compactification with nonperturbative moduli stabilization has been analyzed in Ref.~\cite{BDKM}.  The scenario of interest is a finite throat containing a holomorphically-embedded stack of D7-branes (or Euclidean D3-branes), and this configuration is approximated by a noncompact conifold containing D7-branes wrapping a noncompact divisor (see Fig.~\ref{fig:BDKMMM}).

The four-cycle $\Sigma$ wrapped by the D7-branes
is defined by the holomorphic embedding condition
\beq
h(z_\alpha) = 0\, ,
\eeq
the simplest examples being the Kuperstein embedding~\cite{Ouyang2004, Kuperstein2003a},
\beq
z_1 = \mu\, ,
\eeq
and the Ouyang embedding~\cite{Ouyang2004},
\beq
z_3+i z_4 = \mu\, .
\eeq
The warped volume of the four-cycle, $V_\Sigma = \int \d^4 y \sqrt{-g} e^{-4A}$, determines the gauge coupling of the D7-brane theory.  D3-branes are local sources for $\Phi_+$ and hence contribute corrections to $V_\Sigma$.
If the D7-brane gauge theory generates a gaugino condensate superpotential, then the dependence of the four-cycle volume on the D3-brane position introduces a dependence of the superpotential on the D3-brane position \cite{BDKMMM},
\beq
\label{6_9}
W_{\rm np}(z_\alpha) = {\cal A}(z_\alpha) e^{- a \rho}\, ,
\eeq
where $\rho$ is the K\"ahler modulus associated with the overall volume of the four-cycle and ${\cal A}(z_\alpha)$ is given by the embedding function $h(z_a)=0$,
\beq
\label{6_10}
{\cal A}(z_\alpha) = {\cal A}_0 \, h(z_\alpha)^{1/N_c}\, ,
\eeq
with $N_c$ the number of D7-branes and ${\cal A}_0$ a constant proportional to $N_c^2$.
The full superpotential is then
\beq
\label{equ:WW}
W= W_0 + W_{\rm np}(z_\alpha,\rho)\, ,
\eeq
with $W_0$ the Gukov-Vafa-Witten flux superpotential \cite{Gukov2000}, which is constant after stabilization of the complex structure moduli.
Finally, the K\"ahler potential is \cite{DeWolfe2003,STUD} $K = - 3 \ln[\rho + \bar \rho - \gamma k(z_\alpha, \bar z_\alpha)] $ (with $\gamma$ a constant, {\it{cf.}} \cite{BDKM}).
In terms of the above ${\cal N}=1$ data, one can derive the supergravity F-term potential $V_{F}[\rho,\bar \rho, z_\alpha,\bar z_\alpha]$
in the usual way.

\subsection{Flux Representation of Nonperturbative Superpotentials}
\label{sec:BDKM}

We will now demonstrate that for any specified superpotential $W(z_\alpha)$ for a D3-brane in the conifold, {\it cf.}~Eqn.~(\ref{equ:WW}), there exists a noncompact supergravity solution in which the
Born-Infeld plus Chern-Simons
potential of a probe D3-brane precisely equals the F-term potential $V_{F}[W(z_\alpha),K(z_\alpha,\bar z_\alpha)]$ computed in four-dimensional supergravity with $W(z_\alpha)$ and $K(z_\alpha,\bar z_\alpha)$ as the input data.\footnote{In this discussion we will suppress the dependence on $\rho$ for brevity, but one should keep in mind that $\rho$ is taken to be a dynamical field in the computation of $V_{F}$.} That is, given a superpotential $W(z_\alpha)$, we will provide a noncompact
supergravity solution containing suitable IASD fluxes that {\it{geometrizes}} this superpotential.

At leading order in an expansion in $\frac{\gamma k}{ \sigma} = \frac{\phi^2}{3 M_{\rm pl}^2}$, where $\sigma \equiv \frac{1}{2}(\rho + \bar \rho)$, the F-term potential is~\cite{BDKM}
\beq
\label{equ:Fterm}
V = \frac{\kappa^2}{12 \sigma^2} \frac{e^{-2a\sigma}}{\gamma} \left(g^{\alpha \bar \beta} {\cal A}_\alpha \bar {\cal A}_{\bar \beta} +  2 a\gamma (a\sigma + 3) {\cal A} \bar {\cal A} -  a \gamma( \bar {\cal A} g^{\alpha \bar \beta} k_{\bar \beta} {\cal A}_\alpha + c.c.)  \right) \ + \ {\rm harmonic}\, .
\eeq
We now exhibit the ten-dimensional supergravity solution in which a probe D3-brane experiences the four-dimensional supergravity F-term potential (\ref{equ:Fterm}) \cite{BDKM}.
To accomplish this we turn on fluxes in the first two series of \S\ref{sec:flux}.
We turn on $(1,2)$ flux of the form
\beq
\Lambda_1 = \nabla \nabla f_1 \cdot \bar \Omega\, ,
\eeq
with $f_1$ a holomorphic function. In addition, we turn on a non-primitive $(2,1)$ flux of the form
\beq
\Lambda_2 = \partial f_2 \wedge J\, ,
\eeq
with $f_2$ a holomorphic function.
As we have shown in \S\ref{sec:spectrum}, this leads to the potential\footnote{Notice that $({\rm Re} f_1)^2$ and $\frac{1}{2}|f_1|^2$ are equal up to harmonic terms.}
\beq
\label{equ:PhiMinus}
\Phi_- = \frac{V}{T_3} = \frac{g_s}{32} \left[ g^{\alpha \bar \beta} \nabla_\alpha f_1 \overline{\nabla_\beta f_1} + 2 |f_2|^2 \right] \ + \ {\rm harmonic}\, .
\eeq
Comparing Eqn.~(\ref{equ:PhiMinus}) to Eqn.~(\ref{equ:Fterm}) suggests the matching conditions
\bea
f_1 &=& c_1 {\cal A}\, ,  \\
f_2 &=& c_2 {\cal A} + c_3 k^\beta {\cal A}_\beta\, ,
\eea
which yields
\beq
\Phi_- = \frac{g_s}{32} \left[ |c_1|^2 g^{\alpha \bar \beta} {\cal A}_\alpha \bar {\cal A}_{\bar \beta} + 2|c_2|^2 {\cal A} \bar {\cal A} +2 (c_2 \bar c_3\, {\cal A}\,  \overline{k^\beta} {\cal A}_\beta + c.c.) \right] \, ,
\eeq
once we drop the term proportional to $|c_3|^2$, which is subleading in the $\gamma k/\sigma$ and $(a \sigma)^{-1}$~expansions employed in Eqn.~(\ref{equ:Fterm}).\footnote{The $\gamma k/\sigma$ expansion can be seen to be an expansion in powers of $\frac{\phi}{M_{\rm pl}}$, with $\phi$ the canonically-normalized scalar describing D3-brane motion. As shown in \cite{Baumann2007}, this quantity is bounded from above by $\frac{2}{\sqrt{N}}$ in an $AdS_{5}\times X_{5}$ geometry with D3-brane charge $N$, and hence is a suitable expansion parameter.}

Choosing
\bea
c_1 &=&  c\, , \\
c_2 &=& c \, \sqrt{a \gamma (a \sigma +3)} \, ,\\
c_3 &=& - c \, \frac{a\gamma}{2 c_2} \, ,
\eea
with
\beq
c^2 \equiv \frac{\kappa^2}{ \sigma^2} \frac{e^{-2a\sigma}}{\gamma T_3} \frac{8}{3 g_s}\, ,
\eeq
we recover Eqn.~(\ref{equ:Fterm}).
The omitted harmonic terms can, of course, be adjusted by adding a harmonic piece to $\Phi_-$, no matter the flux background. Notice
also that
\beq
\left|\frac{c_3}{c_2}\right| = \frac{1}{2(a\sigma +3)} \ll 1\, .
\eeq
Finally, we remark that in the decompactification limit ($M_{\rm pl} \to \infty$ and $\gamma \to 0$),
Series~I flux dominates over Series~II flux and the
supergravity F-term potential reduces to the rigid supersymmetry potential.

\section{Towards a D7-brane Geometric Transition}
\label{sec:speculations}

As remarked in the
introduction,
an important open question
in compactifications with nonperturbatively-stabilized K\"ahler moduli
is the computation of nonperturbative contributions to the open string effective action.  One can argue very generally that every divisor supporting nonperturbative effects will contribute to the potential for a mobile D3-brane \cite{Ganor}.  When a coordinate chart containing the divisor is available, it is straightforward to determine the superpotential, and this has been done from a variety of perspectives in toroidal orientifolds \cite{BHK} and in noncompact cones \cite{BDKMMM}.\footnote{It has also been argued that this result can be encoded in a generalized complex geometry \cite{Koerber2007a}.}  In a more generic F-theory compactification, the result of \cite{Ganor} still applies \cite{TolyaProgress}, but it is difficult to translate
this fact into a meaningful contribution to the potential for a D3-brane 
far from a given divisor.

It is therefore natural to pursue an understanding of nonperturbative effects on D7-branes that more efficiently describes the potential for a D3-brane interacting with multiple distant divisors, each bearing nonperturbative effects.  We have suggested in the preceding section that, for the purpose of determining the D3-brane potential, nonperturbative effects on D7-branes can be represented by appropriately-chosen IASD fluxes.
This result, however, is not completely satisfying, as the fluxes were 
engineered to achieve the desired result, rather than emerging from the 
inclusion of local source terms in the equation of motion.  We will now establish a much more interesting fact: {\it{nonperturbative effects on D7-branes source IASD fluxes}}.\footnote{We thank Juan Maldacena and Gonzalo Torroba for discussions on this topic.}
In particular, in a background containing D7-branes wrapping a rigid four-cycle, gaugino condensation in the D7-brane theory corrects the ten-dimensional equation of motion for the fluxes,
by introducing a source term localized to the four-cycle. The ten-dimensional solution in the presence of this source contains IASD fluxes 
that geometrize, in the spirit of \S\ref{sec:4dSUGRA}, the nonperturbative superpotential of the D7-brane theory.

For noncompact configurations, Series~I flux suffices to geometrize the entire scalar potential, while for compact spaces, Series~II flux is also necessary, as shown in \S\ref{sec:4dSUGRA}. In this section we restrict attention to the sourcing of Series~I flux by gaugino condensation, but we emphasize that the results obtained below are valid for compact spaces,
with the proviso that in such cases the D3-brane potential is not fully determined by Series~I flux.

We make the further simplifying assumption that any dilaton gradients are small, in the sense of \S\ref{sec:perturbation}.  This can be consistent with the classical backreaction of D7-branes, {\it{e.g.}} four D7-branes coincident with an O7-plane have vanishing total charge and tension, and hence source neither dilaton monodromies nor a deficit angle. We 
expect that the inclusion of dilaton gradients 
constitutes a purely technical complication, and 
we anticipate that our conclusions can be extended to general F-theory backgrounds.
Although a calculation with non-trivial dilaton is beyond the scope of the present work, in Appendix~\ref{sec:running} we have paved the way for such a computation.


We begin in \S\ref{fermion} by outlining our general strategy, explaining how worldvolume couplings of the gaugino can serve as source terms, and indicating the anisotropic regime in which our methods are
most reliably applied.  Then, in \S\ref{coupling}, we identify the particular coupling of the gaugino that is responsible for sourcing $G_{(1,2)}$ flux.  We provide an indirect argument, via AdS/CFT, for the existence of such a coupling (\S\ref{sec:qftside}), and then confirm this
by direct computation in ten dimensions (\S\ref{sec:bulkside}).  We show that the resulting $G_{(1,2)}$ flux is precisely what is required to geometrize the nonperturbative superpotential in field theory.

\subsection{4D Fermion Bilinears as 10D Sources} \label{fermion}

First, we observe that the ten-dimensional equations of motion for the fluxes in principle include source terms\footnote{We thank James Gray for helpful discussions of this point.} proportional to open string fermion bilinears.  These terms are dropped when considering classical solutions, but nonperturbative effects can induce an expectation value for {\it e.g.}~the D7-brane gaugino bilinear, $\langle\lambda\lambda\rangle \neq 0$.  Then, ten-dimensional closed string fields that couple to $\lambda\lambda$ will obtain D7-brane-localized source terms when gaugino condensation occurs.  For example, a coupling of the schematic form\footnote{Here and in the following, 
$\lambda \lambda$ is shorthand for ${\rm Tr}(\lambda_\alpha \lambda^\alpha)$.}
\begin{equation}
\delta S \ \sim \ \int \d^8 \xi\, \sqrt{-g} \, G_3\, \lambda\lambda\,
\end{equation} could plausibly source bulk fluxes.

\begin{figure}[h!]
    \centering
        \includegraphics[width=0.8\textwidth]{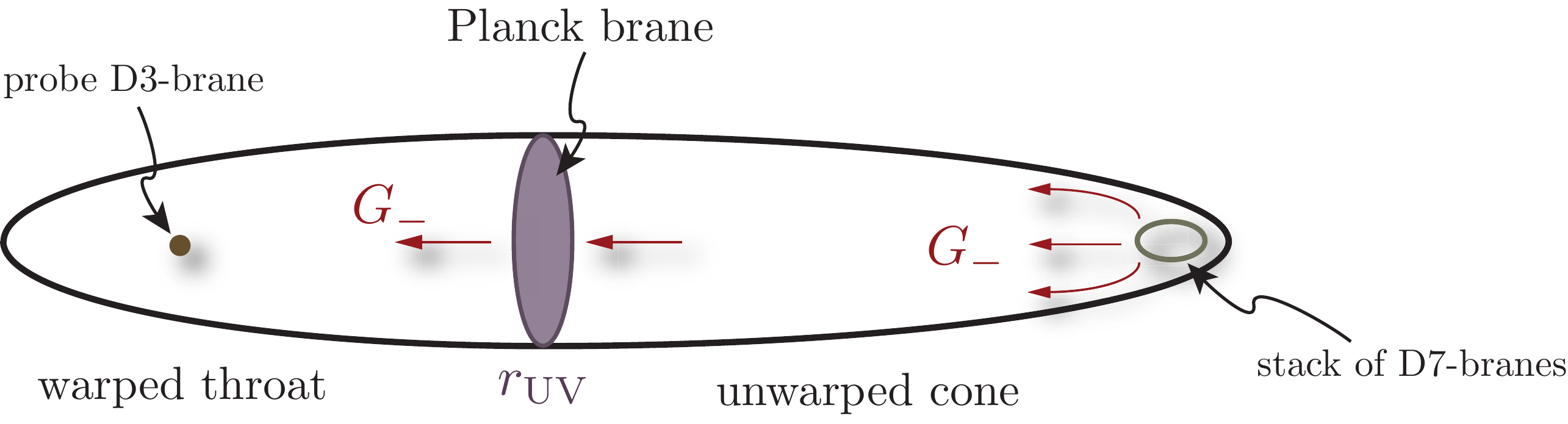}
    \caption{\small \sl Schematic of a configuration in which nonperturbative effects on D7-branes wrapping a small four-cycle are represented, at long distances, by IASD fluxes $G_-$.}
    \label{fig:Pezzo}
\end{figure}

Let us emphasize that we are proposing to solve the {{\it ten-dimensional}} equations of motion for the fluxes, incorporating fermion expectation values that are nonvanishing as a result of {{\it four-dimensional}} nonperturbative physics.\footnote{For a related investigation in the heterotic string, see \cite{Frey2005}.}
This proposal, albeit unconventional, is plausibly consistent, as we will explain in an example.  Consider a stack of D7-branes wrapping the base of a complex cone over a del Pezzo surface (see Fig.~\ref{fig:Pezzo}).  
Gaugino condensation in the D7-brane gauge theory leads to a nonperturbative superpotential that generates a potential for a D3-brane probing the cone.
At radial distances that are large compared to the size of the base, there should be a local supergravity solution describing this system.  Any non-locality caused by the four-dimensional nonperturbative effects should be confined to the region near the collapsed surface, and at larger distances the net effect of gaugino condensation should be to source corrections to the ten-dimensional supergravity solution.  Equivalently, when the four-cycle is small compared to the remainder of the internal space, then the nonperturbative effects  will arise at an energy scale that is large compared to the lowest Kaluza-Klein mass of the internal space.  In such a case, it is reasonable to study the ten-dimensional background fields incorporating four-dimensional nonperturbative effects.  We will restrict our attention to anisotropic configurations of this sort.




\subsection{D7-brane Couplings to Flux} \label{coupling}


We will now identify a specific D7-brane worldvolume coupling that causes gaugino condensation to source fluxes.
In \S\ref{sec:qftside} we will begin by using AdS/CFT to identify the corresponding coupling in field theory.  Then, in \S\ref{sec:bulkside}, we will demonstrate that the well-known tree-level coupling of D7-brane gauginos to flux \cite{Camara2004} serves to source $G_{(1,2)}$ fluxes that geometrize the gaugino condensate superpotential.
The latter result is more general and applies to any Calabi-Yau geometry; specifically, it is not restricted to asymptotically AdS spaces.

\subsubsection{Gauginos as a Source for Flux: Field Theory Perspective}
\label{sec:qftside}


\begin{figure}[h!]
    \centering
        \includegraphics[width=0.6\textwidth]{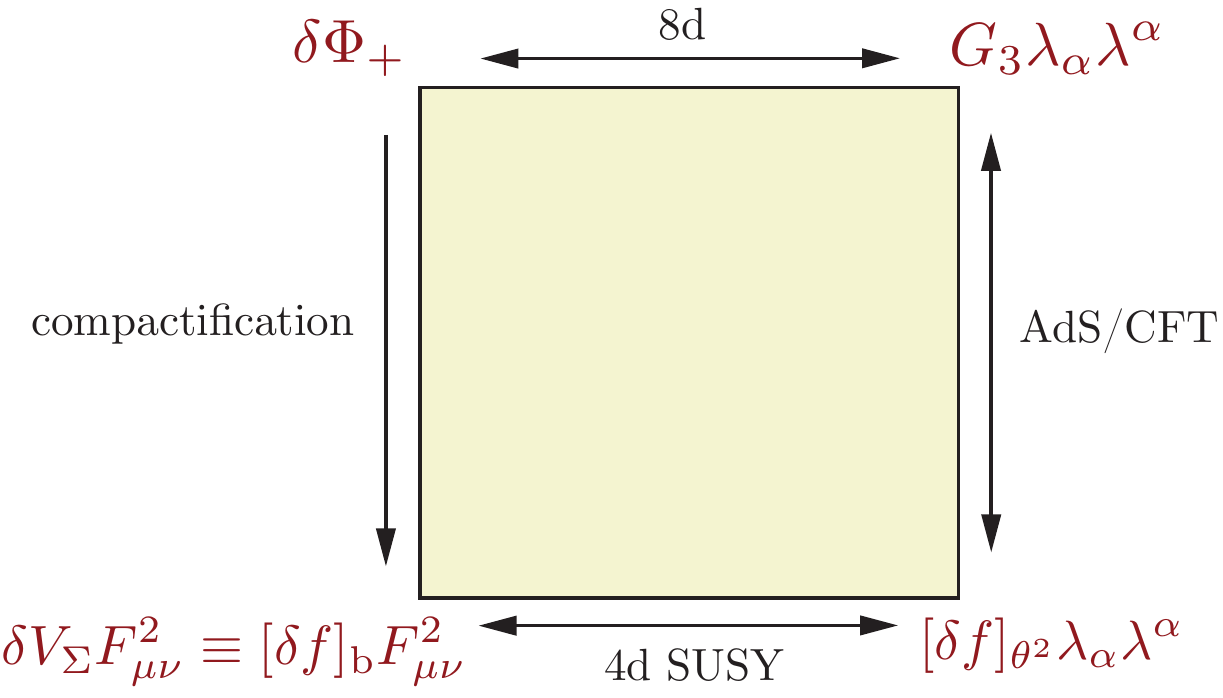}
    \caption{\small \sl 
Indirect argument that a nonvanishing
D7-brane gaugino bilinear $\lambda_{\alpha}\lambda^{\alpha}$ 
sources $G_{(1,2)}$ flux. The D7-brane gauge coupling depends on the D3-brane position via the perturbation $\delta \Phi_+$ \cite{BDKMMM} (left edge).  Four-dimensional supersymmetry relates this to a gaugino mass (bottom edge), which in turn is related by AdS/CFT to a flux-gaugino coupling (right edge).}
    \label{fig:square}
\end{figure}

Four-dimensional supersymmetry requires that the gauge coupling function $f$ of the D7-brane gauge theory is a holomorphic function of 
the chiral superfields $\Phi$,\footnote{To facilitate comparison with the results of  \cite{IntriligatorSeibergLectures} we absorb a factor of $1/32\pi^2$ in $W^\alpha W_\alpha$.}
\beq
{\cal{L}}\  \ \supset \   - \int d^2 \theta\, f(\Phi) W^{\alpha}W_{\alpha}+{c.c.}
\eeq
Expanding the superspace integral, one finds the usual gauge kinetic term, 
as well as the gaugino mass term
\beq
\label{equ:lambda}
 \lambda\lambda \times \int d^2 \theta\, f(\Phi) = F_\Phi {\partial f\over \partial \Phi} \, \lambda\lambda \, .
\eeq
Notice that the D7-brane gaugino mass depends on
the F-components of the fields appearing in $f$; this is just the familiar statement that
in four-dimensional supersymmetry, the gaugino mass term is the F-component of the gauge kinetic function, $F_f \equiv [f]_{\theta^2}$ (bottom edge of Fig.~\ref{fig:square}).

Eqn.~(\ref{equ:lambda}) may also be viewed as a source for the F-term $F_\Phi $, and can be 
represented as a superpotential contribution $\int d^2\theta\, W(\Phi)$ with $W(\Phi)$ satisfying
\beq
\label{equ:superfromlambda}
{\partial W\over \partial \Phi}= - {\partial f\over \partial \Phi}\lambda\lambda\ .
\eeq
This is in agreement with the well-known expression for the gaugino condensate and its relation to the superpotential \cite{IntriligatorSeibergLectures},
\beq
W={N_c}\lambda\lambda \quad {\rm and} \quad\lambda\lambda \sim e^{-f/N_c}\ .
\eeq

Let us now specialize to the warped conifold in order to make use of the AdS/CFT dictionary of \S\ref{sec:cft}.  The general chiral superfields $\Phi$ are therefore replaced by the local coordinates $z_{\alpha}$ (related to gauge-invariant combinations of $A, B$).  We argued in \S\ref{sec:cft} that the F-term of the chiral fields  is dual
to $G_{(1,2)}$ flux of Series~I.
From Eqn.~(\ref{phiminus}) we find that the superpotential $W$ and the defining function $f_{1}$ of the Series~I flux 
are related by
\bea
\label{equ:qftf_1}
\frac{\partial W}{\partial z_\alpha} = \zeta\, {\partial f_1\over \partial z_\alpha}\ ,
\eea
where we have defined the dimensionful parameter
\beq
\zeta \equiv T_3 \sqrt{\frac{g_s}{32}}\, .
\eeq
Comparing this with Eqn.~(\ref{equ:superfromlambda}),
we see that the gaugino condensate
$\langle\lambda \lambda\rangle$ 
in field theory must source $G_{(1,2)}$ flux in the bulk.

In fact, we can be more precise in predicting the form of the sourced flux.
From \cite{Ganor,BDKMMM} we know how $f$ depends on the chiral superfields.  For a D7-brane embedded along a divisor that is defined 
by the holomorphic equation $h(z_\alpha)=0$, the coupling constant is
\beq
\label{equ:f}
 f = 2\pi \rho -\log(h(z_\alpha))\, .
\eeq
Combining the above results,
we obtain a nontrivial prediction for the harmonic function $f_1$ 
(which defines the Series~I flux),
\beq
\label{equ:f1fromgauginos}
{\partial f_1\over \partial z^\alpha}= \zeta^{-1} \lambda\lambda\, {\partial \log h\over \partial z^\alpha}\, .
\eeq

Our argument uses four-dimensional supersymmetry
and the AdS/CFT dictionary
to
deduce a coupling in four dimensions between the gauginos of D7-branes in a warped throat,
and CFT fields dual to $G_{(1,2)}$ fluxes in that throat. 
We expect that this coupling in the four-dimensional theory arises from the dimensional reduction of a corresponding eight-dimensional
coupling between D7-brane fermions and $G_3$. 
Moreover, it is reasonable to expect, and we shall demonstrate, 
that such a coupling is present even in spaces that are not asymptotically AdS.
In the next section we will demonstrate that the coupling in question is nothing but the tree-level 
gaugino mass term of \cite{Camara2004}.


\subsubsection{Gauginos as a Source for Flux: Bulk Perspective}
\label{sec:bulkside}

The tree-level couplings between D7-brane worldvolume fermions and the bulk fields have been studied in \cite{Camara2004,Grana2003a} by expanding the Born-Infeld plus Chern-Simons 
actions around the
D7-brane location.  Ignoring the trivial Minkowski part and assuming constant dilaton, the unique tree level coupling involving $G^*_{(0,3)}\lambda\lambda$ is \cite{Camara2004},
\begin{equation}
\label{D7action}
{\cal{L}} = 16 \, c_0\, \zeta \int\limits_{\Sigma} \sqrt{g}\, G_{3} \cdot \Omega \, \bar\lambda \bar \lambda \ +\ c.c. 
\end{equation}
Here $c_{0} \sim {\cal O}(1)$ is a dimensionless coefficient that could be obtained from a careful uplifting to eight dimensions of the four-dimensional action of \cite{Camara2004}. In the following we set $c_0 \to 1$, but we emphasize that the matching of the overall coefficient in the analysis below would require the computation of $c_0$.
Adding this local coupling (\ref{D7action}) to the bulk action modifies the flux equation of motion, and, as we shall show, links the gaugino expectation value $\langle \lambda \lambda \rangle$ to the bulk flux $G_{(1,2)}$.

Before we proceed, let us recall some facts about divisor delta-functions.\footnote{We thank Luca Martucci for discussions.}
The divisor $\Sigma$ wrapped by the  D7-branes is defined via the holomorphic embedding equation $h=0$.
The Poincare-Lelong Equation then defines a delta-function two-form \cite{Griffiths},
\beq
\label{delta}
\delta^{(2)}_{\alpha\bar\beta}= \frac{1}{\pi}\partial_\alpha\partial_{\bar\beta}\, {\rm Re}(\log h)\, ,
\eeq
which localizes the integral of any four-form $F^{(4)}$ onto $\Sigma$,
\beq
\int\limits_{M} F^{(4)}\wedge \delta^{(2)}=\int\limits_\Sigma F^{(4)}\, .
\eeq
Eqn.~(\ref{delta}) is a generalization of the scalar delta-function, which may be obtained from (\ref{delta}) by contraction with $g^{\alpha\bar \beta}$,
\bea
\label{delta2}
\delta^{(0)} =  g^{\alpha \bar \beta} \delta_{\alpha \bar \beta}^{(2)} = \frac{1}{2\pi} \nabla^2{\rm Re}(\log h)\ .
\eea
Eqn.~(\ref{delta2}) was derived for conifold geometries in \cite{BDKMMM}.

We use Eqn.~(\ref{delta}) to transform the local action on $\Sigma$, Eqn.~(\ref{D7action}), into an
integral over the whole compactification manifold $M$,
\beq
\int\limits_\Sigma \sqrt{g} \, G_3\cdot \Omega\ \bar\lambda \bar \lambda= \frac{1}{2!}\int\limits_\Sigma J\wedge J  \ (G_3\cdot \Omega\ \bar\lambda \bar \lambda) = {1\over 2!}\int\limits_M J\wedge J\wedge \delta^{(2)} \ (G_3\cdot \Omega\ \bar\lambda \bar \lambda)\ .
\eeq
Using that ${1 \over 2!} J\wedge J\wedge \delta^{(2)} = {1\over 3!}J \wedge J \wedge J\, \delta^{(0)}$, we can then write (\ref{D7action}) as
\beq
\label{D7act}
\frac{1}{3!}\int\limits_M
J\wedge J\wedge J\ (G_3\cdot \Omega\ \bar\lambda \bar \lambda\ \delta^{(0)})= \int\limits_M  G_3\wedge \Omega\ ( \bar\lambda \bar \lambda\ \delta^{(0)})\ .
\eeq
Next, we compare the variation of (\ref{D7act}) with respect to $C_2$ and $B_2$,
\beq
16\, \zeta \int\limits_M \delta G_3\wedge \Omega\ ( \bar\lambda \bar \lambda\ \delta^{(0)})\ +\  c.c. \ ,
\eeq
to the variation of the bulk supergravity action,
\beq
- \frac{g_s}{4 \kappa_{10}^2} \int\limits_M \delta G_3\wedge \bar \Lambda \ + \ c.c.\ ,
\eeq
where $\kappa^2_{10} = \frac{1}{2} (2\pi)^7 \alpha'^4 = \pi/ T_3^2$.
We obtain the modified equation of motion
\beq
\label{eomrhs1}
\d\Lambda =  \d \Bigl(\, \frac{2\pi}{\zeta}\ \bar\Omega\, \lambda \lambda\, \delta^{(0)}\, \Bigr)\ , \qquad {\rm with} \quad \star_6 \Lambda = - i \Lambda\, .
\eeq
To solve this equation we consider non-trivial IASD 
fluxes from Series~I, II, and III. The source term on the r.h.s.~of~(\ref{eomrhs1}) is a singular $(1,3)$ form, and therefore the defining functions $f_1$, $f_2$, and $f_3$
must be harmonic everywhere except on the divisor $\Sigma$. Consulting Eqns.~(\ref{equ:s3-12}) and (\ref{equ:s2-12}), 
we see that Series~III flux does not lead to a non-trivial $(1,3)$ form,
and that Series~II flux leads to a combination of $(1,3)$ and $(2,2)$ forms. The only remaining possibility,
therefore, is Series~I flux (\ref{equ:Series1}) with
\beq
\label{equ:xyX}
\d \Lambda =\d \Bigl(\frac{1}{2}  \nabla^2 f_1 \wedge \bar \Omega \Bigr)\, .
\eeq
Comparing Eqns.~(\ref{eomrhs1})  and (\ref{equ:xyX}), we obtain
\beq
\label{eomrhs1_1}
\nabla^2 f_1=\frac{4\pi}{\zeta} \lambda \lambda\, \delta^{(0)} \, .
\eeq
The gaugino condensate $\lambda\lambda$ depends on the location of any D3-branes $\widehat z_\alpha$, but is a constant with respect to the bulk coordinates $z_\alpha$.
Solving Eqn.~(\ref{eomrhs1_1}) then yields
\bea
\label{eq:12fluxinbulk}
f_1= 2 \zeta^{-1} \lambda \lambda\  {\rm Re} (\log h(z_\alpha))\, ,
\eea
in agreement with Eqn.~(\ref{equ:f1fromgauginos}).
This, in fact, is the bulk dual of the field-theory mechanism of \S\ref{sec:qftside}. 

To calculate the potential for a D3-brane we first calculate the flux $G_{(1,2)}$ using Eqn.~(\ref{eq:12fluxinbulk}) and the 
form of the 
embedding function $h(z_\alpha)$. 
Then we calculate $\Phi_-$ using Eqn.~(\ref{phiminus}). At this step $\Phi_-$ depends on the coordinates $z_\alpha$ in the bulk,
and also on the location of the D3-branes through $\lambda\lambda(\widehat z_\alpha) \propto \frac{1}{N_c} h(\widehat z_\alpha)^{1/N_c}$, 
{\it cf.}~(\ref{6_10}) \cite{BDKMMM}.
Finally, we compute the potential for the D3-brane by evaluating the D3-brane probe action $T_3\Phi_-$
at the location of the D3-brane.  The resulting potential 
is given by Eqn.~(\ref{equ:Vr1}), with $W$ the gaugino condensate superpotential~(\ref{6_9}).   Therefore, the scalar potential takes precisely the same form when calculated in four-dimensional supersymmetry or from the ten-dimensional probe action in the flux background (\ref{eq:12fluxinbulk}) sourced by gaugino condensation.


We remark in passing that our result is valid for any number of D3-branes. To illustrate this point, 
let us
consider a conifold region that contains
$N_{c}$ D7-branes wrapping the four-cycle $h(z_\alpha)=\mu-z_1$,  as well as $N_f$ well-separated D3-branes.
We further assume that any adjoint chiral multiplets living on the D7-branes have 
obtained a high-scale mass from appropriate ISD flux, so the low-energy effective theory on the D7-branes in
the absence of probe D3-branes is pure ${\cal N}=1$ supersymmetric
$SU(N_{c})$ QCD.
The D3-branes at finite separation from the D7-branes correspond to massive flavors in the ${\cal N}=1$ super-Yang-Mills theory on the D7-branes, and in this theory with mass matrix $m$ for the flavors, the superpotential is \cite{IntriligatorSeibergLectures}
\begin{equation}
\label{equ:Wformanyd3}
W=N_c\lambda\lambda \ \propto \ {\rm det}(m)^{1/N_{c}}={\rm det}\left( \mu\, \mathbb{I}_{N_f\times N_f}-\widehat{\mathbb{Z}}_1\right)^{1/N_{c}}\, ,
\end{equation}
in agreement with \cite{Marchesano2009, BDKMMM}.
The above superpotential describes an interaction among the D3-branes sitting at locations specified by the eigenvalues of the matrix $\widehat{\mathbb{Z}}_1$.

\vskip 6pt
We note that our proposal for the long-distance supergravity description
of the D7-brane nonperturbative effects has certain similarities to the well-known geometric transition~\cite{Vafa,MaldacenaNunez, KS}, for which
it is now firmly established
that nonperturbative dynamics on a stack of D5-branes wrapping a small curve can be captured by a modified background geometry and suitable $G_3$ flux.
However, at present our only method of probing the supergravity solution sourced by the D7-brane nonperturbative effects is by examining the potential of a probe D3-brane.
Therefore, our speculations are necessarily limited for the time being to the modes of $\Phi_-$ (and hence, IASD fluxes) that nonperturbative effects on D7-branes may produce.  It would be interesting to identify additional probes of the geometry around the D7-branes~\cite{Future}.

Another ten-dimensional description of nonperturbative effects was given in \cite{Koerber2007a}, where a generalized complex geometry was argued to encode the nonperturbative superpotential in the decompactification limit. In our language this would correspond to the presence of
Series~I flux. It would be interesting to understand the relation between the flux solution described in this paper and the configuration studied in \cite{Koerber2007a}; it is conceivable that the ten-dimensional equations of motion relate these results.

\subsection{Comments on Noncommutativity}

Our proposal provides a direct connection between the works of Cecotti {\it et~al.}~\cite{Cecotti2009} and Marchesano and Martucci~\cite{Marchesano2009}.  In \cite{Cecotti2009} it was observed that the noncommutative superpotential induced on coincident D3-branes by three-form fluxes \cite{Myers1999,Camara2004} can solve a problem in F-theory model-building by increasing the rank of the Yukawa matrix. However, in \cite{Marchesano2009} it was argued that a noncommutative superpotential should not appear at tree level in a no-scale compactification, but should instead arise from nonperturbative effects on D7-branes.
We have argued here that
nonperturbative effects on D7-branes can be represented in ten-dimensional supergravity by suitable IASD fluxes.  Therefore, our analysis provides a concrete link between these approaches.
To illustrate this point we briefly sketch how one can arrive at the same answer for the
noncommutativity parameter from either perspective.  Marchesano and Martucci derived the noncommutativity parameter on a stack of $N_f$ D3-branes in the presence of nonperturbative effects on $N_c$~D7-branes~\cite{Marchesano2009}\footnote{For simplicity, \cite{Marchesano2009} considered the case $N_c =1$, corresponding to replacing the effect of gaugino condensation on D7-branes by that of Euclidean D3-branes.  To compare our result to \cite{Marchesano2009} we also take $N_c =1$.}
\bea
\label{theta1}
\theta_{\alpha\beta} \propto \epsilon_{\alpha\beta\gamma}h^{N_f-1}\partial_{\gamma}h\ ,
\eea
where as before $h=0$ defines the four-cycle wrapped by the D7-branes (or
by
a Euclidean D3-brane),
and special complex coordinates have been chosen such that the metric is canonical, $g_{\alpha\bar\beta}=\delta_{\alpha\bar\beta}$.
On the other hand, Cecotti {\it et~al.}~\cite{Cecotti2009} (see also \cite{Camara2004}) showed that the noncommutativity parameter arises from $(1,2)$ flux,
\bea
\label{theta2}
\nabla_\alpha\theta_{\bar\beta\bar\gamma} \propto \Lambda^{(1,2)}_{\alpha\bar\beta\bar\gamma}\, .
\eea
Our conjecture for relating nonperturbative effects on D7-branes to bulk IASD fluxes directly relates the results in (\ref{theta1}) and (\ref{theta2}).
In our approach, the $(1,2)$ flux of (\ref{theta2}) is sourced by 
gaugino 
condensation on D7-branes
in the presence of the $N_f$ D3-branes,
and hence depends on their locations $\widehat z_\alpha$. Namely, for coincident D3-branes
$\lambda\lambda \propto h^{N_f}(\widehat z_\alpha)$,
and hence $\partial_\alpha f_1\propto h^{N_f}\partial_\alpha \log h(z_\beta)$. Comparing the expression (\ref{equ:Series1}) for the $(1,2)$ flux,
$\Lambda^{(1,2)}_{\alpha\bar\beta\bar\gamma}= \nabla_\alpha \nabla_\sigma f_1 g^{\sigma \bar \zeta} \epsilon_{\bar \zeta \bar \beta \bar \gamma}$, with (\ref{theta2}) we conclude that
\bea
\theta_{\alpha\beta} \propto \epsilon_{\alpha\beta\gamma}\partial_{\gamma}f_1\ ,
\eea
in agreement with (\ref{theta1}).
It would be interesting to extend this analysis
to include the effects of a running dilaton by using the expression for the Series~I flux from Appendix~B,
thereby generalizing the results of~\cite{Cecotti2009} and \cite{Camara2004}. 

\section{Conclusions}
\label{sec:conclusions}

We have provided a comprehensive
analysis of the structure of the scalar potential for a D3-brane in a noncompact conifold background subject to arbitrary ultraviolet deformations, in an expansion around the ISD solution 
with $G_{-} = \Phi_- = 0$.  This strongly constrains the form of the effective action
of a D3-brane in a finite warped throat attached to a compact space.  Our general analysis of ultraviolet deformations provides the gravity-side version of a Wilsonian treatment of the D3-brane action, in that the perturbations of the supergravity solution are dual to Planck-suppressed contributions to the D3-brane action.  As is clear from our description of these results
in the dual CFT, strong coupling makes a complete treatment on the field theory side
extremely difficult.

Our
investigation yielded a general solution for IASD fluxes in the singular conifold,
and can be extended to arbitrary Calabi-Yau cones given the scalar harmonics on the corresponding Sasaki-Einstein bases.  For a given flux background, it is straightforward to find the scalar potential, and for $G_{(1,2)}$ and $G_{(2,1)}$ fluxes parametrized by a holomorphic function, we gave the result in closed form. In the remaining cases we provided an integral formula involving the flux density and the (known) Green's function, which is easily evaluated in any case of interest.
Assembling these results, we presented the spectrum of radial scaling dimensions for the leading correction terms, Eqn.~(\ref{equ:OurAnswerAtLast}).

We then systematically matched our non-normalizable flux solutions to sources for dual operators in the corresponding CFT.  This provided a nontrivial check of the completeness of our results, as well as a useful perspective for comparing to four-dimensional field theory analyses.  We showed that for two out of three towers of chiral operators, the scalar potential computed in the field theory matches that found in supergravity. For the third tower of chiral operators, we argued that knowledge of the perturbed K\"ahler potential would be necessary to compare to the supergravity result, but we did provide a consistency check relating the scalar potential to the perturbations of the metric.

Next, we
observed that the D3-brane
potential specified by any given superpotential $W(z_\alpha)$ on the conifold can be {\it{geometrized}}, in the sense that there exists a ten-dimensional supergravity solution in which a probe D3-brane experiences the supergravity F-term potential dictated by $W(z_\alpha)$.
We have verified that for the rigid supersymmetry terms, this correspondence persists even in the presence of large perturbations of the metric and dilaton sourced by D7-branes.

Finally, we showed that gaugino condensation on D7-branes wrapping a four-cycle {\it sources} IASD flux in ten dimensions.  For the purpose of computing the potential of a space-filling D3-brane, these fluxes serve as a useful dual representation of the nonperturbative effects.  Using AdS/CFT, we argued that a suitable gaugino coupling to flux must exist, then demonstrated that in fact the well-known tree-level gaugino mass term \cite{Camara2004} serves to source IASD flux that is dual, in the above sense, to the nonperturbative superpotential.
This observation is similar to the D5-brane geometric transition \cite{Vafa, MaldacenaNunez, KS},
but we leave a thorough investigation of this intriguing possibility as a subject for future work~\cite{Future}.

Our results have a range of applications. An important problem of general interest in string theory is to determine Planck-suppressed contributions to the effective action in sectors that are sensitive to Planck-scale physics. In inflation, one needs to compute these contributions in order to ensure that they do not spoil the inflationary dynamics, while in particle physics scenarios with high-scale supersymmetry breaking, one needs to determine the soft terms from gravity mediation in order to assess the flavor structure.
Direct computation of these effects in general compact models is prohibitively complicated.  In this work, by first decoupling gravity in a noncompact configuration, then systematically reincorporating compactification effects, we
have provided an approach in which the structure of the Planck-suppressed contributions can be computed.

\subsection*{Acknowledgements}
We are grateful to M.~Berg, M.~Cheng, O.~DeWolfe, A.~Frey, S.~Gandhi, J.~Gray, D.~Green, J.~Heckman, K.~Intriligator, Z.~Komargodski, L.~Leblond, J.~Maldacena, D.~Marsh, L.~Martucci, M.~Mulligan, E.~Pajer, L.~Rastelli, N.~Seiberg, A.~Sever, E.~Silverstein, G.~Torroba, S.~Trivedi, H.~Tye, B.~Underwood, and G.~Xu for helpful discussions.

D.B. and L.M. thank the Center for Advanced Conifold Studies at the Stanford
Institute for Theoretical Physics for hospitality while some of this work was performed.
D.B. thanks the theory groups at Cambridge, Princeton, Stony Brook, McGill and UMass, Amherst for hospitality and the opportunity to present this work.
A.D. thanks the theory group at Cornell University for hospitality while part of this work was done.
I.R.K. thanks the KITP for hospitality during part of this work.
D.B., A.D., S.K., and L.M. thank the Aspen Center for  Physics, where some of these results were obtained, for providing a stimulating environment. L.M.
thanks the theory groups at Harvard and the KITP for hospitality while  this work was performed, as well as the Institute for Advanced Study 
and the organizers of the 2010 CERN Winter School for hospitality while it was completed.
The research of D.B. was supported by the NSF under the grants PHY-0855425, AST-0506556 and AST-0907969.
The research of A.D. was supported by the Stanford Institute for
Theoretical Physics, by DOE grant DE-FG02-90ER40542,
and in part by
the grant RFBR 07-02-00878 and the Grant for Support of Scientific Schools NSh-
3035.2008.2.  During the course of this research, S.K. was supported by the NSF under grant PHY05-51164 at KITP, by the Stanford Institute for Theoretical Physics,
and by the DOE under contract DE-AC02-76SF00515 at SLAC. The research of I.R.K. was supported in part by the NSF under grant PHY-0756966.   The research of L.M. was supported by the Alfred P. Sloan Foundation and by the NSF under grant PHY-0757868.

\newpage
\appendix

\section{IASD Fluxes in Calabi-Yau Cones}
\label{sec:IASD}

In this appendix we
derive
the solutions for imaginary anti-self dual  (IASD) flux perturbations on a general Calabi-Yau cone in terms of the scalar harmonics on the Sasaki-Einstein base. These results provide the primary input for computing the flux-induced potential felt by probe D3-branes.

\subsection*{Preliminaries}

We are searching for closed, IASD three-forms $\Lambda$,
\beq
\label{equ:Leom}
\d \Lambda = 0 \, , \qquad \star_6 \Lambda = - i \Lambda\, .
\eeq
We construct these forms using the K\"ahler form $J$ and the holomorphic $(3,0)$ form $\Omega$, as well as harmonic functions $f$.
The K\"ahler form is a $(1,1)$ form with components
\beq
J_{\alpha \bar \beta} = i g_{\alpha \bar \beta}\, ,
\eeq
where $g_{\alpha \bar \beta}$ is the K\"ahler metric
\beq
g_{\alpha \bar \beta} = \partial_\alpha \partial_{\bar \beta} k\, .
\eeq
The holomorphic $(3,0)$ form has components
\beq
\Omega_{\alpha \beta \gamma} = q\, \epsilon_{\alpha \beta \gamma}\, ,
\eeq
where $q$ is a holomorphic function satisfying $q \bar q = \det g$.
We will make use of the identity
\beq
\qquad \frac{1}{2!}   \Omega_{\alpha \beta \gamma} \bar\Omega_{\bar \alpha \bar \beta \bar \gamma} g^{\beta \bar \beta} g^{\gamma \bar \gamma} = g_{\alpha \bar \alpha}\, .
\eeq

Using these ingredients we construct IASD three-forms:
\begin{itemize}
\item
We write the non-primitive $(2,1)$ component of the three-form $\Lambda$ as the exterior product of the K\"ahler form $J$ and a holomorphic one-form $P$,
\beq
\label{equ:21}
\Lambda^{(2,1)} = P^{(1,0)} \wedge J^{(1,1)}\, , \qquad {\rm or} \qquad
\Lambda_{\alpha  \beta \bar \gamma} = i P_{[\alpha} g_{\beta ] \bar \gamma}\, .
\eeq
\item We write the $(1,2)$ component of the three-form $\Lambda$ as the contraction of the anti-holomorphic $(0,3)$ form $\bar \Omega$ and a holomorphic $(2,0)$ form $P$,
\beq
\label{equ:12}
\Lambda^{(1,2)} = P^{(2,0)} \cdot \bar \Omega^{(0,3)}\, , \qquad {\rm or} \qquad
\Lambda_{\alpha \bar \beta \bar \gamma} = P_{(\alpha \sigma)} g^{\sigma \bar \zeta} \bar \Omega_{\bar \zeta \bar \beta \bar \gamma}\, .
\eeq
Only the symmetric part of $P$ gives an IASD form.
\item
Finally, the holomorphic $(3,0)$ form $\Omega$ is IASD.
\end{itemize}

The forms $P$ will be constructed out of derivatives of harmonic functions $f$.

\subsection*{Series I: Three-Form of Type $(1,2)$}

We construct the two-form $P$ in Eqn.~(\ref{equ:12}) out of
mixed
covariant derivatives of a harmonic function~$f$,
\beq
P_{(\alpha \sigma)} = \nabla_\alpha \nabla_\sigma f \, ,
\eeq
where $\nabla^2 f= g^{\rho \bar \zeta} \nabla_\rho \nabla_{\bar \zeta} f =0$.
Below we prove that the resulting three-form is closed. Hence, we obtain the first three-form,
$\Lambda_{\rm I} = \Lambda^{(1,2)} $,
\beq
\label{equ:Lambda1}
\fbox{$\displaystyle
\Lambda_{\rm I} = \nabla \nabla f \cdot \bar \Omega$}\, ,
\eeq
or
\beq
\label{equ:Lambda1A}
\Lambda_{\alpha \bar \beta \bar \gamma} = \nabla_{\alpha}\nabla_\sigma f\, g^{\sigma \bar \zeta} \bar \Omega_{\bar \zeta \bar \beta \bar \gamma}\, .
\eeq

\vspace{0.5cm}
\small
 \hrule \vskip 1pt \hrule \vspace{0.3cm}
\noindent
{\sl The three-form $\Lambda_{\rm I}$ defined in Eqn.~(\ref{equ:Lambda1}) is closed.}
\newline
\noindent
{\bf Proof}:\newline
We aim to show that $\d \Lambda_{\rm I} = (\partial + \bar \partial) \Lambda_{\rm I} = 0$.
First, we consider the anti-holomorphic derivative
\beq
(\bar \partial \Lambda_{\rm I})_{\alpha \bar \beta \bar \gamma \bar \delta}
= \nabla_{[ \bar \delta }  \nabla_{\alpha}\nabla_\sigma f\, g^{\sigma \bar \zeta} \bar \Omega_{\bar \zeta | \bar \beta \bar \gamma]}
\equiv T_\alpha\, \bar \Omega_{\bar \beta \bar \gamma \bar \delta} \, ,
\eeq
where
\beq
T_\alpha \propto g^{\sigma \bar \zeta} \nabla_{\bar \zeta}  \nabla_\alpha \nabla_\sigma f \, .
\eeq
We can exchange the order of $\nabla_{\bar \zeta}$ and $\nabla_\alpha$ because the difference is proportional to
\beq
g^{\sigma \bar \zeta} R^\rho_{\ \sigma \bar \zeta \alpha} \nabla_\rho f = R_{\alpha \bar \beta}\, g^{\rho \bar \beta} \, \nabla_\rho f\, ,
\eeq
which vanishes because the cones we consider are Calabi-Yau manifolds with vanishing Ricci tensor, $R_{\alpha \bar \beta} = 0$.
Hence, we find
\beq
(\bar \partial \Lambda_{\rm I})_{\alpha \bar \beta \bar \gamma \bar \delta}  \propto \nabla_\alpha \nabla^2 f \, \bar \Omega_{\bar \beta \bar \gamma \bar \delta} = 0\, ,
\eeq
where we used that $f$ is harmonic, $\nabla^2 f=0$.

Similarly, we can show that the holomorphic derivative of $\Lambda_{\rm I}$ vanishes,
\beq
\label{equ:B18}
(\partial \Lambda_{\rm I})_{\alpha  \beta \bar \gamma \bar \delta} = \nabla_{[\alpha} P_{\beta] \zeta} \bar \Omega^\zeta_{\ \bar \gamma \bar \delta} = 0\, .
\eeq
To see this we use $\nabla_{[\alpha} P_{\beta] \zeta} = \nabla_{[\alpha} \nabla_{\beta]} \nabla_\zeta f$, which vanishes because the holomorphic covariant derivatives $\nabla_\alpha$ and $\nabla_\beta$ commute; their commutator is the Riemann tensor, which has no nontrivial components with only holomorphic indices if the metric is K\"ahler.
We have therefore shown that $\d \Lambda_{\rm I} = (\partial + \bar \partial) \Lambda_{\rm I} = 0$.
\hfill   $\blacksquare$
\vspace{0.2cm}  \hrule \vskip 1pt \hrule
 \vspace{0.5cm}

\normalsize

\subsection*{Series II: Three-Form of Type $(1,2)+(2,1)_{\rm NP}$}

Our next
ansatz for the two-form in Eqn.~(\ref{equ:12}) is
\beq
\label{equ:P12}
P_{(\alpha \sigma)} = \nabla_{\bar \zeta} \nabla_{ (\alpha} f \omega^{\bar \zeta}_{\ \sigma)}\, ,
\eeq
where
\beq
\omega^{\bar \zeta}_{\ \sigma} \equiv \Omega^{\bar \zeta}_{\ \sigma \rho } k^\rho \, , \qquad k^\rho \equiv g^{\rho \bar \xi} \partial_{\bar \xi} k\, .
\eeq
The resulting three-form,
\beq
\label{equ:s2-12}
\Lambda^{(1,2)}_{\rm II} = \partial \bar \partial f \wedge \bar \partial k + \frac{1}{2} J \wedge \bar \partial(\partial_\rho f k^\rho) -\frac{1}{2} \nabla^2 f\, J\wedge \bar\partial k\, ,
\eeq
is {\it not} closed. However, it becomes closed when an appropriate $(2,1)$ piece is added,
\beq
\Lambda^{(2,1)}_{\rm II} = \partial \Bigl(f + \frac{1}{2} \partial_\rho f k^\rho \Bigr) \wedge J\, ,
\eeq
so that we have
$\Lambda_{\rm II} \equiv \Lambda^{(1,2)}_{\rm II} + \Lambda^{(2,1)}_{\rm II}$,
\beq
\label{equ:Lambda2}
\fbox{$\displaystyle
\Lambda_{\rm II} =( \partial +\bar \partial ) \Bigl(f + \frac{1}{2} \partial_\rho f k^\rho \Bigr) \wedge J + \partial(\bar \partial f \wedge \bar \partial k) $}\, .
\eeq

\vspace{0.5cm}
\small
 \hrule \vskip 1pt \hrule \vspace{0.3cm}
\noindent
{\sl The three-form $\Lambda_{\rm II}$ defined in Eqn.~(\ref{equ:Lambda2}) is  closed.}
\newline
\noindent
{\bf Proof}:\newline
We aim to show that $\d \Lambda_{\rm II} = (\partial + \bar \partial) \Lambda_{\rm II} = 0$. We will begin by showing that $k^\sigma$ is holomorphic, {\it i.e.}
\beq
\partial_{\bar \zeta} k^\sigma = \nabla_{\bar \zeta} k^\sigma = g^{\sigma \bar \rho } \nabla_{\bar \zeta} \nabla_{\bar \rho} k = 0\, .
\eeq
To prove this, it suffices to prove that
\begin{equation}
\nabla_{m}\nabla_{n}k=g_{mn} \,,
\end{equation}  where $m,n$ are {\it{real}} indices. This result is easily checked for any cone with metric $\d s^2 = \d r^2 +r^2 \d\Omega_{X_{5}}^2$ and K\"ahler potential $k=\frac{r^2}{2}$.

We now show that $\bar \partial \Lambda^{(1,2)}_{\rm II}$ vanishes. Since $k^\sigma$ is holomorphic and $\Omega$ is covariantly constant, we can
extract $\omega^{\bar \zeta}_{\ \beta}$ from the covariant derivative of Eqn.~(\ref{equ:P12}). We find
\beq
\bar \partial_{[\bar \delta} \Lambda_{\alpha | \bar \beta \bar \gamma]} =  T_\alpha \, \bar \Omega_{\bar \beta \bar \gamma \bar \delta}\, ,
\eeq
where
\beq
\label{equ:xX}
T_\alpha \equiv \frac{1}{2} \nabla_{\bar \zeta} \nabla_{\bar \sigma} \nabla_{ \alpha} f\, \Omega^{\bar \zeta \bar \sigma}_{\ \ \rho} k^\rho + \frac{1}{2} g^{\rho \bar \sigma} \nabla_{\bar \sigma} \nabla_{\bar \zeta} \nabla_{ \rho} f\,  \omega^{\bar \zeta}_{\ \alpha}\, .
\eeq
Commuting $\nabla_{\bar \zeta}$ and $\nabla_{\bar \sigma}$ with $\nabla_\alpha$ in the first term we find
\beq
\nabla_{\bar \zeta} \nabla_{\bar \sigma} \nabla_\alpha f = \nabla_\alpha \nabla_{\bar \zeta} \nabla_{\bar \sigma} f - R^{\bar \rho}_{\ \bar \zeta \bar \sigma \alpha} \nabla_{\bar \rho} f\, .
\eeq
The result is symmetric in the $\bar \zeta$ and $\bar \sigma$ indices because
\beq
R^{\bar \rho}_{\ \bar \zeta \bar \sigma \alpha}   = \partial_\alpha \Gamma^{\bar \rho}_{\bar \zeta \bar \sigma}.
\eeq
Since $\Omega^{\bar \zeta \bar \sigma}_{\ \ \rho} $ is antisymmetric in $\bar \zeta$ and $\bar \sigma$, the first term in Eqn.~(\ref{equ:xX}) vanishes.
The second term vanishes because $f$ is harmonic.

Before we calculate $\partial \Lambda^{(1,2)}_{\rm II}$, let us rewrite Eqn.~(\ref{equ:P12}) in a slightly different form
\beq
\label{equ:Pas}
P_{(\alpha \sigma)} = \nabla_{\bar \zeta} \nabla_{ (\alpha} f \omega_{\ \sigma)}^{\bar \zeta} = \nabla_{\bar \zeta } \nabla_{\alpha} f \omega^{\bar \zeta}_{\ \sigma} + \nabla_{\bar \zeta} \nabla_{[\alpha} f \omega^{\bar \zeta}_{\ \sigma]}\ .
\eeq
After contracting Eqn.~(\ref{equ:Pas}) with $\bar \Omega$ as in Eqn.~(\ref{equ:12}) the first term gives
\beq
\label{equ:ft}
\nabla_{\bar \beta} \nabla_\alpha f k_{\bar \gamma} - \nabla_{\bar \gamma} \nabla_\alpha f k_{\bar \beta}\, ,
\eeq
while the second term should be proportional to $P^{(0,1)} \wedge J$ for some appropriate one-form $P^{(0,1)}$,
\beq
\label{equ:nz}
\nabla_{\bar \zeta} \nabla_{[\alpha} f \omega^{\bar \zeta}_{\ \sigma]} \bar \Omega^{\sigma}_{\ \bar \beta \bar \gamma} = P_{\bar \beta} g_{\alpha \bar \gamma} - P_{\bar \gamma} g_{\alpha \bar \beta}\, .
\eeq
To find $P^{(0,1)}$ we contract Eqn.~(\ref{equ:nz}) with $g^{\alpha \bar \beta}$. Because $f$ is harmonic we find
\beq
P_{\bar \gamma} = - \frac{1}{2} \nabla_{\bar \gamma}( \partial_\rho f k^\rho)\, .
\eeq
If we combine this result with Eqn.~(\ref{equ:ft}) the resulting $\Lambda_{\rm II}^{(1,2)}$ acquires the
simple form
\beq
\Lambda^{(1,2)}_{\rm II} = \partial \bar \partial f \wedge \bar \partial k + \frac{1}{2} J \wedge \bar \partial(\partial_\rho f k^\rho)\, .
\eeq
We are now ready to calculate the holomorphic derivative $\partial \Lambda^{(1,2)}_{\rm II} $,
\beq
\partial \Lambda^{(1,2)}_{\rm II}  = \partial \bar \partial \left( f + \frac{1}{2} \partial_\rho f k^\rho \right) \wedge J\, .
\eeq
This does {\it not} vanish, so we need to introduce a $(2,1)$ piece
\beq
\Lambda^{(2,1)}_{\rm II}  = \partial \left( f + \frac{1}{2} \partial_\rho f k^\rho \right) \wedge J\, ,
\eeq
for which obviously $\partial \Lambda^{(2,1)}_{\rm II}  = 0$. We see that
\beq
\partial \Lambda^{(1,2)}_{\rm II}  +  \bar \partial \Lambda^{(2,1)}_{\rm II}   = 0\, .
\eeq
We have therefore shown that $\d \Lambda_{\rm II} = (\partial + \bar \partial) \Lambda_{\rm II} = 0$, where $\Lambda_{\rm II} \equiv  \Lambda^{(1,2)}_{\rm II}  +  \Lambda^{(2,1)}_{\rm II} $.
\hfill  $\blacksquare$
\vspace{0.2cm}  \hrule \vskip 1pt \hrule
 \vspace{0.5cm}

\normalsize

 \subsection*{Series III: Three-Form of Type $(1,2)+(2,1)_{\rm NP}+(3,0)$}

Our next
ansatz for the two-form in Eqn.~(\ref{equ:12}) uses two derivatives acting on a harmonic function $f$,
\beq
\label{equ:Pas3}
P_{(\alpha \sigma)} = \nabla_{\bar \zeta} \nabla_{ \bar \rho} f \omega_{\ \alpha}^{\bar \zeta} \omega_{\ \sigma}^{\bar \rho}\, .
\eeq
The resulting $(1,2)$ form,
\beq
\label{equ:s3-12}
\Lambda^{(1,2)}_{\rm III} = \bar \partial (\bar \partial f \cdot \omega) \wedge \bar \partial k \, ,
\eeq
is {\it not} closed, but the three-form becomes closed when appropriate $(2,1)$ and $(3,0)$ pieces are added,
\bea
\Lambda^{(2,1)}_{\rm III} &=& (\bar \partial h \cdot \omega) \wedge J\,, \\
\Lambda^{(3,0)}_{\rm III} &=& (2h + k^\xi \partial_\xi h) \Omega\, ,
\eea
where
$h \equiv 3 f + k^{\rho} \partial_\rho f$.

Our final closed, IASD three-form therefore is $\Lambda_{\rm III} \equiv \Lambda^{(1,2)}_{\rm III} + \Lambda^{(2,1)}_{\rm III} + \Lambda^{(3,0)}_{\rm III}$,
\beq
\label{equ:Lambda3}
\fbox{$\displaystyle
\Lambda_{\rm III} =(2h + k^\xi \partial_\xi h) \Omega +(\bar \partial h \cdot \omega) \wedge J + \bar \partial (\bar \partial f \cdot \omega) \wedge \bar \partial k$}\, .
\eeq

\vspace{0.5cm}
\small
 \hrule \vskip 1pt \hrule \vspace{0.3cm}
\noindent
{\sl The three-form $\Lambda_{\rm III}$ defined in Eqn.~(\ref{equ:Lambda3}) is closed.}
\newline
\noindent
{\bf Proof}:\newline
We aim to show that $\d \Lambda_{\rm III} = (\partial + \bar \partial) \Lambda_{\rm III} = 0$.

Contracting Eqn.~(\ref{equ:Pas3}) with $\bar \Omega$ we get
\beq
\Lambda_{\alpha \bar \beta \bar \gamma} = \omega_{\ \alpha}^{\bar \zeta} \left( \nabla_{\bar \zeta} \nabla_{\bar \beta} f k_{\bar \gamma} - \nabla_{\bar \zeta} \nabla_{\bar \gamma} f k_{\bar \beta} \right)\, .
\eeq
The derivative $\bar \partial \Lambda^{(1,2)}_{\rm III}$ vanishes because $\omega^{\bar \zeta}_{\ \alpha}$ and $k_{\bar \gamma}$ can be extracted from the differentiation and $\nabla_{\bar \delta} \nabla_{\bar \zeta} \nabla_{\bar \beta} f$ will vanish after
antisymmetrization with respect to $\bar \delta$ and $\bar \beta$ (see the discussion following Eqn.~(\ref{equ:B18})).

Next we consider $ \partial \Lambda^{(1,2)}_{\rm III}$. To simplify the expression we contract $(\partial \Lambda)_{\alpha \beta \bar \gamma \bar \delta}$ with $\bar \Omega^{\alpha \beta}_{\ \ \bar \rho}$
\beq
\label{equ:134}
\frac{1}{2} \bar \Omega^{\alpha \beta}_{\ \ \bar \rho} (\partial \Lambda)_{\alpha \beta \bar \gamma \bar \delta} = g^{\alpha \bar \alpha} \nabla_{\alpha} (\nabla_{\bar \alpha} \nabla_{\bar \gamma} f k_{\bar \delta} k_{\bar \rho}) - \nabla_{\alpha} (\nabla_{\bar \rho} \nabla_{\bar \gamma} f k^\alpha k_{\bar \delta}) - [\bar \gamma \leftrightarrow \bar \delta]\, .
\eeq
We expand the covariant derivatives and notice that many terms cancel.
The first term in Eqn.~(\ref{equ:134}) is
\beq
\label{equ:B44}
g^{\alpha \bar \alpha} \nabla_{\alpha} (\nabla_{\bar \alpha} \nabla_{\bar \gamma} f k_{\bar \delta} k_{\bar \rho})  = g^{\alpha \bar \alpha}  \nabla_\alpha \nabla_{\bar \alpha } \nabla_{\bar \gamma} f k_{\bar \delta} k_{\bar \rho} + \nabla_{\bar \delta} \nabla_{\bar \gamma} f k_{\bar \rho} + \nabla_{\bar \rho} \nabla_{\bar \gamma} f k_{\bar\delta}\, .
\eeq
The first term in Eqn.~(\ref{equ:B44})
vanishes because the metric is Ricci-flat and $f$ is harmonic. The second term vanishes after
antisymmetrization with respect to $\bar \gamma$ and $\bar \delta$, and only the third term
survives.
Similarly, the second term in Eqn.~(\ref{equ:134}) gives
\beq
\label{equ:B45}
\nabla_{\alpha} (\nabla_{\bar \rho} \nabla_{\bar \gamma} f k^\alpha k_{\bar \delta}) =
- k^\alpha \nabla_\alpha \nabla_{\bar \rho} \nabla_{\bar \gamma} f k_{\bar \delta} - 4 \nabla_{\bar \rho} \nabla_{\bar \gamma} f k_{\bar \delta}\, .
\eeq
The first term in Eqn.~(\ref{equ:B45}) can be simplified as follows:
we will use the fact that K\"ahler manifolds, for which
\beq
\partial_\alpha (\nabla_{\bar \beta} \nabla_{\bar \gamma} k) = 0\, ,
\eeq
have the property
\beq
k^\alpha R_{\alpha \bar \beta \gamma \bar \delta} = 0\, .
\eeq
Therefore, we see that $k^\alpha \nabla_\alpha$ and $\nabla_{\bar \beta}$ commute. Eqn.~(\ref{equ:B45})
then becomes
\beq
- \nabla_{\bar \rho} \nabla_{\bar \gamma} (\partial_\alpha f k^\alpha) k_{\bar \delta}  - 4 \nabla_{\bar \rho} \nabla_{\bar \gamma} f k_{\bar \delta}\, .
\eeq
Together with Eqn.~(\ref{equ:134}) this gives
\beq
\label{equ:B48}
\frac{1}{2} \bar \Omega^{\alpha \beta}_{\ \ \bar \rho} (\partial \Lambda^{(1,2)}_{\rm III})_{\alpha \beta \bar \gamma \bar \delta}   = - \nabla_{\bar \rho} \nabla_{\bar \gamma} (3f + \partial_\alpha f k^\alpha) k_{\bar \delta} + [\bar \gamma \leftrightarrow \bar \delta]\, .
\eeq
To cancel this term we introduce the $(2,1)$ form
\beq
\label{equ:L21}
\Lambda^{(2,1)}_{\alpha \beta \bar \gamma} = \partial_{\bar \zeta} h \left(\omega^{\bar \zeta}_{\ \alpha} g_{\beta \bar \gamma} - \omega^{\bar \zeta}_{\ \beta} g_{\alpha \bar \gamma} \right)\, ,
\eeq
for some harmonic function $h$. We get
\beq
\frac{1}{2} \bar \Omega^{\alpha \beta}_{\ \ \bar \rho} (\bar \partial \Lambda^{(2,1)}_{\rm III})_{\alpha \beta \bar \gamma \bar \delta}  =\nabla_{\bar \gamma} (\nabla_{\bar \rho}h k_{\bar \delta}) - \nabla_{\bar \gamma}(\nabla_{\bar \delta} h k_{\bar \rho}) -  [\bar \gamma \leftrightarrow \bar \delta]\, .
\eeq
The second term vanishes after
antisymmetrization so that we find
\beq
\frac{1}{2} \bar \Omega^{\alpha \beta}_{\ \ \bar \rho} (\bar \partial \Lambda^{(2,1)}_{\rm III})_{\alpha \beta \bar \gamma \bar \delta}  = \nabla_{\bar \rho} \nabla_{\bar \gamma} h k_{\bar \delta}  -  [\bar \gamma \leftrightarrow \bar \delta]\, .
\eeq
If we choose $h \equiv 3f+ \partial_\alpha f k^\alpha$ this cancels Eqn.~(\ref{equ:B48}).
However, the (2,1) form (\ref{equ:L21}) produces a non-trivial $(3,1)$ form after holomorphic differentiation,
\beq
T_{\bar \delta } \equiv \frac{1}{6} \bar \Omega^{\alpha \beta \gamma} (\partial  \Lambda^{(2,1)}_{\rm III})_{\alpha \beta  \gamma \bar \delta}  = g^{\alpha \bar \alpha } \nabla_\alpha (\nabla_{\bar \alpha} h k_{\bar \delta}) - \nabla_\alpha( k^\alpha \nabla_{\bar \delta} h)\, .
\eeq
Because $h$ is harmonic the result is simply
\beq
T_{\bar \delta } = - \nabla_{\bar \delta} (2h+ k^\alpha \partial_\alpha h)\, .
\eeq
To cancel this term we introduce the $(3,0)$ form
\beq
\Lambda^{(3,0)}_{\rm III} = (2 h + k^\alpha \partial_\alpha h) \Omega_{\alpha \beta \gamma}
\, .
\eeq
We have therefore shown that $\d \Lambda_{\rm III} = (\partial + \bar \partial) \Lambda_{\rm III} = 0$, where $\Lambda_{\rm III} \equiv \Lambda_{\rm III}^{(1,2)}+  \Lambda_{\rm III}^{(2,1)} +   \Lambda_{\rm III}^{(3,0)} $.
\hfill $\blacksquare$
\vspace{0.2cm}  \hrule \vskip 1pt \hrule
 \vspace{0.5cm}
\normalsize

\newpage
\section{Including Dilaton Gradients}
\label{sec:running}

We established in \S\ref{sec:10dSUGRA} that small perturbations of the dilaton and metric do not contribute to the leading-order flux-induced potential for a D3-brane.  However, many interesting compactifications arising from F-theory contain substantial dilaton gradients that cannot be treated perturbatively.
If D7-branes are the only source for these dilaton gradients\footnote{The IASD flux $G_{-}$ discussed in this section sources a running dilaton
if the combination $G_{+}\cdot G_{-}$ is non-zero, but this effect is subleading in our expansion scheme (see \S\ref{sec:perturbation}).} then the axio-dilaton will be holomorphic, so that the equation of motion (\ref{equ:flux0}) for the three-form flux reduces to
\beq
\label{equ:flux20}
\d \Lambda + \partial \phi \wedge (\Lambda + \bar \Lambda)=0 \, ,
\eeq
where the exterior derivative has been split into holomorphic and anti-holomorphic parts, $\d = \partial + \bar \partial$.
Moreover, the internal space will not be Ricci-flat, but will obey\footnote{In this appendix we set $\kappa_{10} \equiv 1$.}
\begin{equation}\label{Ricci}
R_{\alpha\bar\beta}= \partial_{\alpha}\phi \, \partial_{\bar\beta}\phi\, ,
\end{equation}
where $R_{\alpha\bar\beta}$ is the Ricci tensor of the internal space.  Finally, the $\Phi_-$ equation of motion takes the form
\begin{equation}
\label{equ:PhiEoM1}
 \nabla^2 \Phi_- \ =\ \frac{e^{\phi}}{96} \, |\Lambda|^2  \, ,
\end{equation} where we call attention to the non-constant prefactor $e^{\phi}$.

In a general compactification, it is challenging to determine the metric and dilaton in the presence of D7-brane sources ({\it{cf.}}~{\it e.g.}~Ref.~\cite{Burrington2005}).  However, we will now show that some of our considerations can be extended to D7-brane backgrounds {\it{without}} determining the metric and dilaton explicitly.  First, we will generalize the fluxes in Series~I to D7-brane backgrounds, solving Eqn.~(\ref{equ:flux20}).  Then, generalizing \S\ref{sec:global}, we will prove that the global supersymmetry interactions encoded in an arbitrary holomorphic superpotential $W(z_a)$ can be geometrized by these fluxes. That is, in the globally-supersymmetric theory arising on a D3-brane probing a noncompact cone containing D7-branes, for any specified superpotential $W(z_a)$ there is a solution of the ten-dimensional equations of motion in which
fluxes in Series~I give rise to this superpotential.  Let us remark that for warped cones with gauge theory duals, the AdS/CFT correspondence guarantees that
such a supergravity solution exists for any specified superpotential in the gauge theory; we will specify the fluxes in this solution, in terms of the metric and dilaton.

We begin by presenting a solution of the flux equation of motion (\ref{equ:flux20}) in the presence of non-constant dilaton perturbations, which we do not assume to be small.  Given any holomorphic function $f$, we turn on the $(1,2)$ flux
\begin{equation}\label{g12p}
\Lambda_{\alpha\bar\beta\bar\gamma}=  g_s\, e^{-\phi}\, \nabla_{\alpha} \nabla_{\sigma}f \,g^{\sigma\bar\rho}\,\bar\Omega_{\bar\rho\bar\beta\bar\gamma}  \,,
\end{equation} and the $(3,0)$ flux
\begin{equation}\label{g30p}
 \Lambda_{\alpha\beta\gamma}=  g_s \overline{\nabla_{\sigma}f \,g^{\sigma\bar\rho}\, \nabla_{\bar\rho}e^{-\phi}}\,\Omega_{\alpha\beta\gamma} \,.
\end{equation}
For constant dilaton, $\phi = const = - \ln g_s$, this reduces to the flux in Series~I, Eqn.~(\ref{equ:Lambda1A}).
For notational simplicity we now set $g_s \equiv 1$ (or absorb it into the function $f$).
One easily verifies that this combination of fluxes solves Eqn.~(\ref{equ:flux20}).

\vspace{0.5cm}
\small
 \hrule \vskip 1pt \hrule \vspace{0.3cm}
\noindent
{\sl The fluxes in Eqns.~(\ref{g12p}) and (\ref{g30p}) satisfy Eqn.~(\ref{equ:flux20}).}
\newline
\noindent
{\bf Proof}:\newline
We now explain how we constructed the fluxes in (\ref{g12p}) and (\ref{g30p}) that solve (\ref{equ:flux20}).
First we establish a few useful identities.
We assume that the background manifold is complex, the metric $g_{\alpha \bar \beta}$ is K\"ahler, the axio-dilaton $\tau = C + i e^{-\phi}$ is holomorphic and the metric is related to the dilaton through ({\it cf.}~Eqn.~(2.11) in Ref.~\cite{Burrington2005}),
\beq
\label{newg}
\det g_{\alpha \bar \beta} = q \bar q\, e^{-\phi}\, ,
\eeq
where $q$ is the same holomorphic function appearing in $\Omega_{\alpha\beta\gamma}=q\epsilon_{\alpha\beta\gamma}$. Because the metric is K\"ahler the following identity is satisfied
\beq
\label{KID}
\partial_\alpha \left( g^{\alpha \bar \beta} \det g \right) = \partial_\alpha \left(g^{\alpha \bar \beta} q e^{-\phi} \right) = 0\, .
\eeq
 (This identity ensures that $\nabla^2 = 2 g^{\alpha \bar \beta} \partial_\alpha \partial_{\bar \beta}$.)
Holomorphicity of $\tau$ implies the following identities
\bea
\partial_\alpha \tau &=& - 2 i \partial_\alpha \phi e^{-\phi} \, , \\
\partial_{\alpha} \partial_{\bar \beta} e^{-\phi} &=& 0\, , \label{phiID}\\
\partial_\alpha \partial_{\bar \beta} \phi &=& \partial_\alpha \phi \partial_{\bar \beta} \phi\, . \label{phiID2}
\eea
The flux equation of motion is then Eqn.~(\ref{equ:flux20}), and the expression (\ref{Ricci}) for the Ricci tensor follows from $R_{\alpha\bar \beta}=-\partial_{\alpha}\partial_{\bar \beta}\ln \det g$ and (\ref{newg}).  We are now ready to prove that Eqn.~(\ref{equ:flux20}) is solved by Eqns.~(\ref{g12p}) and (\ref{g30p}).

Our goal is to construct an IASD three-form $\Lambda$ that satisfies (\ref{equ:flux20}) and reduces to the IASD (1,2) form of Appendix~\ref{sec:IASD} for constant dilaton,
\beq
\label{equ:phi_const2}
\Lambda_{\alpha \bar \beta \bar \gamma}^{\phi=const} = \partial_{\alpha} \left( \partial_\sigma f g^{\sigma \bar \zeta} \bar q \epsilon_{\bar \zeta \bar \beta \bar \gamma} \right)\, .
\eeq
Clearly, this form is closed if the dilaton is constant (see Appendix~\ref{sec:IASD}).
Now we will try to modify (\ref{equ:phi_const2}) to account for a running dilaton. Let us first notice
that a (1,2) form will produce (3,1), (2,2) and (1,3) terms in Eqn.~(\ref{equ:flux20}). If we leave Eqn.~(\ref{equ:phi_const2}) unchanged, then $\partial \Lambda^{\phi=const} =0$ and the (2,2) term $\partial \phi \wedge \Lambda^{\phi=const}$ is not canceled.
With this in mind we change Eqn.~(\ref{equ:phi_const2}) by an overall factor of $e^{-\phi}$,
\beq
\label{equ:phi_run}
\Lambda_{\alpha \bar \beta \bar \gamma}^{(1,2)} = \partial_{\alpha} \left( \partial_\sigma f g^{\sigma \bar \zeta} \bar q \epsilon_{\bar \zeta \bar \beta \bar \gamma} \right) e^{-\phi}\, .
\eeq
Now the (2,2) piece $\partial \phi \wedge \Lambda^{(1,2)}$ is canceled by $\partial \Lambda^{(1,2)}$. However, the form in Eqn.~(\ref{equ:phi_run}) is not closed with respect to $\bar \partial$, leading to (1,3) and (3,1) terms in Eqn.~(\ref{equ:flux20}). To cancel these terms we introduce a (3,0) form
\beq
\Lambda_{\alpha \beta \gamma}^{(3,0)} = \psi \, q \epsilon_{\alpha \beta \gamma}\, ,
\eeq
with function  $\psi$ to be determined.
Letting $\Lambda \equiv \Lambda^{(1,2)} + \Lambda^{(3,0)}$, Eqn.~(\ref{equ:flux20}) implies two constraints
\bea
\partial_{\bar \alpha} \psi\, q + \partial_{\beta}\phi \partial_{\bar \alpha} \left[\, \overline{\partial_\sigma f g^{\sigma \bar \beta} \bar q}\, \right] e^{-\phi} &=& 0 \, , \label{constraint1}\\
-  \partial_{\bar \beta} \left[ \partial_\alpha (\partial_\sigma f g^{\sigma \bar \beta} \bar q) e^{-\phi}\right] + \partial_\alpha \phi \, \bar \psi\, \bar q &=& 0 \, . \label{constraint2}
\eea
Eqn.~(\ref{constraint1}) is easily solved if we notice that
\beq
 \partial_{\beta}\phi \partial_{\bar \alpha}\left[\, \overline{\partial_\sigma f g^{\sigma \bar \beta} \bar q}\, \right] e^{-\phi}= - \partial_{\beta} e^{-\phi}  \partial_{\bar \alpha} \left[ \, \overline{\partial_\sigma f g^{\sigma \bar \beta} \bar q}\, \right] =  \partial_{\bar\alpha} \left[\, \overline{\partial_\sigma f g^{\sigma \bar \zeta} \bar q} \partial_\zeta e^{-\phi} \, \right]\, ,
\eeq
where in the last step we used Eqn.~(\ref{phiID}). Thus
Eqn.~(\ref{constraint1}) is solved by
\beq
\label{j}
\psi = \overline{ \partial_\sigma f g^{\sigma \bar \beta} \partial_{\bar \beta} e^{-\phi}}\, .
\eeq
It turns out that Eqn.~(\ref{constraint2}) is also solved by (\ref{j}). To show this we rewrite the first term in (\ref{constraint2}),
\bea
 \partial_{\bar \beta} \left[ \partial_\alpha (\partial_\sigma f g^{\sigma \bar \beta} \bar q) e^{-\phi}\right]  &=& \partial_{\bar \beta} \partial_\alpha \left[ (\partial_\sigma f g^{\sigma \bar \beta} \bar q) e^{-\phi}\right] - \partial_{\bar \beta} \left[(\partial_\sigma f g^{\sigma \bar \beta} \bar q) \partial_\alpha e^{-\phi} \right] \, , \\
 &=& \partial_{\alpha} \partial_{\bar \beta} \left[ (\partial_\sigma f g^{\sigma \bar \beta} \bar q) e^{-\phi}\right] + \partial_{\bar \beta} \left[ (\partial_\sigma f g^{\sigma \bar \beta} \bar q) e^{-\phi} \partial_\alpha \phi \right]\, . \label{equ:next}
\eea
Using Eqn.~(\ref{KID}) we then notice that the first term in (\ref{equ:next}) vanishes,
\beq
\partial_{\bar \beta} \left[ (\partial_\sigma f g^{\sigma \bar \beta} \bar q) e^{-\phi}\right]  = 0\, ,
\eeq
if $f$ is harmonic.
The second term in (\ref{equ:next}) can be simplified with the help of (\ref{phiID2}),
\beq
\partial_{\bar \beta} \left[ \partial_\alpha (\partial_\sigma f g^{\sigma \bar \beta} \bar q) e^{-\phi}\right] = \partial_\sigma f g^{\sigma \bar \beta} \bar q e^{-\phi} \partial_{\bar \beta }\partial_\alpha \phi = \partial_\sigma f g^{\sigma \bar \beta} \bar q e^{-\phi} \partial_{\bar \beta} \phi \partial_\alpha \phi\, .
\eeq
This proves that (\ref{j})  solves (\ref{constraint2})  and hence, the fluxes in Eqns.~(\ref{g12p}) and (\ref{g30p}) satisfy the equation of motion Eqn.~(\ref{equ:flux20}) in the case of a running dilaton.
\hfill   $\blacksquare$
\vspace{0.2cm}  \hrule \vskip 1pt \hrule
 \vspace{0.5cm}

\normalsize

Next, we consider the potential of a probe D3-brane in the presence
of the above fluxes, specified by a holomorphic function $f$.   This flux is dual to
a perturbation of the superpotential $W\propto f$, and the corresponding potential should be
\begin{equation}\label{equ:Vr}
V=g^{\alpha\bar\beta}\nabla_\alpha W\overline{\nabla_\beta W} \, .
\end{equation}
Now we will show how Eqn.~(\ref{equ:Vr})
will arise from Eqn.~(\ref{equ:PhiEoM1}).
To verify the solution, we must show that
\begin{equation}\label{dilatongoal}
\nabla^2\, [\, g^{\alpha\bar\beta}\nabla_\alpha f \overline{\nabla_\beta f} \, ] =  \frac{1}{2\cdot 3!} e^{\phi} \, |\Lambda|^2  \, ,
\end{equation} with $\Lambda$ given by Eqns.~(\ref{g12p}) and (\ref{g30p}) and $W^2 =  \frac{T_3}{8}  f^2$.
First, we have
\begin{eqnarray}
\label{4in3}
\nabla^2\, [\, g^{\alpha\bar\beta}\nabla_\alpha  f \overline{\nabla_\beta f} \, ] = g^{\rho\bar\sigma}g^{\alpha\bar\beta}\left(2\nabla_{\rho}\nabla_{\alpha}f \overline {\nabla_{\sigma}\nabla_{\beta}f} +
2\nabla_{\bar \sigma}\nabla_\alpha f \nabla_{\rho}\nabla_{\bar \beta} \bar f\right)+\\ \nonumber +\ \
g^{\alpha\bar \beta}\nabla^2 \nabla_{\alpha}f \nabla_{\bar\beta} \bar{f} + g^{\alpha\bar \beta}\nabla_{\alpha}f \nabla^2 \nabla_{\bar\beta}\bar{f}
 \,.
\end{eqnarray}
The second term in (\ref{4in3})---$g^{\rho\bar\sigma}g^{\alpha\bar\beta}\nabla_{\bar \sigma}\nabla_\alpha f \nabla_{\rho}\nabla_{\bar \beta} \bar f$---vanishes because $f$ is holomorphic.
The first term---$g^{\rho\bar\sigma}g^{\alpha\bar\beta} \nabla_{\rho}\nabla_{\alpha}f \overline {\nabla_{\sigma}\nabla_{\beta}f}$---is equal to $\frac{1}{3!} e^{\phi}|\Lambda^{(1,2)}|^2$. We therefore just need to show that the last two terms combine into
\bea
\label{shouldbe}
\frac{2}{3!}e^{\phi}|\Lambda^{(3,0)}|^2= 2 |\partial_\alpha f g^{\alpha \bar \beta }\partial_{\bar \beta }\phi|^2\ .
\eea
To calculate $\nabla^2\nabla_\alpha f$ we use real notation,
\begin{equation}
\nabla^2\nabla_k f=g^{ij}\nabla_i\nabla_j \nabla_k f=g^{ij}(\nabla_k \nabla_i\nabla_j f-R^l_{jik}\nabla_l f)\ .
\end{equation}
The first term vanishes because $f$ is harmonic and the second is equal to  $R_{mk}g^{ml}\nabla_l f$.
Hence, using (\ref{Ricci}),we obtain
\begin{equation}
g^{\alpha\bar \beta}\nabla^2\nabla_\alpha f \overline{\nabla_{ \beta} f}+{\rm c.c.}=g^{\alpha\bar \beta}R_{\bar \sigma \alpha }g^{\bar \sigma \rho }\nabla_\rho f\overline{\nabla_{ \beta} f}\ +{\rm c.c.}=2|\partial_\alpha f g^{\alpha\bar \beta} \partial_{\bar \beta}\phi|^2\ .
\end{equation}
Comparing to (\ref{shouldbe}), we conclude that (\ref{dilatongoal}) is satisfied.

\newpage
\bibliographystyle{physics}
\bibliography{D3AdSCFT}

\end{document}